\documentclass[12pt]{article}
\usepackage{amssymb,amsmath,epsfig}
\begin{document}
\parindent=0pt
\parskip=6pt
\rm

\begin{center}

{\bf \Large Some basic aspects of quantum phase transitions}

\vspace{0.2cm}

{\large Diana V. Shopova$^{\ast}$, Dimo I. Uzunov$^{\dag}$}

\vspace{0.2cm}

{\em CPCM Laboratory, G. Nadjakov Institute of Solid State Physics,\\
Bulgarian Academy of Sciences, BG--1784 Sofia, Bulgaria}

\vspace{0.2cm}

$^{\ast}$ Corresponding author. {\em Email addresses:}\\
 sho@issp.bas.bg (D. V. Shopova), uzun@issp.bas.bg (D. I. Uzunov).

$^{\dag}$ Temporal address: Max-Planck-Institut f\"{u}r Physik komplexer
 Systeme,\\
N\"{o}thnitzer Str. 38, 0117 Dresden, Germany
\end{center}

{\bf Key words:}  Bose-Einstein condensation, critical phenomena, quantum
 effects,\\
 renormalization group, disorder.

{\bf PACS:} 05.70.Fh, 05.70.Jk, 64.60.-i

\begin{abstract}
 Several basic problems of the theory of quantum phase transitions are
 reviewed.
The effect of the quantum correlations on the phase transition properties is
considered with the help of basic models of  statistical physics. The effect of
quenched disorder on the quantum phase  transitions is also discussed.
The review is performed within the framework of the thermodynamic scaling
theory and by the most general methods of statistical physics for the treatment
of phase transitions:  general length-scale arguments,  exact solutions,
 mean field approximation, Hubbard-Stratonovich transformation,
 Feynman path integral approach, and renormalization group in the field
 theoretical variant. Some new ideas and results are presented. Outstanding
 theoretical problems are mentioned.
\end{abstract}

{\bf 1. Introduction}

{\em 1.1. Historical notes}

The experimental and theoretical research of quantum phase transitions and
quantum critical phenomena is a branch of the statistical physics of a rapidly
growing importance for the explanation of essential features  of
low dimensional fermion and spin systems, dilute Bose fluids,
superconductors, quantum Hall systems,
ferroelectrics~\cite{UZ:1993, BK:1994,Son:1997, Sch:1999,
DeCU:1999, Sach:1999, UZ:2000, Cont:2001}.

The quantum statistics has a substantial influence on the critical
behaviour near various (multi)critical points of low temperature and
zero temperature continuous phase transitions as well as
on thermodynamic and correlation properties near equilibrium points
of low-tempetarure first order phase transitions~\cite{LP:1980,
Huang:1987,Land:1980}.
This involves the quantum statistical physics in the field of a particular
type of phase transitions - the quantum phase transitions. The thermodynamics
and the correlation phenomena near such phase transitions are
essentially affected by quantum mechanical effects.

It seems convenient to begin this review with brief historical notes.
 Although the interest in
 quantum effects on the critical behaviour
in many-body systems dated from the dawn of the quantum statistical
 physics, the theory of quantum phase transitions and, in particular, the
 theory of continuous quantum phase transitions
 (alias, quantum critical phenomema)
 received a substantial development after
 the late sixties of the previous century owing to the application of
 fruitful ideas of
 universality and scaling by renormalization group (RG) methods; see, e.g.,
 Refs.~\cite{UZ:1993, Ma:1976, Bin:1993, Stanley:1971, MEF:1967, PV:2002}.
 The contemporary theory  of these
 phenomena has two distinct periods of
 development.

The first, classical period, began in the seventies of the
 previous century with the pioneering
papers by Pfeuty and Elliott~\cite{Pf:1971}, Rechester~\cite{Rech:1971},
Young~\cite{Young:1975},
and the particularly important work of Hertz~\cite{Hertz:1976}.  This period of
a relatively quiet research based on quantum field and statistical methods,
including RG, continued  till 1986-1987 when the number
of papers on quantum critical phenomena
has abruptly increased due to the great interest in
heavy-fermion~\cite{Cont:2001}
and high-temperature superconductors~\cite{Schn:2000},
low-dimensional magnetic systems~\cite{Sach:1999},
quantum Hall effect~\cite{Son:1997}, metal-insulator transition
problems~\cite{BK:1994,Lee:1985}, and dilute Bose gases and Bose fluids
~\cite{UZ:1993, DeCU:1999, Wei:1986, FishW:1989}.

The new period of
research~\cite{BK:1994,Son:1997,Sch:1999, DeCU:1999, Sach:1999,Cont:2001}
after 1986-1987 seems to be much  more intensive
but one should not forget that this renewed research in
an extended area of problems and their applications relies very much on
the classical results obtained during the first period.
We shall partly consider
some classical results together with related developments and applications
accomplished up to now.

{\em 1.2. Scope and aims of the review}

Numerous real systems, for example, magnets, ferroelectrics, quantum Hall
systems, dilute Bose gases, low-dimensional Josephson-junction arrays,
heavy-fermion and high-temperature superconductors
exhibit phase transition lines which extend to low and up to zero-temperature
phase transition points, namely,
to temperatures of a strong quantum statistical degeneration
\cite{UZ:1993, BK:1994,Son:1997, Sch:1999, DeCU:1999, Sach:1999, UZ:2000,
Cont:2001, Schn:2000, Lee:1985, Wei:1986,FishW:1989}.
In this temperature region the quantum statistical correlations can,
in certain cases and
under certain experimental conditions, produce observable effects on the low-
and zero-temperature phase transition.  The task of the theory is to
predict and describe the quantum effect on the thermodynamic
and correlation properties near such phase transitions.
This difficult task remains among the outstanding problems
of the quantum statistical physics, despite the remarkable success achieved in
this field of research during the last 30-35 years.

We shall not expand our discussion over the great variety of low-temperature
phase transitions.  Rather we shall focus our attention on several aspects of
continuous and,
mainly, second order quantum phase transitions, i.e.  quantum critical
phenomena.  For our aims we shall use general thermodynamic arguments and
esults for basic and relatively simple statistical models.

Despite of the fact that in real systems at low temperatures
the first order phase transitions occur more frequently,
 quantum phase transitions of first order
 are less investigated and
 few results are available for them. Perhaps, this is due to the fact that the
 ordering phenomena and thermal fluctuations
 at equilibrium points of discontinuous (first-order) phase transitions
 have finite length scales and the scaling methods cannot be applied.
 That is why the quantum fluctuations at
first order phase transitions should have a stronger effect on the phase
transition properties than at points of continuous phase transitions.
 This interesting topic is also beyond the scope of the paper.

 Our attention will be focussed on the quantum behaviour
 of Bose gases which reveals the main features of
 the quantum phase transitions in various systems. Furthermore,
 our consideration is closely related to the problem of Bose-Einstein
 condensation (BEC) of  noninteracting bosons (ideal Bose gas,
 shortly, IBG) and the superfluidity of interacting bosons (nonideal Bose gas;
 NBG)~\cite{LP:1980,Huang:1987,Land:1980}.
We shall be particularly interested in the properties of usual and
disordered Bose fluids
in a close vicinity of the $\lambda$-point. General problems of quantum phase
 transitions are also reviewed. Along with this review we shall present
 several new aspects of the quantum phase transitions. The new ideas and
 results will be summarized in our concluding remarks.

In Sec.~2 we perform a phenomenological investigation of
 several topics concerning the change (crossover) of the usual
classical (``high-temperature") critical behaviour when
 the critical temperature is lowered to the
 range of the quantum degeneration up to zero. The
lowering of the critical temperature is referred to as ``low-temperature"
 and ``zero-temperature" limiting  cases of the critical behaviour.
Numerous experimental and theoretical studies indicate
that a rather nontrivial and general ``high-low" temperature crossover (HLTC)
exists, namely, that the high-temperature and low-temperature
 critical phenomena are quite different
 from each other. The first fundamental question is whether this
 HLTC is a result of
 low-temperature and zero-temperature limits themselves or is produced by
 quantum effects.

The problem can be solved provided both the universal and nonuniversal
 properties of the low-temperature critical behaviour are thoroughly
 investigated. Note, that some particular properties of the system
 or special experimental conditions can quell the quantum correlations and
then
 the low-temperature critical behaviour will remain totally or partially
classical up to
the  zero temperature. Therefore, the quantum critical behaviour is a
particular case
 of low-temperature and zero-temperature critical behaviour. The classical-to-
quantum dimensional
 crossover (CQC)~\cite{Pf:1971, Young:1975, Hertz:1976} is also discussed as
 a particular case of HLTC. For our aims in Sec.~2 we apply a new scaling
scheme for quantum systems~\cite{DeCU:1999}
 which is an alternative of the standard quantum scaling, developed for the
first time in
 Ref.~\cite{Wei:1986} (see, also,
Refs.~\cite{Sach:1999,Cont:2001,Schn:2000,FishW:1989}).

 In Sec.~3 IBG at constant density and constant pressure is considered.
 The general methods for the investigation of systems of interacting bosons are
reviewed in Sec.~4. In Sec.~5 the main RG results for NBG are discussed.
In Sec.~6 the transverse Ising model (TIM) is considered with the help of the
 mean-field (MF) approximation, the consideration of the Ginzburg critical
region~\cite{UZ:1993,Land:1980}, and RG. The results in Secs.~3, 5 and 6
 confirm the scaling description outlined in Sec.~2. Disorder effects
 are discussed in Sec.~7. In Sec.~8, the notion about the break down of
 universality for quantum critical phenomena and a classification of these
 phenomena based on this notion are established.
 The concluding remarks are presented in Sec.~9. We presume that the reader is
 well acquainted with the basis of the scaling theory of phase transitions and
  RG~\cite{UZ:1993,Ma:1976,Bin:1993,Stanley:1971, MEF:1967}.

There is an overwhelming amount of papers published in this field and we
 cannot enumerate and discuss all of them within a manageable length of this
 review. We have tried to present the most important original papers,
comprehensive review articles and books.

тодор люб

{\bf 2. Phenomenology and quantum phase transitions}

{\em 2.1. General criterion for quantum effects on phase transition}

The quantum effects (alias, quantum correlations, often called also
 quantum fluctuations and even quantum pseudo-interactions) due to the overlap
of particles wave functions exert an influence on the thermodynamic and
correlation properties near
 the phase transition points in many body systems provided the
 de Broglie thermal wavelength
\begin{equation}
\lambda (T) \; = \; \left ( \frac{2\pi \hbar^{2}}{mk_{B}T} \right  )^{\theta} ,
\end{equation}
is greater than the correlation length
\begin{equation}
\xi(T) \; = \; \xi_{0}(T_{c})|t(T)|^{- \nu}\;
\end{equation}
of the thermal fluctuations:
\begin{equation}
\varrho \;\; = \;\; \frac{\lambda}{\xi}\;\; > \;\; 1 \;.
\end{equation}
In Eqs.~(1)~-~(2),
\begin{equation}
0 \leq |t(T)| = \frac{|T-T_c|}{T_c} \ll 1\:,
\end{equation}
defines a broad vicinity of the (multi)critical point $T_c$, in which the phase
transition phenomena occur, $\xi_0 \equiv \xi(T_c)$ is the so-called
``zero temperature correlation length,'' the critical exponent $\nu > 0$
 describes
the behaviour of the correlation length $\xi(T)$ in the phase transiton
region~(4), $\theta > 0$ is the exponent for the thermal wavelength.
 We have to note that usually, $\theta = 1/2$~\cite{Huang:1987} but in order to
comprise all  possible quantum statistical models we consider $\theta > 0$;
 see, e.g., ~\cite{UZ:1993}. In Eq.~(1), the parameter $m$ denotes either
 the real particles mass or an effective mass
 $m_{\scriptsize \mbox{eff}} \sim \hbar^2/c(g)$ of composite bosons (boson
excitations) which represents the effect
 of some interaction constant $g$; below the suffix ``eff" will be omitted
(see also Secs. 3.2 and 4.1).

We must keep in mind that the correlation length $\xi(T)$ describes only
 classical phenonena. The length $\xi_0$ is
 a nonuniversal quantity and can be specified only for concrete systems
 (models)  which means that it cannot be presented in a concrete
 mathematical form before the choice of the concrete model (Hamiltonian) is
done and,  in particular cases, as for example IBG, after the thermodynamic
 analysis of the system is fulfilled.
 That is why,  at this general stage of consideration we do not give a
 mathematical formula for $\xi_0$. But we must emphasize that $\xi_0$
 depends on intrinsic parameters of the system and, especially, on the
 critical temperature $T_c$. The term ``zero-temperature correlation length,''
 that is sometimes used for the scaling amplitude $\xi_0$ of $\xi(T)$ is
 an indication of the simple fact that $t(0)=1$ and, hence,
 $\xi(t=1) =\xi_0$.

It will be shown in Sec.~4.3 that $\theta = 1/z$, where $z$ is the
so-called dynamical critical exponent which describes the intrinsic (quantum)
dynamics of the system~\cite{Hertz:1976}.
It can be stated as a theorem that the relation
$\theta = 1/z$ is valid always when the dynamics of the quantum
system~\cite{Hertz:1976} is not
influenced by other time-dependent phenomena~\cite{Ma:1976, HH:1977}.

It has been already mentioned that $T_c$ is a (multi)critical temperature,
i.e., it denotes the phase transition point of quantum or
classical phase transitions of second and higher order, i.e.,
continuous phase transitions.  For the first order phase transitions, however,
Eqs.~(2)~and~(4) can be also used. In this case $T_c$ is a characteristic
temperature of the system near
the equilibrium phase transition temperature
$T_{eq} \neq T_c$, $(|T-T_{eq}|/T_{eq}) = t_{eq}(T) \sim |t| \ll 1$.  The phase
transition phenomena, including the metastable states are located
in the temperarute domain
$|t_{eq}(T) \sim t_{eq}(T_c)|$~\cite{UZ:1993}.  The infinitesimally small
vicinity $[(t(T) \rightarrow 0]$ of critical points $T_c$ of continuous phase
transitions is often referred to as an ``asymptotic critical region" and the
behaviour in this small domain is usually called an ``asymptotic critical
behaviour."

Phase transitions which satisfy the criterion~(3) in the vicinity of their
equilibrium phase transition points $T_c$ are called {\em quantum phase
transitions}.  This criterion has been introduced for
the first time by M.  Suzuki
~\cite{Suz:1976} for quantum critical phenomena (phenomena at
continuous
quantum phase transitions) and here we find reasonable to extend
it to all quantum phase transitions. The condition~(3) is very similar
to the respective one for quantum fluids, for them there is a requirement for
the thermal length $\lambda$ to exceed the
mean interparticle distance (the lattice constant in crystal bodies):
 $a = \rho^{1/d} =
(N/V)^{1/d}$~\cite{LP:1980,Huang:1987,Hal:1965}; $N$ is the particle number,
$V = (L_1...L_d)$ is the volume of
 the system, and $d \geq 0$ is the spatial dimensionality.

The criterion~(3) is a direct result of the fundamental notion that the phase
transition is a (quasi)macroscopic phenomenon and, hence, phenomena at length
scales shorter than the length scale $\xi$ of the classical fluctuations are
irrelevant~\cite{Suz:1976}.  It is confirmed by the available
studies of quantum statistical models. The criterion~(3) yields
both  the necessary and the sufficient conditions for the appearance of a
quantum
phase transition but the circumstances, under which this can actually happen
 remain hidden in its generality.

It seems at first glance that the condition~(3) can be more easily fulfilled
for first order transitions, where $\xi(T_{eq}) < \infty$. However, the answer
of the question whether the
inequality~(3) can be satisfied even in this case
depends on the values of the parameters $\xi_0$ and
$\lambda_0 = (2\pi\hbar^2/mk_B)^{\theta} = \lambda T^{\theta}$ for the
concrete substance of interest. The same problem exists for the continuous
phase transitions.  Hereafter we shall discuss continuous
phase transitions, where $\xi(T)$ is always infinite.

{\em 2.2. Quantum critical region}

The vicinity~(4) of the critical point
$T_{c}$, where the phase transition phenomena occur and the
Landau expansion of Gibbs free energy in powers
of the fluctuation order parameter $\phi(\vec{r})$ is valid,
includes both the MF domain of description and the Ginzburg
critical region of strong fluctuations very close to
$T_{c}$~\cite{UZ:1993,Land:1980}. This
vicinity of $T_c$, $ \xi(T) \gg \xi_{0}(T_{c})$, where the
 classical (thermal) fluctuations and ordering phenomena occur,
hereforth will be called ``classical transition region'' or,
 shortly, ``transition region''.
  For some types of phase transitions
the transition region may be defined by $|t(T)| < 1$ rather than by the
strong inequality~(4).

While the correlation length $\xi(T)$ does not describe the MF domain of
 ordering and, moreover, this quantity cannot be defined within the standard
 MF approximation, the inequality $\xi(T) \gg \xi_0(T)$, equivalent to
 $t(T) \ll 1$, does include the temperature domain of the phase transition
 phenomena and can be used in our discussion.
We shall distinguish between the Ginzburg critical
 region~\cite{UZ:1993,Land:1980} and the much larger classical transition
 region defined above.
The subdomain of the transition region, where the quantum fluctuations
 have an essential influence
on the phase transition properties will be called a ``quantum
transition (sub)domain." At a next stage of the present consideration, this
 picture will be generalised to include additional thermodynamic parameters.

From Eq.~(1), the
condition~(3) and $\xi_0(T_c)\geq a_0$ -- a condition which is always satisfied
in real systems, we see that a quantum critical phenomenon
may occur only in the regime of quantum degeneration of the system
($\lambda > a$), i.e., near low- and zero-temperature critical points,
where conditions (3) and (4) are simultaneously satisfied.  Here we accept as
a definition that we shall use the term "quantum critical phenomenon" to
indicate the fact that the quantum effects do penetrate in the transition
 region $|t| \ll 1$.

Quantum critical phenomena, that is, quantum effects in the
``transition region", may occur
in condensed matter systems and gases, where the critical line
$T_{c}(X)$ extends over the low-temperature region to the zero-temperature
critical point $T_{c}(X_{0}) = 0$ by variations of an additional
thermodynamic parameter $X$. Depending on the particular
system, the parameter $X$ may represent some auxiliary
intrinsic interactions which suppress the main interaction
responsible for the phase
transition, or other physical quantities (density, pressure, concentration
of impurities, etc.) having the same effect; see Fig.~1, where
such critical lines are depicted for $X_{0} > 0$ and $X_{0} = 0$.  The
slope of the curve $T_c(X)$ and the fact that it is tilted to the left
are irrelevant to the present discussion. In certain systems, the critical
line $T_c(X)$ is tilted to the right and then the maximal value of $T_c$ will
 be at some $X > X_0$ (see Sec. 6.2).

\begin{figure}
\begin{center}
\epsfig{file=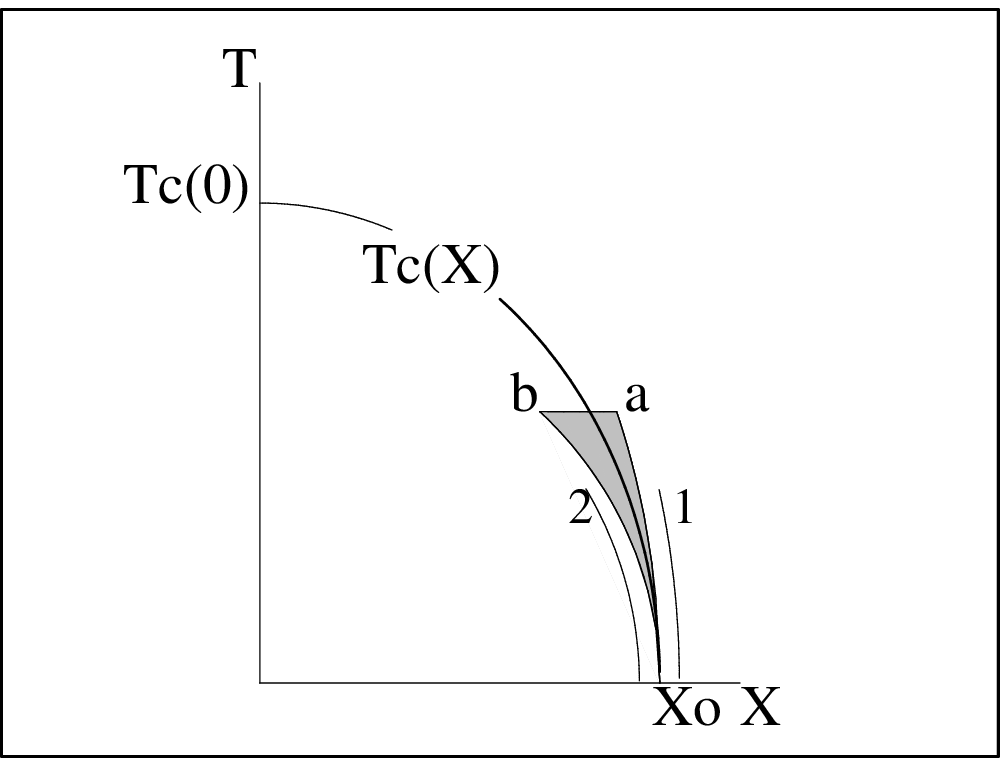, width=7.5cm}
\epsfig{file=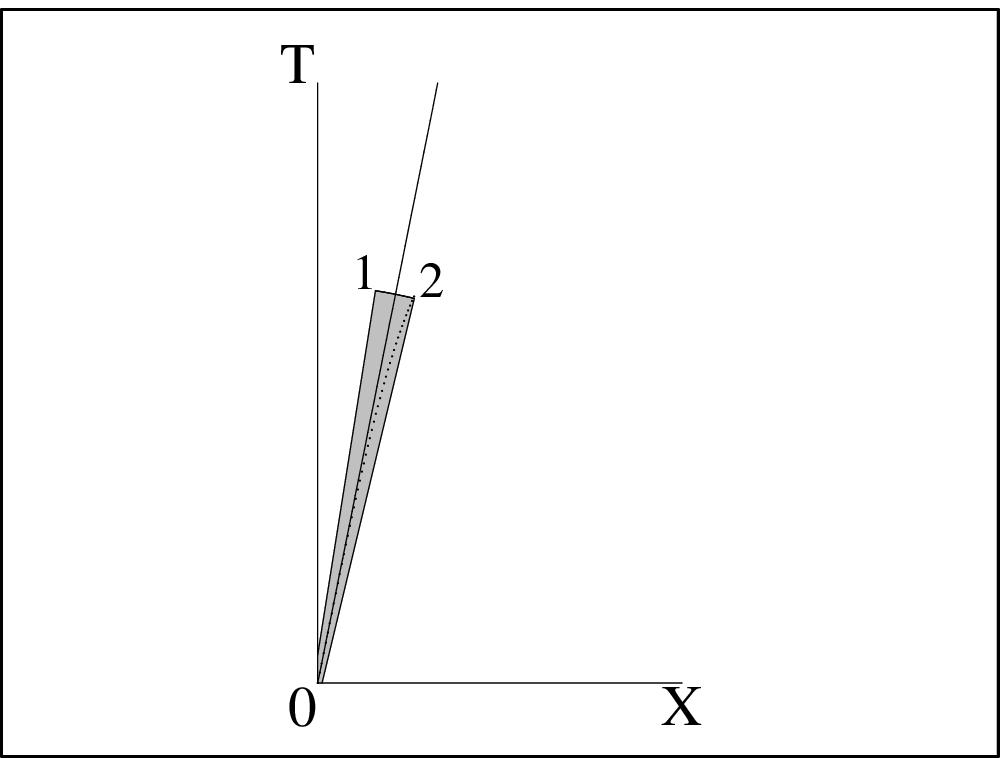, width=7.5cm}\\
(a)\hspace{7.5cm} (b)
\end{center}

\caption{(a) Low temperature part of a critical line with zero temperature
critical point $T_c(X_0) = 0$.  The shaded part ($a-X_0-b$) of the transition
region, marked by the lines 1 and 2, corresponds to a low temperature classical
behaviour.  (b) Low temperature critical line with $X_0 = 0$;
domains ($a-X_0-b$) and $1-0-2$ coincide.}
 \label{QPTf1a.fig}
\end{figure}

We should keep in mind that Eqs.~(2) and (4) describe only the temperature
dependence of the characteristic length $\xi$.  In general, the correlation
length $\xi$
depends on the parameter $X$ as well:  $\xi = \xi(T,X)$.  The dependence of
thermodynamic and correlation quantities on $X$ must be
considered on the same footing as the dependence on $T$.  The reason is
that, as seen from Fig.~1, the phase transitions at $T_{c}(X)$ or,
equivalently, at $X_{c}(T)$, can occur both
by variations of $T$ at fixed $X$ ({\em $T$--driven transitions})
and by variations
of $X$ at fixed $T$ ({\em $X$--driven transition}).
Phase transitions produced by
both $T$-- and $X$--variations are also possible. We suppose that their main
properties can be clarified by the investigation of $T$-- and
 $X$--transitions.

These notes are important
for the evaluation of that part of the transition region where the quantum
effects penetrate, i.e., in the quantum transition region. The inequality~(4)
 defines the transition region of the critical phenomena
produced by $T$--transitions and, hence, the temperature transition
width $[T_{>}(X) - T_{<}(X)]$ along the $T$--axis
 satisfies the condition $[T_{>}(X) - T_{<}(X)] \ll T_c(X)$ for
 each $X = X_c$; here, $T_>(X) > T_<(X)$. As $T_c(X) \rightarrow 0$ for
 $X \rightarrow X_0$,
 the temperature transition width vanishes at the point $(T,X) = (0,X_0)$.
The shape of this temperature transition region is similar to the form of
the shaded domains $a$ - $X_0$ - $b$ and 1-0-2 in Figs.~1a,b.

However, the transition region along the $X$--axis is produced by
 $X$--transitions and this effect should be considered as well.
That is why the notion for the transition region should be generalized to
include the $X-$driven phenomena, too. At any given $T=T_c$, $\xi(T_c,X)
 \rightarrow \infty $ for $X \rightarrow X_c$ because of the general
 criticality assumption: $\xi(T_c,X_c) = \infty$. Therefore, one can represent
 $\xi(T_c,X)$ in a general scaling form, namely,
 $\xi(T_c,X) = \tilde{\xi}_0\tilde{t}^{-\tilde{\nu}}$, where
$\tilde{t}(X) = |X-X_c|/X_c$ is the $X-$distance from $X_c$,
 $\tilde{\nu} \geq 0$ is the respective critical exponent, and
$\tilde{\xi}_0 \equiv \tilde{\xi}_0(X_c) = \xi[T_c, \tilde{t}(0)] > 0$
 is the scaling amplitude. Now one can define the $X-$transition interval
$[X_{>}(T) - X_{<}(T)] \ll X_c(T)$ for each $T = T_c$; here
 $X_{>}(T) > X_{<}(T)$. This interval will
vanish, only if $X_c(T) = 0$ at some critical temperature $(T=T_c)$.
 The latter case is illustrated in Fig.~1a by the high-temperature
 critical point $(T_c, 0)$, and in Fig.~1b, where $X_c(0) = 0$.

The total transition region $(T,X)_{\mbox{\footnotesize t}}$ in the vicinity
 of the critical line $T_c(X)$ includes both $T-$transition
 and $X-$transition widths and can be represented as
 a two-dimensional domain of shape similar to the form of the domain
 confined between the lines 1 and 2 in Figs.~1a,b. The picture
 of the transition region $(T,X)_t$ outlined above will be valid provided the
 scaling amplitude $\xi_0(T_c)$ is finite for all $T_c \geq 0$. Remember that
 $(T,X)_t$ will be a classical transition region till the quantum effects are
ignored in our consideration.

The next question is whether the quantum correlations,
 if properly taken into account, will affect the
 thermodynamic properties in the transition region $(T,X)_t$. When the
 transition region $(T,X)_t$ vanishes, as this is the case for some
 zero temperature critical points, the quantum fluctuations will fill up
 the whole nearest vicinity of the critical point and then we have an entirely
 quantum phase transition. The problem is nontrivial when the classical region
$(T,X)_t$ exists.

 There are three scenarios:

 (i)The quantum effects are relevant
 in the whole region $(T,X)_t$. Then the latter is a quantum
 transition region, where quantum phase transition phenomena occur.

(ii) The quantum effects are relevant in a part (quantum subdomain)
 of the total transition region. This will be enough to include the phase
transition in the quantum ones.

(iii) The quantum effects are irrelevant anywhere in $(T,X)_t$ and, hence,
the phase transition is completely classical.

The variants (i)-(iii) are shown in Fig.~1a,b. The shaded domain
 $a$ - $X_0$ - $b$ in Fig.~1a represents the classical subdomain of the total
transition region $(T,X)_t$, where the classical (thermal) fluctuations
 dominate and the phase transition is completely classical. The unshaded
 domain between the lines 1 and 2 is the quantum transition subdomain.
 Here one can find both cases (i)-(ii) depending
 on whether $ T>0$ or $T = 0$ and, whether one is interested in $T-$ or
 $X-$driven phase transitions. In Fig.~1b the classical domain
 $a$ - $X_0$ - $b$ coincides with the total
transition domain 1-0-2 which is an illustration of the classical variant (i).

In Secs.~2.3-2.4 we shall investigate the variants (i)-(iii)
 with the help of general scaling arguments. For a methodical convenience,
we shall discuss the $T-$ transition region along the $T$--axis and the
 $X$--transition region along the $X$-axis separately, although they are
 a result of one and the same reason -- thermal and quantum fluctuation and
 ordering phenomena.

{\em 2.3. $T-$driven transitions}

According to Eqs.~(1)~and~(2), $\xi \rightarrow \infty$ for
$T \rightarrow T_{c} > 0$,
but $\lambda$ remains finite.  Therefore,
the close vicinity of finite temperature critical points $T_{c} > 0$
will always exhibit a classical behaviour. In particular, this is true for the
so--called asymptotic critical behaviour corresponding to
the infinitesimally narrow distance ($|t| \rightarrow 0$) from $T_{c}$.

Using the criterion~(3)
as well as Eqs.~(1) and~(2), it is easy to show that quantum critical
phenomena
will occur, i.e.,
the quantum effects will penetrate in the (temperature) critical
region
$|t(T)| \ll 1$, if
 \begin{equation}
T_{c} (\xi_{0}/\lambda)^{1/\nu} \; <  \; |(T - T_{c})| \; \ll \;T_{c} \; .
\end{equation}
The conditions~(5) are well defined for $ T_{c} > 0$, including
the zero-temperature
limit $ T_{c} \rightarrow 0$, provided $ T \rightarrow 0$ too,
so that $|t| < 1$ is satisfied. The size of the quantum subdomain
$(\Delta T)_Q = |T^\ast - T_c|$ above and below $T_c$ is given by
$(\Delta T)_Q = T_c^{(1 + \theta/\nu)} (\xi_0/\lambda_0)^{1/\nu}$, where
$\lambda_0 = \lambda(T)T^{\theta}$. The size of this temperature interval
 depends on the behaviour of the function $\xi_0(T_c)$.

 Obviously, the quantum portion~(5) of the
transition region will exist, if
 \begin{equation}
\xi_{0}(T_{c}) \; <  \; \lambda (T),
\end{equation}
which corresponds to a moderate $(\lambda > a)$ or strong
$(\lambda \gg a)$ quantum degeneration for $(\xi_{0} > a)$ and
$(\xi_{0} \gg a)$, respectively. The condition (6) can be written in the form
$T^{\theta} < [\lambda_0/\xi_0(T_c)]$.

The inequality~(6) may bring more information about
the quantum criticality corresponding to $T$--driven transitions,
if we remember that within the present description ($|t| \ll 1$ or $|t|< 1$),
we can freely substitute
$T$ with $T_{c}$ in all formulae except that for $t(T)$.  Conversely,
the substitution of $T_{c}$ with $T$ will be also allowed in the whole
transition
domain~(4), if the nominator of $t(T)$ is not affected.  Besides, the latter
variant of the theory seems to be more closely connected with the original
 form of the free energy~\cite{UZ:1993,Land:1980}, which is derived by
 standard statistical methods from the microscopic Hamiltonian of the
 system (see also Sec. 4).

The further consideration depends on the way, in which we shall treat the
nonuniversal length $\xi_{0}(T_c) > a$. When $T_{c}$ is decreased to some
extent, depending on the specific properties of a given system,
$\xi_{0}(T_c) $ grows and this is described by the relation
\begin{equation}
\xi_{0}(T_{c}) \; =  \; \xi_{\footnotesize 00}T_{c}^{- \nu_{0}}\;,
\end{equation}
where $ a < \xi_{\footnotesize 00} < \infty $ is a new (low-temperature)
scaling amplitude and
$ 0 \leq \nu_{0} \leq \infty $ is a new (low-temperature) exponent;
$ \nu_{0}$ must be nonnegative number
 because of the criticality at low $T_{c}$.

The value of exponent $\nu_0$ and the range of temperatures, where the
low-temperature dependence~(7)
will be valid, can be evaluated
from the properties of system.  The quantities
$\xi_{\footnotesize 00} > a$ and $\nu_0 > 0$
appear as a result of a gradual low-temperature crossover in the critical
behaviour owing to $T_{c}$ lowering.

The temperature, $T$,
is not sufficient for the description of the low-temperature
crossover and the entire investigation
can be performed in terms of both relevant thermodynamic parameters $T$
and $X$.  The low temperature critical behaviour limit should be given by
the ratio
$[T_{c}(X)/T_{c}(0)] \ll 1 $ or, equivalently, by the ratio
$[g(X)/g_0] \ll 1$, where $g_0 = g(0)$ is the interaction responsible
for the $T$--driven transition and $g(X)$ describes the effective decrease
of this interaction owing to the auxiliary parameter $X$; in certain cases,
$g(X) \sim T_c(X)$.  These remarks,
involving the term ``interaction" in our phenomenological analysis, are not
restricted to systems of interacting particles only; see Sec.~3, where
the phase transition is produced rather by the Bose statistics and global
thermodynamic
constraints than by direct interparticle interactions.

Having in mind these notes we continue our analysis of the condition~(6)
by substituting
$\lambda (T)$ with $\lambda (T_{c})$ and $\xi_{0}(T_{c})$ with the scaling
form~(7):
\begin{equation}
\frac{\xi_{00}}{\lambda_{0}} \; <  \; T_{c}^{ \nu_{0} - \theta}\;.
\end{equation}
When the critical temperature
is lowered enough, the condition~(8) will be broken unless
\begin{equation}
\theta \;  >  \; \nu_{0}\;.
\end{equation}
If the inequality (9) is satisfied, the quantum critical phenomena
will occur at low-temperatures in certain part (5)
of the temperature
interval (4) and, moreover, for $T_{c} \rightarrow 0$, they will prevail
in the asymptotic critical behaviour $[t(T) \rightarrow 0]$, too.

If the criterion(9) is fulfilled, the classical fluctuations will
 be completely
irrelevant to the critical behaviour at $T_{c} = 0$.  Their effect at
$T_{c} \sim 0$ (extremely low-temperature critical points) will be restricted
in a negligibly narrow vicinity of $T_{c}$, which is practically
inaccessible to experiments. In both cases, $T_{c} \sim 0$ and $T_{c} = 0$,
the experiment should observe quantum critical phenomena only.
For $\theta = \nu_{0}$,
the quantum correlations will have an effect on the
pre-asymptotic low-temperature critical behaviour outside the small distance
$T_{c}(\xi_{\footnotesize 00}/ \lambda_{0})^{1/ \nu}$ from $T_{c}$, provided
$(\xi_{\footnotesize 00}/\lambda_{0}) <1 $, i.e., in case of convenient
nonuniversal
properties of the particular system.  Quantum critical phenomena
will not exist at all, if $\theta < \nu_{0}$.  In this case the critical
behaviour ramains classical.

The same results can be obtained by an alternative consideration
which is consistent with our discussion about the possible
change of factors $T$ with $T_{c}$, and vice versa. The scaling
law~(2) has the following alternative definition:
 \begin{equation}
\xi(T) \; = \; \bar{\xi}_{0}(T)|\bar{t}(T)|^{- \nu}\;,
\end{equation}
where $\bar{t}(T) = (T - T_{c})/T$. The quantity $\bar{\xi}_0 (T)$
 defined by Eq.~(10) can be called ``zero$-T_c$ correlation length'';
 $\bar{t}(T_c =0) = 1$. In the high-temperature and,
to some extent, in the low temperature range of critical
temperatures $T_{c}$,
Eq.~(10) will give the same leading scaling dependence as
Eq.~(2) because the corrections to the scaling are of
the order $O(|t|^{1 - \nu})$. In the limit $T \rightarrow 0$ however
the quantity $\bar{t}(T)$ is not singular, whereas the correlation length
$\xi(T)$ should tend to infinity at $ T_c=0$. This means that
the decreasing of the temperature will unavoidably yield a HLTC. The latter
 consists in a gradual temperature-driven transformation of
 the $\bar{t(T)}-$singularity in Eq.~(10) to a new singularity --
the divergence of the amplitude $\bar{\xi}_0(T)$ at $T_c = 0$.

Let us suppose
that $\bar{\xi}_{0}(T)$ obeys a scaling law with respect
to $T$ having the form~(7).  Then
one can immediately rederive a condition for
temperature $T$ which will be identical to the condition~(8) for
$T_{c}$ and, hence, the criterion~(9) will be straightforwardly confirmed.
Note, that the new (low temperature) singulatiry may arise with respect to
the vasiable $\tilde{t}(X)$ rather than towards $T$ (see an example in
 Sec.~6.2).

The singularity~(7) is a mere indication for an existence
of a scaling law for $\xi(T)$ of the type
$\xi(T) \sim \bar{\xi}_0(T) =\bar{\xi}_{00}/T^{\nu_{0}}$ corresponding
to zero critical temperatures $(T_{c} = 0)$. In the low-temperature
limiting case, the critical behaviour singularities are developed as
singularities of the scaling amplitude while at high temperatures
this scaling amplitude
is a slow varying function of temperature. It is worth noting that
the present phenomenological consideration, including
the ansatz~(7) for $\xi_0(T_c)$ and the related one for $\xi_0(T)$, is
 supported by results from RG calculations for particular
systems~\cite{Car:1997} and experiments on ferroelectric phase
transitions~\cite{Sam:1981, Sam:1988}.

{\em 2.4. $\boldmath X$--driven transitions}

In general, the properties of $X-$driven transitions are quite
different from those of ~$T-$driven transitions. In particular,
  essential differences may be expected in
the high-temperature $( X \sim 0)$ and low-temperature
$( X \sim X_{0})$ ranges of temperatures.
By substituting $T$ with $X$ and, of
course, $T_{c}$ with $X_{c}$ in~(2)
and~(4), one can perform the phenomenological analysis (Sec.~2.3) in
terms of  the variable
$X$. This analysis directly yields a form of scaling law for
the correlation length $\xi(T,X)$ which is
identical to that related with $T$--transitions
in the high-temperature region where $X \sim 0$.
The only difference may come from the values
of critical exponents which describe the scaling laws with respect
to $t(T)$ and $\tilde{t}(X)$. There is no reason to suppose that the exponents
towards $t(T)$ and $\tilde{t}(X)$ should be equal.

In fact, the critical behaviour
in the low-temperature limit described in Sec.~2.3 has a similarity
with high-temperature $X$--transitions because $X_{c} \sim 0$ at high
temperatures. A conformity of type $T \leftrightarrow X $ concerning the
form of
scaling laws and the way, in which the low-temperature
limit (for $T$--transitions) and the
high-temperature limit (for $X$--transitions) are developed from
the corresponding scaling
amplitudes certainly exists. However, except for particular systems, a total
correspondence including the values
of respective critical exponents and scaling amplitudes cannot
be expected; note, that the length $\lambda (T)$ has no analog in terms
of $X$. In contrast, there is no such similarity between the
low-temperature $T$-- and $X$--transitions.

For $X$--transitions, the criterion~(3) gives
 \begin{equation}
|\tilde{t}(X)| \; > \; \left [\frac{\tilde{\xi}_{0}(X_{c})}{\lambda (T)}
 \right ]^{1/\nu},
\end{equation}
where all quantities are defined by the change of $T$ with $X$
in~(2) and~(4).  Having in mind Eq.~(1) and the fact that the critical region
(lines 1 and 2 in Fig.~1) along the
$X$--axis is defined by $|\tilde{t}(X)| \ll 1$, it becomes evident,
that in the low-temperature range $(X \sim X_{0})$, where
$[\tilde{\xi}_{0}(X_c)/\lambda(T) < 1]$,
the quantum subdomain gradually enlarges when the temperature is decreased
or, equivalently, when $X_{c}(T)$ approaches $X_{0}$. In the zero-temperature
limit
$(T \rightarrow 0)$ or, equivalently, for $X_{c}(T) \rightarrow X_{0}$,
this subdomain fills up totally the transition region.

In the range of temperatures, where
$[\tilde{\xi}_{0}(X_c)/\lambda(T)] > 1$, the critical
 behaviour is totally classical.
We should have in mind that the latter condition for a total classical
criticality can be easily satisfied in the temperature range of quantum
degeneration in real systems with a large zero-temperature
correlation length
$(\xi_{0} \gg a)$.

At high temperatures, where $X_{c} \sim 0$,
the scaling of correlation length $\xi$ can be conveniently investigated
by a scaling law with respect to $X$ or $(X - X_{c})$, instead of
$\tilde{t}(X)$.  The
problems in the treatment of $X$--transitions in the high-temperature
limit are
analogous to those for the low-temperature $T$--transitions and the analysis
can be carried out by following the ideas presented in Sec.~2.3.

A quite
special situation should exist when $X_{0} = 0$; see Fig.~1b. In this case
both $T$ and $X$ are equal to zero at the zero-temperature critical point,
where the correlation length $\xi(T,X)$ may exhibit scaling
dependence on both $t(T)$ and $\tilde{t}(X)$.
Obviously, such zero-temperature critical points will offer less  opportunity
for  the quantum critical phenomena observation or, in some cases, this
opportunity
 may not exist at all. To emphasize that the lack of any quantum critical
phenomenon may be expected at such zero-temperature
 critical points, we have depicted in Fig.~1b a classical transition
region which completely fills up the total transition region
$(T,X)_{t}$, confined between the lines 1 and 2.

However, there exist systems, where the critical temperature $T_c$ depends
 on $X$  and tends to zero
for $X\rightarrow X_0 = 0 $ but the correlation length $\xi$ does not show
any divergence with respect to the
parameter $X$. In this case the function $\xi(T,X)$, for all possible $T$ and
$X$,  has the general form $\xi[t(T), T_c(X))]$,
which describes only the scaling law with respect to $t(T)$ as
given by Eq.~(2) or Eq.~(10). The transition region width along
the $T$--axis tends to zero for
$T_c \rightarrow 0$ and a transition region with respect to the parameter
$X$ does not exist at all. Thus the lines
1 and 2 in Fig.~1b will definitely terminate at $T=X=0$; see the dashed line
0--2.  Certainly, the $X$--transitions at low temperatures offer
much more favourable conditions
for quantum critical phenomema, although the practical observation
of such transitions
in certain systems, where the parameter $X$ cannot be gradually
varied, is almost impossible.

We end the discussion of the total and  quantum transition regions
 by two remarks. In some systems at extremely low and zero temperature,
 the quantum transition region might be larger than the total classical region.
Then quantum critical phenomena may happen outside the classical transition
 region. In this case, quantum critical phenomena may occur even if the size
 of the classical transition region tends to zero (the case depicted in
 Fig.1b). The second remark is about the possibility for a further
 generalization of this consideration by assuming that the symbol $X$
 represents more than one thermodynamic parameters: $X = (X_1,...)$.
 This generalization is straightforward.

{\em 2.5. Crossovers}

We have shown that the low-temperature critical behaviour can be either
quantum or classical depending on the specific (nonuniversal) properties of
a particular system.
Besides, there are no general arguments
indicating that high-temperature and low temperature critical properties
should be equivalent. Thus we may suppose that, in general, classical
high-temperature classical low-temperature and quantum critical properties
present three different types  of critical behaviour.
These three types of
critical phenomena can be included in the framework of a quite general notion
for a ``high-low temperature crossover'' (HLTC) of the critical behaviour
which corresponds to the change of
the critical properties (or of some of them) when the thermal length $\lambda$
varies from $\lambda < a$ at high temperatures to $\lambda > a$ (up to
$\lambda \gg a)$ at sufficiently low temperatures,
and vice versa.  A special case of HLTC is the classical $(\varrho < 1)$ to
quantum $(\varrho > 1)$ crossover (CQC) which describes the difference between
 classical high-temperature phenomena and
the quantum critical phenomena at $T \rightarrow 0$~
\cite{Pf:1971,Young:1975,Hertz:1976}.  This CQC is treated by statistical
methods (Secs.~4-7).

The experiment has a finite accuracy and omits very narrow temperature
intervals such as the asymptotic classical regions in the close vicinity of
``almost--zero-temperature" critical points $(T_{c} \sim 0)$. This should
be taken into account in interpretations of experiments. Because of the zero
 temperature unattainability, quantum critical phenomena produced by
zero-temperature
$X$--transitions cannot be observed in a real experiment but certain
low-temperature $X-$transitions might be experimentally investigated.
On account of limitations in the accuracy of the
equilibrium temperature measurement, the very narrow classical region
will remain unattainable for experimental studies and, hence, the experiment
will reproduce results which, to some extent, will give an information
about asymptotic quantum critical phenomena at zero temperature.

{\bf 3. Ideal Bose gas}

{\em 3.1. Preliminary notes}

The model of IBG describes the phenomenon of Bose--Einstein condensation (BEC)
~\cite{UZ:1993,Land:1980,Huang:1987,Land:1980}. BEC of IBG
is important for the undestanding of quantum phenomena in
many areas of physics, in particular, in superfluid
helium liquids~\cite{LP:1980}, excitonic phases in
semiconductors~\cite{Halp:1968}, the recently discovered
BEC of dilute Bose gases in magneto-optical traps~
\cite{Kett:1996, Toun:1997, CW:2002, Dal:1999, Legg:2001}
although in these and other real
systems the interparticle interactions do affect the
thermodynamic and correlation properties.

We find reasonable
to emphasize that the IBG does not contain interparticle interactions but it
possesses two other properties which may cause a phase transition.
Firstly, the quantum-statistical correlations act as attractive
``pseudo-interactions." This can be seen
from the negative sign of the first quantum correction
to the equation of state of the ideal classical (Boltzman) gas coming from
the Bose statistics~\cite{Huang:1987,Land:1980}.  Secondly, the
constraints imposed on the system usually have an effect similar to that of
some interaction. Let us remember that the thermodynamics of IBG is
ruled by constraints (a constant density, or, the less
common constraint of a constant pressure and, why not, a constraint of constant
temperature, which has been never investigated).  These constraints are a
simple experimental requirement and are used to define the chemical potential
as a function of the temperature within the framework of the grand canonical
 ensemble~\cite{Huang:1987, Land:1980}.

Besides, we have to emphasize that the ground states of IBG and NBG
are quite different in their physical properties.  While IBG produces the
originally predicted by A. Einstein BEC, a very small interaction is needed
to induce a superfluid ground state in NBG
as demonstrated theoretically in a rigorous way
by H.  H.  Bogoliubov~\cite{Bog:1947} (see, also, Ref.~\cite{LP:1980}).

Another problem is the description of the phase transitions in Bose gases:
from one side, the phase transition to BEC in IBG at finite and zero
temperatures and, from the other side, the so-called $\lambda-$transition,
i.e.,  the phase transition to a superfluid state in NBG.  These phase
transitions
exhibit quite different properties.  Here we shall review the results
from Refs.~\cite{GB:1968,CG:1968,Lac:1974,Busi:1985}, where BEC in IBG
has been investigated within the framework of the scaling theory of phase
 transitions~\cite{UZ:1993,Ma:1976,Bin:1993}.

{\em 3.2. Thermodynamic equations}

The thermodynamic properties of $d$--dimensional IBG, including BEC,
 are described by equations for the grand canonical
potential $\Omega$ and the number density $\rho = (N/V)$ of bosons
~\cite{LP:1980, Huang:1987}:
\begin{equation}
\Omega \; = \; k_B T \sum_{\vec k} \mbox{ln} \left [ 1 -
 e^{-\beta \varepsilon (k)} \right] \;,
\end{equation}
and
\begin{equation}
\rho \; = \; \frac1 V \sum_{\vec k} \langle a^+_{\vec{k}} a_{\vec{k}} \rangle
\; ,
\end{equation}
respectively.Hete the brackets $\langle \rangle$ denote a statistical
averaging,
$\beta = 1/k_B T$, $a^+_{\vec{k}}$ and $a_{\vec{k}}$ are the creation
and annihilation Bose operators for a plane--wave state of wave vector
$\vec{k} = (k_i;\;i=1, \ldots, d)$,
$V = (L_1 \ldots L_d)$ is the volume of the gas.  The self--energy
$\varepsilon(k) = \varepsilon_0(k) + r$ is represented by
the energy spectrum
$\varepsilon_0(k)
= \hbar^2 k^2 / 2m$ of free (noninteracting) real particles or
excitations, and the chemical potential $\mu = -r \le 0$; $k = |\vec{k}|$.

Throughout the paper we shall use periodic boundary conditions.  The wave
vector components
$k_i =2\pi l_i / L_i$, $(l_i = 0, \pm 1, \ldots $), are given by the spatial
dimensions $L_i$.  They are supposed
to be much larger than any characteristic length of the system, for
example, $L_i \gg (\xi, \lambda)$.  In this case it is usually said that the
dimensions $L_i$ are ``infinite".  The stated condition allows to pass to the
continuum
limit, i.e., from a summation in Eqs.~(12) and~(13) to the corresponding
integration.  The quite uncommon case of composite Bose exitations
due to long--range interactions in
electron and magnetic systems can be included in the consideration by the
generalization $\varepsilon_0(k) = \hbar^2 k^{\sigma} / 2m, (0<\sigma<2)$
of energy spectrum $\varepsilon_0(k)$; see, e.g., Refs.~
\cite{UZ:1993,Jo:1972, UZ:1996}.  A simple dimensional
analysis of the exponent $\beta\varepsilon(k)$ in Eq.~(12) shows that
 $\xi = (\hbar^2/2mr)^{1/\sigma}$.

With the help of the Bose distribution
\begin{equation}
n(k) \; = \; \langle a^+_{\vec{k}} a_{\vec{k}}\rangle \; = \;
 \frac1 {e^{\beta \varepsilon (k)} - 1}\;,
\end{equation}
Eq.~(13) will take the form
\begin{equation}
\rho \; = \; \frac1 V \sum_{\vec k} \frac1 {e^{\beta \varepsilon (k)} - 1}\;.
\end{equation}
The correct investigation of IBG thermodynamics in the continuum limit implies
to take the density $\rho_0 = (N_0/V) = n(0)$ of bosons with zero wave numbers
out the sum~(15). This is important for obtaining a correct description of
the properties below finite temperature critical points $T_c > 0$.
For all other cases this separation is redundant.  As we shall be mainly
interested in the latter case, we shall avoid the mentioned separation;
 see Ref.~\cite{Busi:1985}.

Substituting the $\vec{k}$--summation in Eqs.~(12) and (13) by a
$d$--dimensional integration,
\begin{equation}
\frac{1}{V}\sum_{\vec{k}} \; \rightarrow \; \int \frac{d^d k}{(2\pi)^d} \;
 \equiv \; \int\limits^{\infty}_{0}\;dk\:k^{d-1}\;,
\end{equation}
$K_d = 2^{1-d}/\pi^{d/2}\Gamma(d/2)$, we obtain
\begin{equation}
P \; = \; k_B TA \lambda^{-d}g_{(d/\sigma + 1)}\left ( \frac{r}{k_B T} \right )
\end{equation}
and
\begin{equation}
\rho \; = \; A \lambda^{-d}g_{(d/\sigma)}\left ( \frac{r}{k_B T} \right ) \;.
\end{equation}
\noindent Here $P = -(\Omega/V)$ is the pressure,
$\lambda = (2\pi \hbar^2/mk_B T)^{1/\sigma}$ is the thermal wavelength
with an exponent $\theta = 1/\sigma$, c.f.  Eq.~(1). The function
$g_{\nu}(\kappa)$ and the parameter $A$ which enter in above equations are
given by the expressions,
 \begin{equation}
g_{\nu}(\kappa) \; = \; \frac{1}{\Gamma(\nu)} \int\limits^{\infty}_{0} \;
\frac{x^{\nu - 1} dx}{e^{x + \kappa} - 1}\;,
\end{equation}
and
\begin{equation}
 A \; = \; \frac{2^{1 - d + 2d/\sigma}\Gamma(d/\sigma)}
{\sigma \pi^{d(1/2 - 1/\sigma)}\Gamma(d/2)} \;,
\end{equation}
(for $\sigma = 2, A = 1$).

The BE condensate is a coherent state of a macroscopic number
$N_0 \sim N$ of bosons with a momentum $\hbar k = 0$.  The critical
temperature $T_c$ of the transition to BEC in the momentum
($\vec{k}-$) space is
defined by the equation $[r(T_c)/T_c] = 0$.  The BEC order parameter
$\phi_0 = \langle a^+_0\rangle/\sqrt{V}$
is related to the square root of the number
density $\rho_0 = (N_0/V)$ of the condensate bosons:
$|\phi_0| = \sqrt{\rho_0}$.
In the present problem, the ratio~(3) takes the form
\begin{equation}
\varrho \; =  \; \left (  \frac{4\pi r}{k_B T}\right )^{1/\sigma}.
 \end{equation}
The IBG critical properties can be investigated with the help of
Eqs.~(17) and (18). They depend on the thermodynamic conditions
imposed on IBG, i.e., on the way, in which the chemical potential $\mu = -r$
is determined.

We shall briefly consider three cases:  a constant pressure $P$
~\cite{Lac:1974}, a constant density $\rho$ and spatial
dimensions
$d > \sigma$~\cite{GB:1968,CG:1968}, and a constant
density at dimensions $d < \sigma$~\cite{Busi:1985}.  In all cases the
critical regime is defined by the condition $r \leq k_B T$ and all critical
phenomena are almost $( r \sim k_BT)$ or completely $( r \ll k_BT)$ classical;
 c.f. Eq.~(21).

{\em 3.3. Constant pressure}

The parameter $r$ is calculated from Eq.~(17) as a function of $T$ and $P$.
The result
$\mu(T,P)$ is substituted in Eq.~(18) and, hence, one obtains the equation of
state $f(T,\rho,P) = 0$.  The solution $T_c(P)$ of Eq.~(17) with $r=0$ yields
the critical temperature
\begin{equation}
T_c(P) \;  = \; \left [ \frac{\lambda_0^d P}{\zeta(d/\sigma + 1)Ak_B }
\right ]^{\sigma/(d + \sigma)}\;.
\end{equation}
A characteristic feature of the
phase transition at constant $P$ is that the critical temperature is
finite $(T_c > 0)$ for all dimensions $d > 0$ and $P> 0$.  The zero temperature
critical point exists only in the limit $P \rightarrow 0$.  This phase
transition does not exhibit a crossover to a low-temperature
behaviour when the pressure $P$
tends to zero because the critical exponents remain unchanged.  There is
a crossover, however, from Gaussian exponents at dimensions $d > \sigma$ to
low--dimensional $(d < \sigma)$ Gaussian exponents.  This crossover cannot be
 considered as HLTC or CQC.

{\em 3.4. Constant density}

At a constant density $\rho = (N / V)$, the equilibrium chemical
potential $\mu (T, \rho)$ is obtained as a solution of Eq.~(18).
The equation of state is given by Eq.~(17) after the substitution of
solution $r(T, \rho)$.  Eq.~(18) with $r=0$ yields
\begin{equation}
T_c(\rho) \;  = \; \left [ \frac{\lambda_0^d \rho}{\zeta(d/\sigma)A }
\right ]^{\sigma/d}
\end{equation}
and, therefore, $T_c(\rho) > 0$ for $d > \sigma$, provided $\rho > 0$, whereas
$T_c = 0$ for $d \leq \sigma$; note, that the zeta function $\zeta(d/\sigma)$
tends to infinity when the ratio $d/\sigma$ decreases to unity.

The finite temperature critical behaviour $(T_c > 0)$ is identical
to that of the classical Berlin-Kac~\cite{Berlin:1952} spherical
model~\cite{GB:1968,CG:1968} (for the spherical model
see, e.g., Ref.~\cite{Jo:1972}, and the brief note in Sec.~4.1).
The reason for this behaviour is in the constant
density $\rho$ condition which, as is seen from Eq.~(13), is equivalent to
the(mean) spherical constraint on the variations of mean densities
$\langle a^{+}_{\vec{k}}a_{\vec{k}} \rangle /V$.  The
critical exponents are shown in Table 1, including the case
$d>2\sigma$ when they take Gaussian values~\cite{UZ:1993,Ma:1976}.
All critical exponents are given by their usual notations and
definitions; see, e.g., Refs.~\cite{Ma:1976, MEF:1967}.

{\bf Table 1.} Values of critical exponents ($\rho$ = const).\\

\footnotesize

\begin{tabular}{cccccccc}
\hline
$T_c$ & $d/\sigma$ & $\eta$ & $\alpha$ & $\alpha_s$ & $\gamma$ &
$\tilde{\gamma}$ & $\nu$ \\ \hline
$T_c>0$ & $d\geq 2\sigma$ & $2-\sigma$ & 0 & $(2\sigma-d)/\sigma$ & 1 & -- &
$1/\sigma$ \\
\hline
$T_c>0$ & $\sigma<d<2\sigma$ & $2-\sigma$ & 0 & $(d-2\sigma)/(d-\sigma)$
&
$\sigma/(d-\sigma)$ &--& $1/(d-\sigma)$\\
\hline
$T_c=0$ & 1 & $2-\sigma$ & -1 & -- & $\infty$ & $\infty$ & $\infty$
\\
\hline
$T_c=0$ & $0<d<\sigma$ & $2-\sigma$ & $ -d/\sigma$ & -- &
$\sigma/(\sigma-d)$ & $d/(\sigma - d)$ & $1/(\sigma-d)$\\
\hline
\end{tabular}

\vspace{0.3cm}

\normalsize

The critical exponent $\gamma$ describes the off--diagonal susceptibility
\begin{equation}
 \chi \; = \; -\left ( \frac{\partial \Omega}{\partial h \partial h^*}
 \right )_{h = 0}
 \; \sim \; (1/r) \;,
\end{equation}
where $h$ is a complex number -- a fictitious external field
conjugate
to the Bose operator ($\sum_{\vec{k}} a^+_{\vec{k}}$).  For $T_c = 0$, a new
critical exponent $\tilde{\gamma}$ can be introduced to describe the
density $n(0) \approx k_B T \chi$.  For $d<\sigma$, the values of $\gamma$
from Table~1 and the
relation $\tilde{\gamma} = (\gamma - 1)$ yield: $\tilde{\gamma} =
d(\sigma - d)$ for $d<\sigma$, and $\tilde{\gamma} = \infty$ for $d = \sigma$.
The exponent
$\tilde{\gamma}$ does not exist for $T_c>0$, where the density $n(0) \sim \chi$
is described by the usual exponent $\gamma$.

In Table~1, the exponent $\alpha_s$ is an auxiliary exponent intended to give a
more detailed description of IBG specific heat for $d>\sigma$.
For $d>\sigma$, the specific heat $C(T)$ of IBG and spherical
model can be represented in the form
\begin{equation}
C(T) = C_{0} + C^{\prime} (T - T_c)^{-\alpha_s}\;,
\end{equation}
where $C_0$ is a regular constant.  This behaviour corresponds to the zero
value of the usual specific heat exponent $\alpha $, defined by the scaling law
$C(T) = C^{(0)}|T-T_c|^{-\alpha}$,
where $C^{(0)}$ is the scaling amplitude.  Usually several shapes of $C(T)$
are usually ascribed to the value $\alpha = 0$, namely, a logarithmic
divergence
(ln$|T_c-T|/T_c$), a finite jump as is in the MF theory, and cusps of different
forms.  In order to avoid this arbitrariness, Fisher~\cite{MEF:1967}
introduced the second term in Eq.~(25).
The values of $\alpha_s$ in the Table~1 are
negative and this next--to--leading term in $C(T)$ is always small but the
derivative $dC(T)/dT$ is divergent at $T_c$ and this is an information about
 the cusp shape at $T_c$.

The essential difference between the properties of the phase transition to BEC
in IBG and the usual second order phase transitions can be easily
seen by the comparison of the Landau free energy~\cite{UZ:1993,Land:1980}
$F(\varphi) = Vf(\varphi)$ of a standard second order phase transition,
$f = (r_0\varphi^2 + \varphi^4)$
with the free energy density $f(\varphi)$ of the Bose gas at constant
density~\cite{CG:1968}, given by $f = (r_0 + \varphi^2)^3$, where
 $r_0 \sim t(T)$.
 The simple form of these free energies corresponds
 to the suitable choice of units for $f$ and the order parameter $\varphi$.
 The functions
$f(\varphi)$ are shown in Figs.~2a and 2b
 for several values of the parameter $r_0$. While the well defined
minima of $f$ in Fig.~2a describe the ordered phase of usual second order
phase transitions BEC is described by the inflection points
$(\partial f/\partial \varphi = \partial^2 f/\partial \varphi^2 = 0,\;
\partial^3 f/\partial \varphi^3 > 0)$
 of the $f-$curves
shown in Fig.~(2b). At these inflection points the $f-$curves intersect
 the $\varphi-$axis;
the regions where $f<0$, see Fig.~(2b),
correspond to a positive value of the chemical potential and, hence, are
 unphysical~\cite{CG:1968}.

\begin{figure}
\begin{center}
\epsfig{file=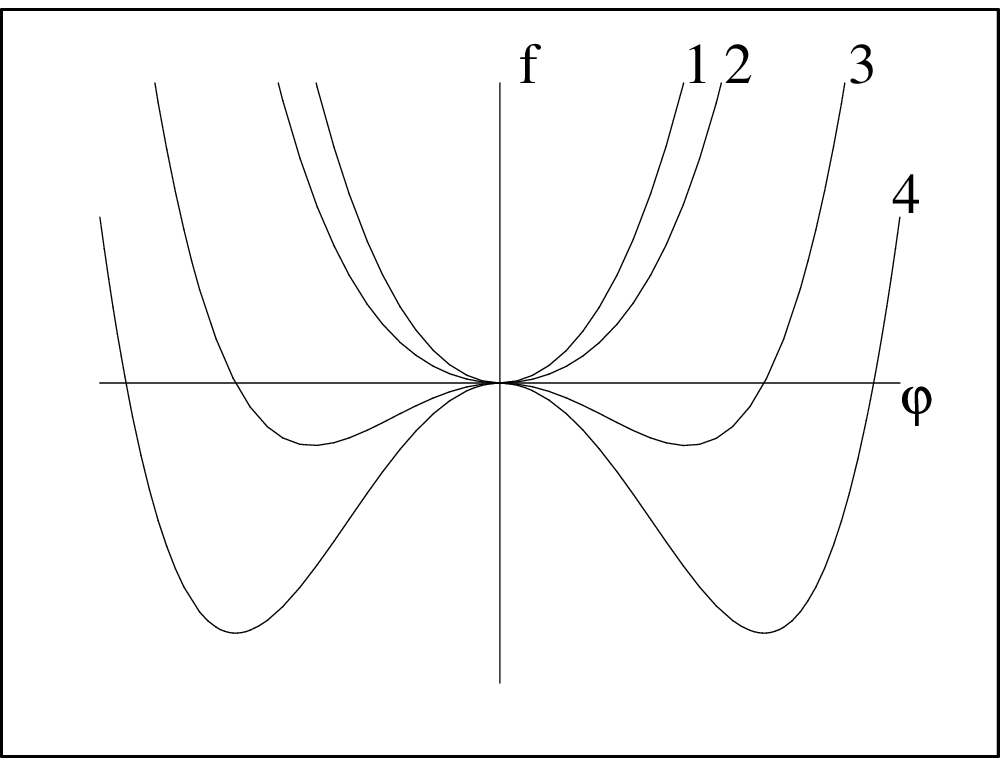, width=7.5cm}
\epsfig{file=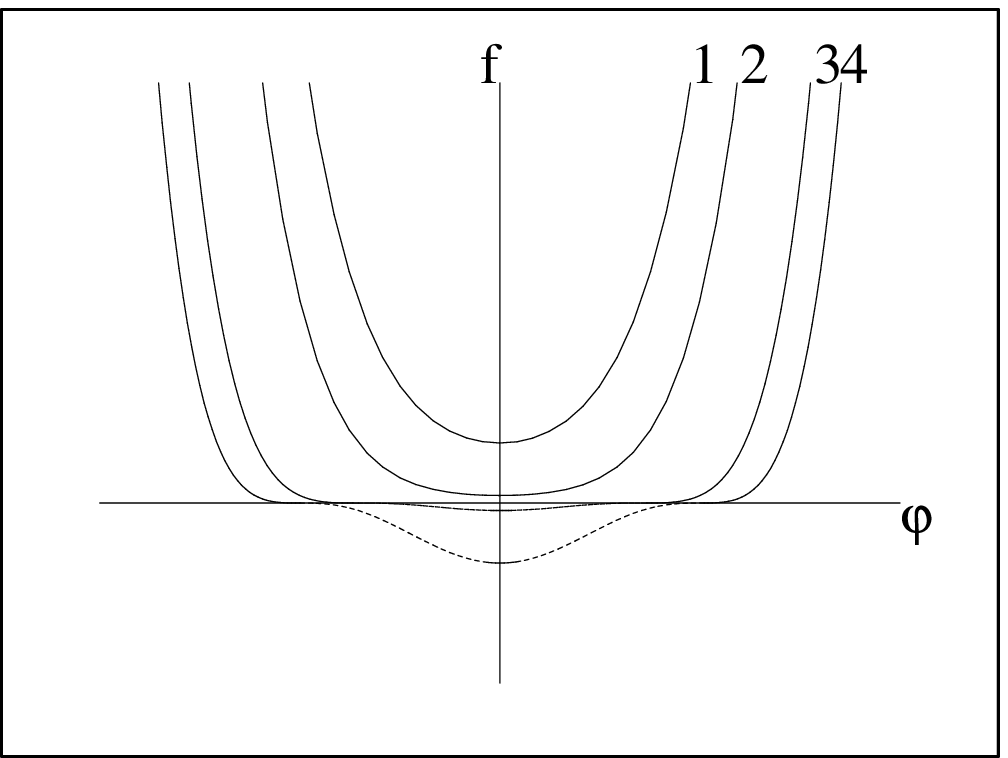, width=7.5cm}\\
(a)      \hspace{7.5cm}             (b)
\end{center}

\caption{(a) The function $f(\varphi)$ for a standard second order phase
transition for: $r_0 = 1.0$ (curve 1), $r_0 = 0.5$ (2), $r_0 = -0.5$ (3),
 $r_0 = -1.0$ (4). (b) The function $f(\varphi)$ for the IBG at constant density
(the curves 1-4 correspond to the same values of $r_0$ as given for Fig.2a).}
 \label{QPTf2.fig}
\end{figure}

Certainly, the properties of IBG
below the condensation point $r_0 < 0$ also correspond to
a continuous phase transition which is quite different from the standard
 second order transition. Note, that the coefficient of the
$\varphi^4-$term in the free energy density of the Bose gas,
 $f = (r_0 + \varphi^2)^3$, is proportional to $t$
and tends to zero when $t\rightarrow 0$.
This behaviour of the fourth-order
 term is quite similar to that of the Landau expansion for tricritical
 phenomena~\cite{UZ:1993}. But the behaviour of IBG differs
 from that at standard tricritical points~\cite{UZ:1993},
 where the temperature dependence of the Landau coefficients is different.

{\em 3.5. Zero temperature condensation}

The zero-temperature $(T_c = 0)$ BEC at a constant density was
investigated
in Ref.~\cite{Busi:1985}.  This condensation is possible at a finite constant
density $(\rho > 0)$
and a low spatial dimensionality $d\le\sigma$.  The results for the critical
exponents at $d=\sigma$ and $0<d<\sigma$ are shown in Table~1.  These
critical exponents describe the scaling laws with respect
to the variations of temperature, for example, $C(T) = C_0/T^{\alpha}$.

It can be seen
from Table~1 that the zero-temperature condensation at a constant
density exhibits quite
unusual critical properties.  The critical exponents $\gamma$ and $\nu$
for $d<\sigma$ can be obtained from the familiar values at
$\sigma<d<2\sigma$ by the change of sign of ($d-\sigma$).
The Fisher exponent $\eta$ has its usual value but the exponent $\alpha$ is
negative rather than zero as is for $d>\sigma$.  For zero-temperature
$T$--driven transitions (Sec.~2.3), as is in our case of
$d \leq \sigma$ spatial dimensions, the negative values
of the critical exponent $\alpha$ are consistent with the Nernst
theorem.  When $d > \sigma$, $\alpha = 0$ and this
theorem is
satisfied in the limit $T_c(\rho) \rightarrow 0$ of an extreme dilution ($\rho
\sim 0$)
because the scaling amplitude $C_0$ tends to zero.

The infinite values of the exponents
$\nu$ and $\gamma$ at $d = \sigma$ indicate the exponential divergence of
the correlation length $\xi = (\hbar^2 / 2mr)^{1/\sigma}$ and the
susceptibility
$\chi = (1/r)$ at $T_c=0$.  These exponential divergences are known from the
mechanism of critical fluctuations in classical systems at their lower
borderline (critical) dimensionality $d_L$.  Here the exponential
divergence of $\xi$ and $\chi$ comes from the same low dimensional effect.
The exponents given in the last two lines of Table~1 can be referred to as
{\it low dimensional spherical exponents}.

The next question is why this zero-temperature
critical behaviour exists.  It is intuitively clear
that the low dimensional ($d \le \sigma$) IBG should have a ground state
of a total $\vec k$--space condensation ($N_0 = N$) at $T=0$ which means that
the solution
$T_c=0$ of the equation $[r(T_c)/T_c]=0$ should exist.  For $T\sim0$,
$(r/k_B T)\ll 1$ and, hence, the contributions to the sum~(15) are given by
terms
with low energies $\epsilon(k) \sim r$, i.e., with small wave numbers $k$.

There is an onset of BEC at $T \sim 0$, where the bigger part of
bosons are at states $\varepsilon(k) \sim 0$ but the macroscopic
condensation
occurs only at $T=0$~\cite{Busi:1985}.  At $T = 0$ the order parameter
$\phi_0$ jumps from zero to $|\phi_0| = \rho^{1/2} $ and all $ N $ bosons enter
in
the ground state $(k = 0)$.
This peculiar critical behaviour results from the constant
density $\rho$ condition (the spherical constraint).

The same behaviour can be obtained from the classical spherical model.
Although
the $\vec k$--space condensation is a pure quantum effect due to the Bose
statistics,
the critical behaviour around critical points ($T_c \ge 0$) is totally ruled by
classical fluctuation effects described by the classical spherical model.

The IBG critical behaviour can be exactly derived from thermodynamic
Eqs.~(12) and (15) because this model is exactly solvable.
Sometimes the RG analysis of such exactly solvable models is also useful,
for example, in studies of the crossover of the critical behaviour of IBG
to that of NBG~\cite{Wei:1986}.

Note that a zero-temperature BEC $(T_c=0)$ is possible also in the limit
$\rho \rightarrow 0$ (extreme dilution) for the case of constant density, and
 in the limit $P \rightarrow 0$ for the case of constant pressure for each
spatial dimensionality $D \geq 0$. As mentioned in Sec. 3.3 at these onset
transitions there is no HLTC or CQC.

{\em 3.6. Classical and quantum critical regimes}

The critical behaviour of IBG in a close vicinity ($r \ll k_BT$)
of phase transition points $(T_c \geq 0)$ to BEC of IBG is definitely
classical.  All our calculations have been performed
by keeping the leading power in $(r/ k_BT) $ in the corresponding series for
Bose integrals~(19) in powers of $(r/ k_BT) < 1 $.  This
approximation has been used to determine the leading scaling behaviour, namely,
the scaling laws.  This  essentially
classical expansion gives the critical regime ($\mu \sim 0$).

The transition region (Sec.~2.2) where the phase transition phenomena occur,
can be defined for
a broad interval of temperatures around $T_c \geq 0$, i.e., by $r < k_B T$.
In this region but not very near $(r \ll k_BT)$ to $T_c$ we must take into
account some next--to--leading terms in the same series expansion
 of the Bose integral~(19) in powers of $r/k_B T$. These secondary terms have a
quantum origin and,
therefore, taking several of them as corrections to the leading scaling powers,
we shall obtain the quantum corrections to the main scaling behaviour.

Thus the classical region {\it (a--0--b)} discussed in Sec.~2 can be
defined by the inequality $r \ll k_B T$.  For both $T_c(P)$ and $T_c(\rho)$
this region will look like the shaded wing $1-0-2$ in Fig.~1b; see the dashed
line $0-2$.  The  quantum region will be outside this vicinity of critical
line, in the
rest part of the critical domain $(r < k_B T)$ discussed in Sec.~2.

The parameters $P$ and $\rho$ rule the critical temperature but they
do not participate in the scaling laws.  Thus they have not a direct  effect
on the critical region size.  This is a particular IBG
property. The size of the classical critical region is reduced with
 the decrease of $P$ or $\rho$, i.e., of $T_c$. This is valid  for both
widths of classical $(r \ll k_BT)$ and total $(r < k_B T)$
critical regions along the $X-$axis; $X = (P,\rho)$.
In fact, at fixed $X$
we should consider the inequalities $r < k_B T_c(X)$ and $r \ll k_B T_c(X)$.
When $T_c(X)$ is lowered the critical regions width along the $X$--axis will
decrease  to zero for $X \rightarrow 0$.

The variations of the overall density $\rho$ in the ground state
  $[T = T_c(\rho) = 0]$ of the low-dimensional $(d \leq \sigma)$ IBG
 does not produce critical effects, because
 the condensate density is equal to $\rho$ (all particles
in IBG are in the condensate). If the zero-temperature critical point,
 containing the complete BE condensate of IBG is
approached by $T$--variations in the disordered phase, the  observed critical
phenomena will be mainly classical  on account of the real phase transition
to BEC at $T_c = 0$.

The phase diagram of the zero-temperature BEC in low dimensional systems at
 a constant density (Sec.~3.4) is very simple because
the critical line coincides with the $\rho$--axis.  The critical line
$[T_c(\rho) =0, 0 < \rho < \infty]$ can be approached only by
 $T$--variations and the critical phenomena are purely classical.

Finally we shall note that the criterion~(9) is not satisfied for the above
mentioned transitions. There is no real CQC in IBG because of
the lack of fluctuation interactions but we can consider the crossover from
the high-temperature to the zero-temperature at a constant density $\rho > 0$
caused by the change of the dimensionality from $ d > \sigma$ to $d < \sigma$.
This formal dependence describes the difference
between the high--dimensional ($d > \sigma$) critical phenomena at finite
critical temperatures and those of  zero-temperature BEC at a constant density
$\rho > 0$  and  low dimensionality ($d < \sigma$).

{\bf 4. Interacting bosons}

{\em 4.1. General methods}

The interacting bosons can be investigated by propagator expansions within
the Green functions method~\cite{LP:1980,Kad:1962,AGD:1963,Mahan:1981}.
 The infinite perturbation series  are truncated on the basis of several
 approximations. The perturbative
approach breaks down in a close (asymptotic) vicinity of the critical point
because of the strong fluctuation interactions.  In this situation, after
accomplishing the MF analysis, the strong fluctuation interactions in the
Ginzburg fluctuation region~\cite{UZ:1993,Land:1980} around the critical point
and their effects on the asymptotic critical behaviour are investigated by
 RG~\cite{UZ:1993,Ma:1976,Bin:1993}.

 The RG method consists of a suitable choice of a length-scale transformation
combined with the so-called
loop expansion~\cite{UZ:1993, Bin:1993} (see also Sec.~4.2).
In the framework of the general RG approach the MF approximation is equivalent
to the so-called ``tree  approximation'' -
 the lowest order theory within the loop expansion.
Certain aims as, for example, the evaluation of the size of the fluctuation
effects by a calculation of the Ginzburg critical
region size or investigations of qualitative features of fluctuation effects
in the pre-asymptotic transition region, can be achieved by suitable
 perturbation calculations without RG (see, e.g., Ref.~\cite
{Pf:1971,Rech:1971}, and Sec.~6.3).  This outline of general investigation
methods of critical
phenomena is common for both classical and quantum statistical models.

For the interacting bosons we must use the Hamiltonian $H[\hat{\psi}(\vec{r})]$
in terms of second--quantized Bose field operators $\hat{\psi}(\vec{r})$
which for practical calculations are expanded in terms of annihilation and
creation operators mentioned in Sec.  3.1.  Within this formulation which
gives the opportunity to use the powerful Green functions method the quantum
effects are ``hidden'' in the commutation relations for the field
operators $\hat{\psi}(\vec{r})$ and, hence, in the time ordering procedure
under the trace for partition function and statistical averages.  For finite
temperatures the ``time'' ordering procedure, i.e., the account of quantum
effects, is performed with the help of an auxiliary
``time'' $\tau$, reciprocal to the temperature:
 $(0 \leq \tau \leq \hbar\beta)$.
 This quantity is called  an ``imaginary time'' because of the relation
 $\tau = it$ with the real time $t$.

Alternatively, the theory can be formulated by the Feynman path
integrals~\cite{Cash:1968,Popov:1983,Neg:1988, Nagaosa:1999}. With the help of
the coherent state
representation the grand canonical partition function can be written as
a functional integral over a ($c-$number) field $\phi(x)$ depending on a
($d + 1$)--dimensional vector $x = (\tau, \vec{r})$ in restricted
``time''--space:
$ [(0 \leq \tau \leq \hbar\beta), \vec{r} \in V]$.  So, we may introduce a
$(d + 1)$--dimensional ``volume'' $V_{d + 1} = (\beta L_1...L_d)$, where $\beta
(= L_0)$ is the finite size (``thickness'') of the time--space ``hyperslab''.
Therefore, the quantum effects bring
an extra dimension $\hbar\beta$ along the new ($\tau-$) axis. At sufficiently
high temperature (classical limit), the $\tau-$
dimension collapses to zero and the behaviour of the system is classical.
 Of course, the same classical limit can be achieved by setting $\hbar = 0$.
At sufficiently low temperatures the $\tau-$ dimension is large and
 becomes infinite at the absolute zero $(T=0)$.
At such temperatures the classical limit
$(\hbar \rightarrow 0)$ does not produce reliable results.

The field $\phi(x)$ represents the order parameter, which describes
 the ordering below the phase transition temperature. The equilibrium ordering
is given by the statistical average $<\phi(x)>$. The fluctuation phenomena
 are represented
by the fluctuation part $\delta\phi(x) = [\phi(x) - <\phi(x)>]$ of the
(nonequilibrium) field $\phi(x)$. In usual cases,
 for example, when the phase transition is a result of
 a spontaneous breaking of the global discrete or
 continuous symmetry, the average
$<\phi(x)>$ does not depend on $x$, i.e. we have a uniform equilibrium order
 parameter. For the sake of simplicity we shall often use this example.

The complex Bose field $\phi(x)$ is sometimes
referred to as a ``classical field''. This term points to the difference from
the field operators which obey commutation rules.  In fact,
the classical fluctuation field is $\phi(0, \vec{r}) \equiv \phi(\vec{r})$
 and corresponds to the classical limit $(\hbar = 0)$.
The $\tau$--variations of the field $\phi(x)$ are created by quantum
fluctutions. The way, in which the quantum fluctuations take part in this
 picture is very similar to that, in which the classical fluctuations appear as
spatial variations of the field $\phi(x)$.

Furthermore, in order to describe
 a large variety of systems one may consider
$\phi(x)$ as a $(n/2)-$component complex vector field:
 $\phi(x) = \{\phi_{\alpha}(x), \alpha =1,
 ...,n/2\}$, where the compoments $\phi_{\alpha}$ are complex functions of $x$
and $n$ is an even positive number depending on the symmetry of the
system, or, more precisely, on the symmetry of the possible ground states
 (orderings). The number $n$ is often called a  ``symmetry index'' or
a ``number of order parameter components.''

For certain problems~\cite{UZ:1993,Ma:1976,Bin:1993} the field
theories of $n-$component real fields are equivalent to the theories
of $(n/2)$-component complex fields.  The classical systems are usually
described by real scalar or vector fields whereas the quantum systems are
represented either by real field components or by complex field components of
twice less number (the choice depends on the aims of the particular
 investigation). Thus one may compare
the results for quantum systems with a symmetry index of ordering $n/2$ to the
results for classical systems with a symmetry index $n$.  Moreover,
this correspondence can be extended to all integer numbers $n > 0$.  Then a
complex field with $n' = n/2$ components will correspond to a classical Ising
system for $n = 1$, i.e., $n' = 1/2$.  A Heisenberg ferromagnet ($n = 3$) will
be described by a complex vector field of $n' = 3/2$ components.

On account of the analytical relation $\tau = it$ the purely quantum
fluctuations reveal the intrinsic quantum
dynamics of the system.  The dependence of physical quantities on real
time $t$ is obtained by an analytical continuation ($\tau \rightarrow it$) of
the results from calculations to the real time axis $t$.  While in classical
systems the static and dynamic phase transition properties are considered
separately, in the framework of the quantum statistical mechanics these
phenomena have a unified treatment.
The intrinsic quantum dynamics contained in the quantum statistical
correlations is important for the dynamical critical
properties of Bose systems; see, e.g.,
Refs.~\cite{Abe:1974,AH:1974,KS:1974,SI:1974}.  Note, that this
intrinsic quantum dynamics can be substantially changed
under the action of external time-dependent
potential(s)~\cite{Ma:1976,HH:1977}.

Adopting periodic boundary conditions along the $\tau$--axis,
$\phi(\tau,\vec{r}) = \phi(\tau + \hbar\beta, \vec{r})$, and having
 in mind the periodic boundary conditions along the spatial axes (Sec. 3.2)
 we can expand the field components $\phi_{\alpha}(x)$ in Fourier series
\begin{equation}
\phi_{\alpha}(x)\;  =\;  \frac{1}{\sqrt{\beta V}} \sum_{q}\:
e^{iqx}\phi_{\alpha}(q)\;,
\end{equation}
where the $(d + 1)$--dimensional vector $q = (\omega_l, \vec{k})$ is given
by the Matsubara frequencies $\omega_l = 2\pi l k_B T/\hbar$, where
 $l = (0, \pm 1, ...)$; $qx = (\omega_l\tau + \vec{k}.\vec{r})$.
Setting $\omega_l = 0$ in Eq.~(26) we neglect the quantum fluctuations
and the amplitudes $[\phi(0,\vec{k})/\sqrt{\beta} \rightarrow \phi(\vec{k})]$
will describe classical fluctuations only.
In the $x$--representation this corresponds to the limit
 $\beta \rightarrow 0$, in which the system size (``thickness'')
along the $\tau$--direction tends to zero.

The functional formulation of Bose systems in terms of ($c-$number)
 complex functions $\phi(x)$ is performed directly for the original
 microscopic second-quantized Hamiltonian because the commutation relations
 of the Bose operators $\psi(\vec{r})$ have a direct classical limit
~\cite{LP:1980,Cash:1968,Popov:1983}.
The microscopic Hamiltonians of fermionic and spin systems are
 transformed to effective Bose Hamiltonians with the help
 of Hubbard-Stratonovich transformations and Feynman path integration
~\cite{BK:1994,Hertz:1976,Popov:1983,Neg:1988,Nagaosa:1999}.
 The effective field Bose Hamiltonians have the same meaning and role in the
 theory of quantum phase transitions as the effective field Hamiltonians for
classical systems~\cite{UZ:1993, Bin:1993, UZ:1996}. In both cases of
 classical and quantum systems the effective field Hamiltonians is not
 exact counterpart of the respective original Hamiltonians but
they provide a correct description of the (quasi)macroscopic phenomena
 which is enough for a reliable treatment of phase transition problems.
The lack of an entire correspondence with the microscopic model in
nontrivial cases of interacting particles is a result of certain forms of
 coarse graining which are unavoidable product of the known
 theoretical techniques of derivation of effective field models.

As our interest is focussed on standard second order phase transitions
we shall consider $\phi^4$--effective Bose ``Hamiltonians"
$\cal{H}$ of fermionic and
spin systems, as well as, the  same type microscopic Hamiltonians of
genuine Bose systems.  In fact, the ``Hamiltonian" $\cal{H}$
that we shall investigate is
related to the action $\cal{S}$~of the system by $\cal{H} = - \cal{S}$;
here we follow a conventional terminology used in a number of papers;
 see, e.g., Ref.~\cite{Hertz:1976}.

A quite general quantum effective Hamiltonian
$({\cal{H}} = \beta H)$ comprising a number of systems
can be written in the form
\begin{equation}
{\cal{H}}[\phi] \; = \; \sum_{\alpha,q}G^{-1}_0(q)|\phi_{\alpha}(q)|^2 \;+ \;
\frac{v}{2\beta V}\sum_{\alpha,\beta;q_1,q_2,q_3}\;
\phi_{\alpha}^{\ast}(q_1)\phi^{\ast}_{\beta}(q_2)\phi_{\alpha}(q_3)\phi_{\beta}
(q_1 + q_2 - q_3)\;,
\end{equation}
where $v > 0$ is the interaction constant.  The (bare) correlation
(Green) function $G_0(q) = \langle |\phi_{\alpha}(q)|^2\rangle_0$ is given by
\begin{equation}
 G_0^{-1}(q) \;  = \; i\omega_l + \varepsilon(k)\;
\end{equation}
for interacting real bosons and transverse $XY$ model~\cite{GB:1977,Gold:1979}.
For other effective Bose models of quantum systems
the bare correlation function $G_0(q)$ can be written in the
form~\cite{Hertz:1976}
\begin{equation}
 G_0^{-1}(q) \;  = \; \frac{|\omega_l|^{m}}{k^{m'}} + ck^{\sigma} + r\;
\end{equation}
with positive exponents $m,m'$, $0 < \sigma \leq 2$, and Landau parameters $c >
0$ and
$r \sim t(T)$ or $r \sim t(X)$.  The parameter $c$ can be always presented
as $c = \hbar^2/2m(g)$ where the effective mass $m(g)$
of the composite bosons (Bose excitations) depends on some interaction constant
$g$.

 The microscopically formulated Bose Hamiltonians contain
an upper cutoff $\Lambda = (\pi/a)$ corresponding to the Brillouin zone ends
$(-\pi/a) < k_{i} \leq (\pi/a)$. An example of such Hamiltonians is considered
in Sec.~5.  The effective quasimacroscopic Hamiltonians derived from
microscopic Hamiltonians contain only
large--scale spatial fluctuations of the field $\phi(q)$ and, hence,
the corresponding cutoff $\Lambda$ is relatively small: $ 0 < \Lambda \ll
(\pi/a)$~\cite{UZ:1996}.

 An example of a quasimacroscopic (effective field)
 Hamiltonian, derived by a Hubbard-Stratonivich transformation from the
 microscopic TIM Hamiltonian, is discussed in Sec.~6. As far as we are
 discussing critical phenomena
$(\xi \gg a, \lambda \gg a)$ the precise value of the cutoff
$\Lambda$ is not important.  It is, however,
important to have the conditions $\xi > (1/\Lambda)$ and
$\lambda > (1/\Lambda )$ satisfied and
then, the relevant phenomena included into consideration.

The statistical treatment implies the calculation of the {\it generalized}
grand canonical partition function
\begin{equation}
{ \cal{Z}}(T,\tilde{a}) \;  = \; \int {\cal{D}}\phi\:
{\Large e}^{-{\cal{H}}[\phi(x)]}\;,
\end{equation}
which gives the Gibbs free energy
\begin{equation}
\Omega = -\beta\mbox{ln}{ \cal{Z}}(T,\tilde{a})\;.
\end{equation}
In Eq.~(30), $\tilde{a}$ denotes the Hamiltonian parameters $(c,r,v)$,
 $\int{\cal{D}}\phi$ denotes the functional
integration over all allowed field $\phi(x)$ configurations
\begin{equation}
 \int \prod^{n/2}_{\alpha = 1}\:
\prod_{x \in V_{(d + 1)}}d\phi_{\alpha}^{\ast}(x)d\phi_{\alpha}(x)\;,
\end{equation}
or, in the $\vec{k}$--space,
\begin{equation}
 \int\prod_{\alpha;q}d\phi_{\alpha}^{\ast}(q)d\phi_{\alpha}(q)\;
\end{equation}
over all allowed Fourier amplitudes $\phi_{\alpha}(q)$.
Constraints on the field configurations lead to a critical behaviour change.
Within the present functional formulation the constraint of a constant density~
(13) is given by the mean spherical condition
\begin{equation}
 \frac{nN}{2} \; = \; \; \sum_{\alpha, q} \langle
|\phi_{\alpha}(q)|^2\rangle\;.
\end{equation}
Therefore, the mean square of the $Nn/2$ dimensional complex vector
$\tilde{\phi} = [\phi_{\alpha}(q)]$
(with components given by all possible $\alpha$ and $q$) describes a sphere of
radius $Nn/2$.
In the $\phi^4$--Hamiltonian of the nonideal Bose gas (NBG) given by Eqs.~(27)
and (28) with
$r = - \mu$, this constraint can be taken into account in the so--called
large--$n$ limit ($n \rightarrow \infty)$~\cite{UZ:1993,Ma:1976}.

Let us note, that for certain problems~\cite{BU:1987} the
effect of constraints like~(34) can be taken into account in Eq.~(27) by
adding an auxiliary
$\phi^4$--interaction term:  $\sim \tilde{u}\phi^4$.  In contrast to the
$u$--interaction terms
$\phi_{\alpha}^{\ast}(q_1)\phi^{\ast}_{\beta}(q_2)\phi_{\alpha}(q_3)\phi_{\beta
}(q_1 + q_2 - q_3)$
in Eq.~(34), the auxiliary interaction is given by two summation $q$--vectors:
$\phi_{\alpha}^{\ast}(q_1)\phi^{\ast}_{\beta}(q_2)\phi_{\alpha}(q_1)
\phi_{\beta}(q_2)$~\cite{BU:1987}.

The practical calculations are carried out by a substitution of the summation
over the wave vector components $k_i$ with an integration as shown by
Eq.~(16), provided the corresponding dimensions $L_i$ satisfy the criterion
$L_i \gg \xi$ for a quasi--infiniteness.  The summation over the
Matsubara frequencies $\omega_l$ can be substituted with the integration
 \begin{equation}
\frac{1}{\beta}\sum_{\omega_l} \; \rightarrow \; \int\limits^{\infty}_{-\infty}
 \frac{d\omega}{(2\pi)}\;,
\end{equation}
provided the temperature $T$ is low enough or, in an exact mathematical sense,
if $T \rightarrow 0$.

In order to determine the condition, under which
the integration (35) can be used without substantial errors
in the final results, let us consider the model~(27) corresponding to the
correlation function $G_0(q)$ given by Eq.~(29) with $m' = 0$ and $m = 1$.
Then the $\omega_l$-dependent modes $\phi(\omega_l \neq 0, \vec{k})$ in
Eq.~(27)
will yield relevant contributions to the partition function, if they have
a relatively
high statistical weight.  This may happen provided $ \omega_1 = 2\pi k_B T < r$
near the critical point $ r \sim 0$.  In terms of characteristic lengths
$\lambda = (4\pi c/k_B T)^{1/\sigma}$, and $\xi = (c/r)^{1/\sigma}$ we have
$\lambda > (8\pi^2)^{1/\sigma}\xi$ which is consistent with the criterion~(3);
here we have used Eq.~(2) and $c = \hbar^2/2m$.

If the criterion~(3) is satisfied,
we can substitute the summation over $\omega_l$ with an integration as shown
by the rule~(35) without introducing a substantial error in the calculations.
If the criterion~(3) is not satisfied, the frequencies $\omega_l$ can be
neglected and the quantum fluctuations ignored.

There exists a well developed theory~\cite{Popov:1983,Hugenholtz:1959,
DeDominicis1:1964,DeDominicis2:1964,Hohenberg:1965,Toyoda:1982} of the
$\lambda-$transition in $^4$He which can be applied to other Bose fluids with
critical temperatures $T_c >0$.  We shall not dwell on this important topic.
Our attention will be  concentrated
on the case $T_c \rightarrow 0$ when the asymptotic scaling is
influenced by quantum effects.

{\em 4.2. Notes about renormalization group}

A number of problems in the scope of this article
are usually solved by RG in the $q-$space, i.e., by field theoretical
variants of RG~\cite{UZ:1993,Ma:1976,Bin:1993,PV:2002}.  However, because of
the simultaneous presence of more than one
effect (quantum and thermal fluctuations, disorder, anisotropy or
gauge-field effects) a comprehensive RG investigation of quantum systems
can be performed by widely applicable and not extremely sophisticated
variants of RG such as, for example, the $\epsilon-$ or
the $(1/n)-$expansion~\cite{UZ:1993,Ma:1976,Bin:1993}.  The latter expansion is
convenient for the calculation of critical exponents but is quite hard as a
method of investigation of the RG transformation~\cite{Ma:1976}.

In the $\epsilon = (d_U - d)-$expansion the difference $(d_U-d$) is used as a
small paremeter; $d_U$ is the so-called
upper critical (borderline) dimensionality, above which the fluctuations are
irrelevant and the system is described by the MF approximation.  For the
classical variant ($\omega_l = 0$) of the Hamiltonian~(27) we have $d_U =
2\sigma$ but for other Hamiltonians $d_U$ may have
another value~\cite{UZ:1993}.  Within the $\epsilon-$expansion the physical
quantities can be calculated as series in powers of $\epsilon$ and the order
of accuracy of the calculation (first, second, etc.  order in $\epsilon$)
coincides with the order of the loop expansion which has been taken into
account~\cite{UZ:1993,Bin:1993}, that is, the calculated order in
$\epsilon$ exactly corresponds to the respective order in the loop expansion.
The method works for dimensionalities
$d > d_L$, where $d_L$ is the so-called lower critical (borderline)
dimensionality, below
which the ordering is destroyed by the fluctuation effects; for the
Hamiltonian~(27) with $(\omega_l = 0)$ we have $d_L = \sigma$~\cite{UZ:1993}.

The $\epsilon-$ expansion has been widely used
in investigations of critical phenomena and in the remainder of this paper
we shall discuss results obtained by this method.
Note, that the $\epsilon-$expansion is asymptotic and
all reliable predictions about the possible types of fixed points (FPs) of the
RG differential equations~\cite{UZ:1993,Bin:1993}, or, equivalently, of the
Wilson-Fisher recursion relations~\cite{UZ:1993,Ma:1976,Wilson1:1971,
Wilson2:1971, Wilson:1972, WF:1972}
are obtained within the one-loop approximation (equivalent to the first
order in the $\epsilon-$ expansion).

The next order of the theory, i.e., the
two-loop order, which is equivalent to the second order in the
$\epsilon-$expansion,
provides better quantitative results for the critical exponents,
corresponding to the FPs,
but this and even the higher orders in the loop expansion
do not essentially contribute
to the understanding of the main qualitative features of the system,
such as the possible types of critical behaviour and their
stability properties.  The same is valid for the precise determination of the
domains of attraction of simultaneously existing stable FPs of complex
Hamiltonians describing more than one interaction effect; this point has been
discussed in Ref.~\cite{LMU:1987} on the basis of a quite complex model.

The above notes
justify our approach to the classical and quantum critical phenomena
in complex systems with
competing effects, for which we shall often use results in first order
of $\epsilon$. We would like to emphasize an important
but not widely accepted feature of the modern RG
methods, namely, that the lowest (one-loop) results reveal the entire picture
of the possible critical phenomena whereas the
higher-order considerations in the loop expansion are useful for
other aims:  the improvement of numerical results for the critical exponents
and also the investigation of the asymptotic nature of the RG
perturbation-like method(s).

Bearing in mind that the most often used
expansions in the theoretical physics are asymptotic we should not take the
asymptotic nature of the $\epsilon-$expansion as a great disadvantage,
or, as a reason for an unreliability of the $\epsilon-$results.  The
experience accumulated by the research done during the last 20 years
firmly indicates that the $\epsilon-$ expansion results are meaningful
and quite useful,
 provided they are consistently analyzed and interpreted~\cite{LMU:1987}.

The main types of critical behaviour which can be described by a concrete
model (Hamiltonian) are given by the FPs of the respective RG equations.
Different FPs describe  different types of critical behaviour and,
hence, different critical exponents.  These exponents are obtained from the
relevant and irrelevant stability exponents of the FPs.
If a FP appears of order $O(\epsilon^2)$ but does not
exist within the one-loop (first order in $\epsilon$) RG equations, it
cannot be accepted as a new reliable object of investigation and the reason
lies in the asymptotic nature of the $\epsilon-$expansion.

At relatively high spatial dimensionalities $d > d_U$ which usually do
not correspond to real systems the RG equations have one stable FP, the
so-called Gaussian FP (GFP).  This FP describes the MF
behaviour for $d > d_U$.  The same GFP is unstable for $d<d_U$ and the stable
critical behaviour in this domain of spatial dimensionalities, which
includes the real system
dimensionalities $(d =1,2,3)$, is described by
another stable FP.  This is
usually a nontrivial FP which gives $\epsilon-$corrections to the MF
values of the critical exponents, i.e.,  this FP yields a nontrivial (non-MF)
critical behaviour.  Sometimes such  FP is called the Wilson
FP~\cite{Wilson1:1971,Wilson2:1971,Wilson:1972} but the most often used terms
are either  Heisenberg FP (for systems with a continuous symmetry, $n > 1$) or
Ising FP (for systems with a discrete symmetry, $n = 1$). When the
system has a XY symmetry $(n = 2)$ the nontrivial FP is called XY FP.
GFP has zero coordinates $(r_G =v_G = 0)$ in the $(r,v)$ parameter space of the
Hamiltonian but the nontrivial FPs have nonzero coordinates which are given
to first or higher order in $\epsilon$, depending on the order of the loop
approximation used in the particular investigation.

The possible types of critical behaviour predicted by RG are classified in
universality classes.  For a given structure of the Hamiltonian, one can
define the universality classes by the couple $(d,n)$, that is, the critical
behaviour depends on the dimensionality $d<d_U$ and the symmetry index
(number of
components of the order parameter field $\phi$) $n$. But one must keep in mind
that the change of the mathematical structure of the Hamiltonian usually leads
to another double series ($d,n$) of universality classes. The equivalence of
the critical behaviour of many substances,
described by the same Hamiltonian structure and
the couple $(d,n)$ is recognized as a property of the universality of
critical phenomena.  The MF behaviour at $d>d_U$ is superuniversal because it
does not depend on $d$ and $n$.

Below the upper critical dimensionality $d_U$ more than one FPs may exist
and be stable in different domains (domains of attraction)
of the Hamiltonian parameter space.
Usually this happens for complex systems with competing
effects.  The Hamiltonians of such systems include more than one interaction
term.  The simultaneous stability of more than one FP is a manifestation
of the simultaneous action of more than
one fluctuation interaction or the presence of other effects which interfere
with the main fluctuation interaction; the latter is usually represented by a
$\phi^4-$term in the Hamiltonian as shown in Eq.~(27).
This situation is quite common in the RG
analysis of Hamiltonians with competing effects.  Now one is faced with the
hard task to determine precisely the FPs domains of attraction
which is often necessary for the correct prediction of the critical
behaviour.  We have already mentioned that for quite important cases this
cannot be reliably done even in orders higher than the first order
in $\epsilon$~\cite{LMU:1987}.  In other cases, the effects competing the main
fluctuation interaction $(\phi^4)$ are hidden in additional modes like the
modes $\phi(\omega_l,\vec{k})$ with $\omega_l \neq 0$ of the quantum
 fluctuations, and under
certain circumstances, these modes may have an essential effect on
the critical behaviour.

The RG predictions are uncertain when there are no stable FPs.  This may
happen for a definite domain of the Hamiltonian parameter space or, in
extreme cases (see Sec.~7), for the whole parameter space.  The lack of
stable
FP is usually interpreted as a lack of a stable (multi)critical behaviour of
usual type.  Sometimes, the absence of stable FP can be
considered either as a signal for a first order phase transition or
as a more particular type of a continuous phase transiton.  The reliable
prediction of the phase transition type in this situation requires
additional heuristic arguments and investigations outside the scope
 of the RG method; see, e.g., Refs.~\cite{UZ:1993,Ma:1976,Bin:1993,PV:2002}.

{\em 4.3. Classical-to-quantum dimensional crossover}

We have already mentioned that the $d$--dimensional
quantum systems resemble $D = (d + 1)$--dimensional classical ones.
This is valid for the model~(29) with $m' = 0$ and $m = \sigma$.
By the formal substitution $k_B T = c^{1/\sigma}/L_0$ in $\omega_l$, where
$L_0 \equiv \lambda = (c^{1/\sigma}/k_B T)$ is the thermal wavelength
corresponding to the model (29), we obtain a $D = (d + 1)$--dimensional
momentum
$q = [q_i; i = 0,1,...,d]$ with components $q_i = (2\pi l_i/L_i)$.  The
representation gives a total formal correspondence between the present
time--space geometry
and a hyperslab with $(d + 1)$ spatial dimensions.  This formal analogy was
revealed for the first time in Refs.~\cite{Suz2:1976,Pfeuty:1976} and
further explored in Ref.~\cite{Law1:1978}.
Using the formal analogy we have shown that
for this effective model the notation~(1) yields the exponent $\theta = 1$.

The temperature dependence $\lambda = \lambda_0/T^{1/\sigma}$, as in IBG, is
valid
for other effective models, for example, for the model~(28) and
that given by Eq.~(29) with $m' = 0$ and $m = 1$.
Therefore, the form of the scaling law
$\lambda = \lambda_0/T^{\theta}$ depends on the values of exponents $m,\; m'$
and $\sigma$:
$\theta = \theta(m,m',\sigma).$ A simple dimensional analysis of Eqs.~(28) and
(29) shows that
$\theta = 1/z$.  The same result $(\theta = 1/z)$ can be obtained by the RG
rescaling  transformation~\cite{UZ:1993,Ma:1976,Bin:1993}.
Then the criterion~(9) becomes $z\nu_0 < 1$.
Usually the (bare) values of exponents
$\theta$ and $\nu = 1/\sigma$ of $\lambda$ and $\xi$, respectively, receive
perturbation corrections which are calculated with the help of RG.

It has been shown by Hertz~\cite{Hertz:1976} with the help of
the above mentioned method as well
by a RG analysis that
\begin{equation}
z \; = \; \frac{\sigma + m'}{m}\;
\end{equation}
and, hence, that the real CQC is given by
\begin{equation}
D \; = \; d + z\;.
\end{equation}
CQC is always associated with the dimensional crossover (37) which
was  revealed before Hertz by Pfeuty and Elliott~\cite{Pf:1971}
and by Young~\cite{Young:1975} on the basis of TIM (for the latter,
$m'=0$, $\sigma = 2$, $m = 2$, and $z = 1$).  Note, that
the upper borderline dimensionality changes from its classical values $d_U$
to the ``quantum" value $d_U^0 = (d_U - z)$ and the lower borderline
 dimensionality becomes $d_L^0 = \mbox{max}[0,d_L - z)]$.

Thus we should expect that the quantum critical phenomena
in $d$--dimensional (quantum) systems are described by the universality
class $(D,n)$ known for $D=(d + z)$--dimensional classical systems.
If the shift $d \rightarrow D$ of
spatial dimensionality $d$ is the only result of quantum effects, we can say
that the quantum critical phenomena
in the particular $d$--dimensional quantum system
(or a class of systems) obey a form
of universality, namely, that they are identical
to the critical phenomena in the corresponding
$D$--dimensional classical system (or class of systems).  Under
the term ``corresponding classical system'' of the quantum model~(27)
we understand
 the system described by the classical variant ($\omega_l = 0$) of Eq.~(27).

This definition of the
universality of quantum critical phenomena allows a comparison
between the critical properties of classical and quantum systems.  There are
other points of view on the definition of the universality for quantum
critical phenomena.
For example, sometimes it is said that the quantum critical phenomena
in a system are universal, if
a classical critical behaviour can be found at the same or another
spatial dimensionality which describes the same critical
phenomena~\cite{GB:1977,Gold:1979, KopK:1983}.  This point is important,
for example,
in the discussion of the extraordinary zero-temperature critical behaviour of
 NBG (Sec.~5).

{\bf 5. Nonideal Bose gas}

{\em 5.1. Preliminary notes}

The first studies~\cite{Singh1:1975, Singh2:1975,Singh1:1976,Singh2:1976}
of NBG by RG were performed in the framework
of the operator formalism~\cite{LP:1980,Huang:1987, Kad:1962, AGD:1963}.
The direct RG applications to the second--quantized NBG Hamiltonian lead to
a scaling law for the commutation relations of the field operators
$\hat{\psi}(\vec{r})$. In Refs.~\cite{Singh1:1975,
Singh2:1975,Singh1:1976,Singh2:1976}
this was interpreted as a rescaling of the boson mass $m$ which grows with the
successive RG transformations.  The fact that
the quantity, to which such a rescaling should be ascribed is the thermal
length $\lambda$
or, equivalently, the factor $(1/mT)$ associated with it,
has become clear later.  However, this circumstance was undoubtedly
irrelevant to the main conclusion of these works, namely, that the finite
temperature
$(T_c > 0)$ BEC should exhibit a universal
critical behaviour corresponding to the universality class $(d,2)$ of
classical XY model.  The latter result has been confirmed
and extensively discussed in other studies; see
 e.g.,~\cite{Stella:1976, Bald:1976,
DeCe:1978, Uzun:1981,Olinto:1985,Olinto:1986, Kor:1987,UW:1985}.
An error in the
calculations~\cite{Singh1:1975,Singh2:1975}
which is irrelevant for finite critical points
($T_c > 0$) but very important for the zero-temperature
critical behaviour at $T_c = 0$) has been pointed out
in Ref.~\cite{Uzun:1981} and noticed also in Ref.~\cite{Wei:1986}.
Other relatively early papers on critical
properties of real bosons systems are mentioned
in Refs.~\cite{UZ:1993, DeCU:1999, Wei:1986}.

The zero-temperature
critical behaviour of  NBG was treated for the first time
in Ref.~\cite{BDe1:1980} in the one-loop approximation and despite of the
correct result for
the critical exponents there are errors in the derivation of the RG equations
and their FPs. In Ref.~\cite{BDe2:1980}
the correct calculation of the self-energy function, the Fisher exponent
$\eta$ and the dynamical critical exponent $z$ was made in the
two-loop approximation.

The Wilson--Fisher recursion
relations~\cite{UZ:1993,Ma:1976} appropriate for a correct and thorough
treatment of the critical
properties of NBG at low and zero temperature critical points
have been derived in Ref.~\cite{Uzun:1981} and in Secs.
5.2--5.6 we shall follow this approach.  The
perturbation series and the diagrammatic representation of perturbation
terms are
standard and we shall not dwell on technical details; for a more
instructive explanation of the  corresponding calculations, see
Refs.~\cite{UZ:1993, Uzun:1981, Uzun:1982}.

Because
of the model particular properties the low-temperature
limit gives an extraordinary opportunity for an
exact summation in all orders of the loop expansion and, therefore, for
obtaining of the low-temperature critical behaviour
without approximations. This was demonstrated for the first
time in Ref.~\cite{Uzun:1981}.
However, the exact result~\cite{Uzun:1981} has been overlooked in subsequent
 papers~\cite{Wei:1986,FH:1988, Weichman:1988, Rasolt:1984} where the
same low-temperature critical
behaviour was rediscovered in the one--loop approximation and
applied to dilute Bose systems.
In Ref.~\cite{FH:1988}, the Gaussian-like critical
 behaviour~\cite{Uzun:1981} to one-loop order (see also Sec.~5.3) was
very conveniently interpreted  as a form of ``quasi-universality.''
 In Ref.~\cite{Wei:1986, Rasolt:1984}
 the RG recursion relations~\cite{Uzun:1981}
were re-established in the form of one-loop RG differential equations
and solved by the Rudnick--Nelson~\cite{RN:1976} method of intergation.

More recently, the simple structure of the perturbation
series discussed below, was found also in the
nonrelativistic scalar field theory and appropriately called the ``scaling
anomaly`` ~\cite{Berg:1992}.
Sachdev and co-workers~\cite{Sach:1994} introduced another appropriate
term -- ``zero scale--factor
universality'' --
for the same exact solution~\cite{Uzun:1981}.  These authors
overlooked the opportunity to take advantage of perturbation series
exact summability (see below) but recently the exact result has been
rediscovered and lengthy explained by one of them without any reference to the
original source (see Ref.~\cite{Sach:1999}).
The same disadvantage characterizes the results of
other papers~\cite{Wei:1986,FH:1988} because the exact
solution~\cite{Uzun:1981} is not familiar to the authors~\cite{PH:PC, MF:PC}.

Here we shall use the functional formulation of  NBG
(see, e.g., Ref.~\cite{Popov:1983})
which seems to be more convenient for RG studies
~\cite{Bald:1976, Uzun:1981}.  We shall be mainly interested
in the zero-temperature critical behaviour of NBG~\cite{Uzun:1981}.
The relation to other model systems with similar properties
and recent research will be also mentioned (Sec.~5.7).
The functional NBG formulation is given by Eqs.~(27) and (28) where the
parameter $r = - \mu$~is related to the (bare) unrenormalized chemical
potential $\mu \leq 0$.  In order to avoid
lengthy mathematical formulae we shall set $\sigma = 2$.  The generalization of
results to
$0 < \sigma \leq 2$ is not difficult.  The critical behaviour is investigated
with respect to variations of the chemical potential $\mu$ from the value
$\mu_c = \mu(T_c)$, corresponding to the critical point (for IBG, $\mu_c =
0$).  A treatment of the phase transition in NBG at a constant density
with the help of variations of the quantity $t \sim (T-T_c)$ has been
performed in Ref.~\cite{UW:1985}.

{\em 5.2. Renormalization group equations}

The RG recursion relations within the one--loop approximation
(to first order in $\epsilon = d_U - d)$ are~\cite{Uzun:1981}:
\begin{equation}
k'_i = bk, \;\;\;\;\; \omega'_l \; = \; b^{2 - \eta}\omega_l, \;\;\;\;\;
 m' \;=\; b^{\eta}m\;,
\end{equation}
\begin{equation}
r' \; = \; b^{2 - \eta} \left [ r +
\frac{1}{2}(n+2)(v/\beta)I_1(r) \right ] \;,
\end{equation}
and
\begin{equation}
(v/\beta)' \; = \; b^{4 -d - 2\eta} \left\{ (v/\beta) -
\frac{1}{2}(v/\beta)^2\left[(n+6)I_2(r) + 2\tilde{I}_2(r)\right] \right\} \;.
\end{equation}
Here we shall write in an explicit form the quantities
\begin{equation}
I_1(r) \; = \beta K_d \int\limits^{\Lambda}_{\Lambda/b}dk\:
k^{d-1}n(k)\;,
\end{equation}
\begin{equation}
 I_2(r) \; = \; \left(\frac{\beta}{2}\right)^{2} K_d
\int\limits^{\Lambda}_{\Lambda/b}dk\frac{k^{d-1}}
{\mbox{sh}^2[\beta\varepsilon(k)/2]}\;,
\end{equation}
and
\begin{equation}
\tilde{I}_2(r) \; = \; \frac{\beta}{2} K_d
\int\limits^{\Lambda}_{\Lambda/b}dk\:
k^{d-1}\frac{\mbox{cth}[\beta\varepsilon(k)/2]}{\varepsilon(k)}\;.
\end{equation}
In the above expressions the integration is performed up to an upper cutoff
$\Lambda \ll (\pi/a)$; $b>1$ is the RG rescaling factor~\cite{UZ:1993,Ma:1976}.

In contrast to other RG studies
here the problem for the cutoff $\Lambda$ is not trivial
~\cite{Wei:1986, Uzun:1981}.  The problem arises in the treatment of
the dilute $(d > 2)$--dimensional Bose gas in the
low temperature limit $(T_c \rightarrow 0)$ which corresponds to an extreme
 dilution
$\rho \rightarrow 0$; see Eq.~(23) which indicates that $\Lambda \sim (\pi/a)
\rightarrow 0$ for $\rho = a^{-d} \rightarrow 0$.  It seems at first sight
that the interparticle distance $a$ cannot be used in the
definition of the  finite cutoff $\Lambda$ or
as a length unit of dimensionless characteristic lengths $\xi/a$ and
$\lambda/a$.  For this reason, the fourth (and last)
characteristic length in the problem -- the finite scattering length
$v \sim (\hbar^2/m)a_{sc}$~\cite{LP:1980, Huang:1987} was discussed
as a possible inverse cutoff,
$(1/\Lambda) \sim a_{sc}$,
for the first time in Ref.~\cite{Uzun:1981} and later used also in
Ref.~\cite{Wei:1986}; for $d \neq 3$, $v \sim (\hbar^2/m)a^{(d-2)}_{sc}$.

Let us note, however, that the RG investigation can be reliably performed by
the standard
cutoff $\Lambda = \pi/a$, although $a$ tends to infinity for the
 zero temperature critical
behaviour.  In fact, according to Eq.~(23), $\lambda > a$ for all temperatures
in the interval
$0 \leq T < A\zeta(d/2)T_c(\rho)$ including the close vicinity of $T_c$ where
$\xi > a$, too;
note that $A\zeta(d/2) > 1$.  In the temperature range of interest
the characteristic lengths $\xi$ and $\lambda$ are always
greater than $a$ and, therefore,
there is no danger of shortcomings in the description; see also the brief
discussion after Eq.~(50) in Sec.~5.3.

The summation over the frequencies $\omega_l$ of the internal lines of
perturbation diagrams
formed by the legs of one and the same Hamiltonian is performed with the help
of the rule
\begin{equation}
\sum_{\omega_l}e^{(+0)i\omega_l}G_0(q)\;,
 \end{equation}
where (+0) denotes a positive number which is set equal to zero after the
calculation.  The correlation function
$G_0(q)$ is given by Eq.~(28); c.f.  Refs.~\cite{UZ:1993,Uzun:1981,Uzun:1982},
where the Green function $G_0(q)$ has an opposite sign.

The summation (44) leads to a difference between the integrand $n(k)$ in
$I_1(r)$ and the corresponding integrand in the RG relations presented in
 Refs.~\cite{BDe1:1980} but coincides with the corresponding result in
Refs.~\cite{Singh1:1975,Singh2:1975, Stella:1976}.  Moreover, the integrand in
$\tilde{I}_2(r)$ differs from the corresponding
integrand in the RG equations in Ref.~\cite{Singh1:1975, Singh2:1975}
but coincides with that
in Ref.~\cite{BDe1:1980} as mentioned for the first time in
Ref.~\cite{Uzun:1981}.  The mentioned errors in papers before 1981 do not
have an effect on the results for the critical behaviour in the classical
limit ($\varrho < 1, T_c \neq 0$).  However, in the
low-temperature limiting case ($T_c \rightarrow 0$), these errors lead
to wrong results for the FP coordinates
and the rational extension of the loop expansion in higher than first order.

Owing to this formulation of the RG treatment the relation (38) for
$\omega_l$ should be referred to the temperature:
\begin{equation}
T' \; = \; b^{2 - \eta}T\;.  \end{equation}
For the same reason the mass $m$ should be kept invariant.
We shall proceed with this choice up to the end of this subsection,
where we shall propose a more elegant scheme of scaling.  The mass
$(m-)$ invariance implies
$\eta = 0$.  This is the usual result within the one--loop approximation.
Apart from quite
special cases the exponent value $\eta = 0$ receives $\epsilon$--corrections
of order
$O(\epsilon^2)$, i. e., in the two-- and higher--loop approximations.

The task is to solve RG Eqs.~(38)--(40).  This means to obtain FPs and
investigate their stability.  Eq.~(45) shows two temperature FP values:
$T^{\ast}_C = \infty$ (classical) and $T^{\ast}_Q = 0$ (quantum).  The
infinite FP value $T^{\ast}_C$ of the temperature has nothing in common with
infinite temperatures.  Rather it describes the classical critical behaviour
near finite temperature critical points $T_c > 0$.

Remember that the classical limit of quantum Hamiltonians (27) is taken,
in a strict mathematical sense, by decreasing the interval $[0, \beta]$ of
variations
 of Matsubara time $\tau$ to zero which corresponds to the limit $T
\rightarrow \infty$.
Obviously, the FP coordinate $T^{\ast}_Q = 0$ should correspond to the
low-temperature regime.  As FPs lie on the critical surface $[\xi (T_c) =
 \infty]$ in the Hamiltonian parameter space it is clear that the zero FP
value corresponds to a zero-temperature critical point $(T_c = 0)$.
However, it can be shown that for ``high-temperature" critical points the
parameter $T$ is absolutely redundant together with the relation (45) and the
corresponding infinite FP value of temperature (see Sec.~5.5).
The infinite FP value $T^{\ast}_C = \infty$ merely indicates that the zero-
temperature critical behaviour is unstable towards temperature fluctuations.

Eq.~(38)--(40) can be solved analytically in two limiting cases:

(i) $\beta\varepsilon(k) \ll 1$ (high-temperature behaviour),

(ii) $\beta\varepsilon(k) \gg 1$ (low-temperature behaviour).

These conditions will be discussed in Sec.~5.4.  Here we shall take the
leading terms of
the integrals (41)--(43) for each of these limiting cases and perform the
formal analysis of RG equations.

The limit (i) yields the classical universality as shown in
Refs.~\cite{Singh1:1975,Singh2:1975}.  In this case ($\epsilon = 4 - d$) the FP
value of the temperature is
$T^{\ast}_C = \infty$.  This allows small values of the factor $\beta$ in the
condition (i).
The RG relations have two stable FPs.  For dimensionalities $d > 4$, the
Gaussian FP with
$(r,u)$--coordinates $ r_{\scriptsize G} = u_{\scriptsize G} = 0$
is stable and describes the usual MF behaviour; $u = v/\beta$.  For
dimensionalities $2 < d <4$,
the Heisenberg FP is stable.  The coordinates of HFP are given by
\begin{equation}
r_{\scriptsize H} = -\frac{(n + 2)}{2(n+8)}\frac{\hbar^2\Lambda^2}{2m}\epsilon,
\;\;\;\;\; u_{\scriptsize H} = \frac{16\pi^2}{(n+8)}\left
(\frac{\hbar^2}{2m}\right)^2\epsilon\; .
\end{equation}
As the temperature FP coordinate is infinity the FP value $v_{\scriptsize H}$
of the parameter $v = \beta u$ is equal to zero
because the respective FP value $u_{\scriptsize H}$ of $u$ is finite, as given
by Eq.~(46).  This behaviour of the interaction
parameter ($v$ or $u$) is not strange because
it precisely describes the behaviour of the system at high temperatures
where the modes $\phi(q)$ with
$\omega_l \neq 0$ can be neglected.  Ignoring the quantum fluctuations we
obtain a classical Hamiltonian
with an interaction constant of the form $u = v/\beta$ which plays the role of
an interaction constant in the usual classical Hamiltonians
$({\cal{H}} = \beta H$).

However, this is not the final answer.  As shown
in Sec.~4, we must substitute the modes $\phi(0, \vec{k})$ with the true
classical modes
$\sqrt{\beta}\phi(\vec{k})$ and this yields a factor $\beta \approx (1/k_B
T_c)$ in front of
parameters in the $\phi^2$--part as well as the square of the same factor
in front of the interaction constant.  So, in the classical variant of the
theory the actual interaction constant
is $\beta v$.  Therefore, the parameter $u=(v/\beta)$ is the interaction in the
classical Hamiltonian
in terms of the field $\phi(0, \vec{k})$ and this parameter transforms to $ u
=\beta v$ when the classical
 Hamiltonian is written by the field $\phi(\vec{k})=
  \phi(0, \vec{k})/\sqrt{\beta}$.

This discussion
can be used to explain the infinite value of FP coordinates of temperature:
$T^{\ast}_G = T^{\ast}_H \equiv T^{\ast}_C = \infty$.  The field $\phi(0,
\vec{k})$ has a  rescaling
factor $b$ (for $\eta = 0$). The field crossover $[\phi(0, \vec{k}) =
\sqrt{\beta}\phi(\vec{k})]$ implies that this rescaling
factor should be associated with the proper classical field $\phi(\vec{k})$
 rather than with the
temperature.  This is just what happens in the classical limit (i) where the
relation (45) does not
exist at all.  Therefore, in the limiting case (i) the temperature
is a redundant parameter and the relation (45) can be suspended from the
high-temperature analysis although it remains in the general RG scheme.
Up to the stage, for which the investigation
is performed in terms of the field $\phi(0, \vec{k})$, i.e., in the
low-temperature
(classical or quantum) regime we must consider the relation (45) and
the FP temperature value $T^{\ast}_Q = 0$.

{\em 5.3. Low temperature behaviour}

In the low-temperature
limiting case (ii), Eqs.~(38)--(40) are solved with the help of
$\epsilon = (2 - d)$--expansion.  This expansion reflects the dimensional CQC
which is
given by $D = (d + 2)$.  The upper critical dimensionality $d_U$ is changed
from $d_U = 4$ for the classical case (i)
to $d_U = 2$ for the quantum case (ii)~\cite{Hertz:1976}.
Up to now there is no
evidence that the case (ii)
describes quantum critical phenomena but it is clear that the low-temperature
limit
can be taken within the RG scheme and that it will bring to a new critical
behaviour.  Within RG this limit always exists because of the
lower cutoff $(\Lambda/b) > 0$ which permits the condition (ii)
irrespective of the value of the ratio $(r/k_B T)$; see also Eq.~(21).
For the case (ii) the integral $I_2(r)$ tends exponentially to zero but the
integral
$\tilde{I}_2(r)$ is finite and exhibits a power--law infrared divergence for $d
< 2$.  This
yields the borderline value $d_U = 2$.  Performing standard calculations we
obtain Eqs.~(39) and (40) in a simple form
\begin{equation}
r'\; = \; b^{2}r\:, \;\;\;\;\; v' \; = \; b^{2 - d}v(1 - a_0v)\;,
\end{equation}
where
\begin{equation}
a_0 \; \equiv \; \left [\tilde{I}_2(r)\right]_{\beta\varepsilon(k)
 \rightarrow \infty} \; =
 \; \frac{1}{4\pi}
\int\limits^{\Lambda}_{\Lambda/b}\frac{k\:dk}{\varepsilon(k)}\;.
\end{equation}
The straightforward calculation gives
\begin{equation}
a_0 \; = \frac{m}{2\pi\hbar^2}\mbox{ln}b\;.
 \end{equation}

The relevant parameters are $T$ and $r$.
The variations of these parameters near the FP values
$T^{\ast} = r^{\ast} = 0$ drive the system away from the zero-temperature
critical state.
The inclusion of the temperature $T$ as a second relevant parameter corresponds
to a real physical situation, namely, that the zero-temperature
critical state is approached when both $T$ and $r$ tend to zero.  The zero-
temperature critical behaviour is described by two stable FPs:
the Gaussian FP ($T_{\scriptsize G} = r_{\scriptsize G} = v_{\scriptsize G} =
0$), which is stable for $d > 2$, and the {\it Gaussian--like}
 FP~\cite{Uzun:1981},
\begin{equation}
T_{\mbox{\scriptsize Gl}} \; = \; r_{\mbox{\scriptsize Gl}}\; =\; 0\:,\;\;\;\;
v_{\mbox{\scriptsize Gl}}\; = \; \frac{2\pi\hbar^2}{m}\epsilon \;.
\end{equation}

The equation for $v_{\mbox{\scriptsize Gl}}$ shows that the renormalized value
of
$s$--wave scattering length is $a_{sc} \sim (1/\epsilon)^{1/\epsilon}$;
$\epsilon = (2 - d)$.
The scattering length $a_{sc} \rightarrow \infty$ for the important case
$\epsilon \rightarrow 0$ of $2d$ Bose fluids and, therefore,
the cutoff $\Lambda \sim (1/ a_{sc})$ applied
in Ref.~\cite{Wei:1986} seems to be inconvenient.

\begin{figure}
\begin{center}
\epsfig{file=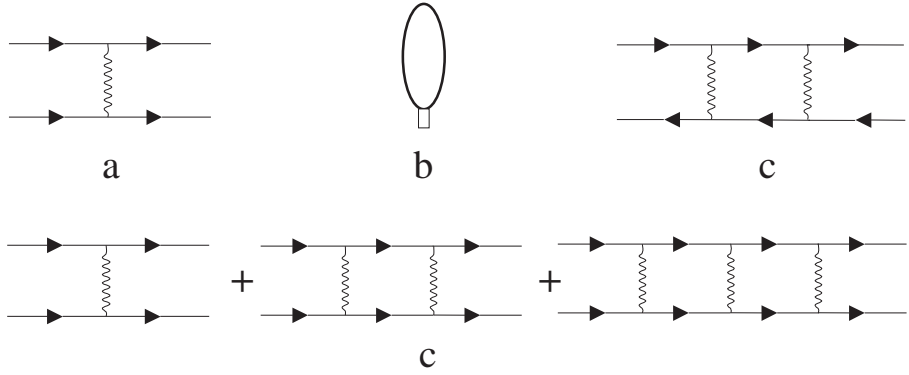, width=12cm}
\end{center}
\caption{(a) A tree diagram denoting the interaction part of the Bose
Hamiltonian. (b) The compact self-energy diagram which is equal to zero in the
limit $T \rightarrow 0$ [the thick loop denotes the full (renormalized) Green
function $G(q)$]. (c) An example of a diagram from the perturbation series for
the interaction vertex which gives a zero contribution in the
zero temperature limit. (d) The infinite ladder series of diagrams which
yields the geometric progression (51).}
\label{QPTf3.fig}
\end{figure}

In the remainder of this Section we shall discuss the nontrivial
Gaussian--like FP which will be called GlFP.
GIFP has Gaussian values
for the main critical exponents:  $\eta = 0$, $\nu = 1/2$, $z = 2$.  But the
interaction parameter gives a correction-to-scaling exponent and this
circumstance does not allow to put this
critical behaviour in the Gaussian universality class (such an incorrect
conclusion has been made in Ref.~\cite{BDe2:1980}).
Note, that the Gaussian universality class is sometimes denoted by $n = -2$
(see, e.g., Ref.~\cite{Gold:1979}).

In the present case the $\epsilon$--analysis can be extended to any order in
$\epsilon$~\cite{Uzun:1981} and this exceptional case will
be discussed below.  The problem has an exact solution~\cite{Uzun:1981}
because of the great simplification of the perturbation series in the limit
(ii).
In this limit,
all self--energy perturbation contributions (see Fig.~3b)
exponentially tend to zero. In each diagrammatic (perturbation) term in the
 self-energy function~\cite{UZ:1993, LP:1980, Popov:1983} all poles in the
 frequency ($\omega$) lie in the complex upper-half plane and, hence, the
 contour of the respective frequency integral can be deformed into the
 lower-half plane to give zero. Therefore, in this limiting case,
 no corrections to the MF values of the critical
 exponents describing the main scaling laws will appear to any order
 of the loop expansion.

 Furthermore, in the same zero-temperature limit (ii)
 the bigger part of perturbation
 contributions to the interaction vertex $v'$ tend to zero, too.
For example, the diagrams of type shown in Fig.3c do
not give contribution.  The only exception is the ladder series shown in
Fig.~3d.  This is the so--called
superconductivity channel for the interaction vertex where the
Green function lines are oriented in one and the same direction; for more
details, see Refs.~\cite{UZ:1993,Uzun:1982}.
The final result of the summation of the ladder in
Fig.~3d is an infinite geometric progression~\cite{Uzun:1981}
\begin{equation}
v' \; = \; b^{2 - d}\frac{v}{1 + a_0v}\;.  \end{equation}

The above consideration proves that within the grand canonical formalism
the dynamical and static critical exponents of main scaling behaviour
described by GlFP
have Gaussian values. This is true for the scaling laws written in terms of
the chemical potential
$|\mu| = r$.  As the parameter $r$ does not receive a renormalization
it remains equal to that of IBG. The zero temperature critical point
 is defined by $[r(T_c)/T_c] = 0$ as is for IBG (Sec.~3).

The FP value $v_{\mbox{\scriptsize Gl}}$ can be obtained from Eq.~(51) to any
order in $\epsilon = (2-d)$.  To do this one should expand the integral (48)
and Eq.~(51) to the corresponding order in $\epsilon$ by having
in mind that the FP value $v_{\mbox{\scriptsize Gl}}(\epsilon)$ have to be
also  expanded.
For the sake of convenience in calculations to second order in $\epsilon$,
the FP value $v_{\mbox{\scriptsize Gl}}(\epsilon)$ is usually written in the
form
\begin{equation}
v_{\mbox{\scriptsize Gl}}(\epsilon) \; = \; \frac{\hbar^2 \epsilon}{mK_d}[1 +
\epsilon\mbox{ln}\Lambda]\;
\end{equation}
Expanding $K_d$ in $\epsilon$ we have
\begin{equation}
v_{\mbox{\footnotesize Gl}}(\epsilon)\; = \;
\frac{2\pi^2}{m}\epsilon \left\{1 + \epsilon \left [\frac{c_E}{2} +
\mbox{ln}\frac{\Lambda}{\sqrt{4\pi}} \right] \right\} \;,
\end{equation}
($c_E$ is the Euler constant).  The FP coordinate $ v_{\footnotesize Gl}$
receives corrections in all orders of the
$\epsilon$ expansion.  Within the one--loop approximation, the stability
exponent $y_v$ associated with the interaction parameter $v$ has
a negative value
$y_v = -\epsilon = (d - 2)$ for $0 < d < 2$.  This exponent does not receive
higher order
$\epsilon$ corrections and, therefore, the value $(d - 2)$ is exact.

This completes the proof~\cite{Uzun:1981}.
The above results have been confirmed by Monte Carlo calculations for
one--dimensional interacting bosons~\cite{Batr:1990}.

We should emphasize that GlFP is  non--Gaussian FP.  It is a product of the
thermal
and quantum fluctuation interactions represented by the interaction parameter
$v$.  In fact,
GlFP is conjugate to actual GFP ($v_{\scriptsize G} = 0$) within the same RG
analysis.  Moreover,
the total critical behaviour includes corrections to the main scaling laws
which are described by
GlFP.  The interpretation~\cite{BDe2:1980} of results within the classical
universality class
($ d + 2, n = - 2$) has no heuristic significance.  As we have
already mentioned this interpretation is incorrect because of the correction-
to-scaling exponent coming from the interaction.  Therefore,
GlFP constitutes
a new universality class of zero-temperature QCP which has numerous
applications, in particular, to fluids of real bosonic atoms and XY magnets
(Sec.~5.7).

{\em 5.4. Breakdown of quantum universality}

GlFP and the critical behaviour described by it are quite unusual.
Let us mention that here the term
``usual zero-temperature FP"  means a zero-temperature FP which can be
obtained from the respective stable finite temperature FP of the same system
after substituting $\epsilon = (d_U - d)$ with $\epsilon = (d_U - d-z_0)$ in
the formulae for the coordinates and the attached critical exponents of
the latter.  Therefore, there is a clear form of conformity between the finite
temperature FP and the respective zero-temperature FP which we temporally
call ``usual" zero-temperature FP. This point of view is
 consistent with the definition of universality of quantum critical
 phenomena given in Sec.~4.3.

If all zero-temperature FP are usual in the
above mentioned sense then one does not need to define new universality
classes for quantum critical phenomena.  The universality classes determined by
the known finite-temperature FPs can be used, after the account of the
dimensional shift due to the CQC, to describe the possible types of
quantum critical phenomena, too.  If this scheme works, we can introduce the
term ``universality"  of quantum critical phenomena by understanding this as a
quantum universality which coincides with the classical universality at a
shifted
effective dimensionality of the space (see also Sec.~4.3).  Then a
zero-temperature FP obeying
the mentioned correspondence with the respective finite-temperature FP can
be called ``FP describing quantum
critical universality" or, shortly, ``quantum universality FP."

Certainly,
the GlFP is not such  usual or quantum universality FP.  It does not
obey the present definition of quantum universality although it comes as a
result of  CQC in  NBG.
According to Fisher and Hohenberg~\cite{FH:1988}
the GlFP describes a ``quasiuniversality'' associated with the existence of
 an almost marginal line in the Hamiltonian parameter space $(T,\mu,v)$
 defined by $T = \mu= 0$ and a nonzero interaction parameter $v$ which can
 take positive values.

This is in contrast with other
zero-temperature FP revealed by J.  Hertz~\cite{Hertz:1976} which are
usual in the above mentioned aspect and obey the quantum universality.
Therefore, there is a number of real systems which do not satisfy the widely
accepted notion that the main effect of the quantum fluctuations on the
zero-temperature critical behaviour is  CQC.   Such systems are the
Bose fluids and the XY magnets (Sec.~7) where the quantum universality
concept is not valid.

{\em 5.5. Discussion of limiting cases}

The limiting cases (i) and (ii) are analyzed under the general condition
$(\xi/\Lambda) \gg
1$.  For a microscopic model, $\Lambda \sim (\pi/a)$ and, hence, we have
$(\xi/a) \gg 1$.  Then the inequality (i) yields $\lambda \ll a$, i.e., the
high-temperature condition.  The values $\lambda \ll a$ have a statistical
meaning in the continuum limit $(V/a^{d}) \rightarrow \infty$.

The high-temperature  condition is certainly satisfied near the infinite FP
value $T^{\ast}_C = \infty$ and well
below it.  The condition (ii) is valid either in the classical low-temperature
critical region
 \begin{equation}
a \ll \lambda < \xi \;,
\end{equation}
or in quantum one, given by
\begin{equation}
a \ll \xi < \lambda \;;
\end{equation}
c.f.,  the general criterion~(3).  The inequalities~(54) and (55) can be
written in terms of
parameters $T$ and $r$ which enter in the RG relations.  The condition (54)
will be valid near and at
low-temperature FPs given by the zero values of $T$ and $r$ provided in the
 zero-temperature limit $T \rightarrow 0$
we have $(r/k_B T) \rightarrow 0$.  In the same limit,
the quantum condition (55) will be valid, if
$(r/k_B T) \rightarrow \infty$.  In particular, to answer the question which of
these two types of the low-temperature critical behaviour is described by
GlFP (50), we must know the value of the
ratio $(r_{\mbox{\scriptsize Gl}}/T_{\mbox{\scriptsize Gl}})$.  If it is
greater than unity, GlFP will describe quantum critical phenomena
but if it is less than unity, the
low dimensional $(d \leq 2)$ low-temperature critical behaviour will be
classical.

A standard example is given by the system of interacting bosons  at  a constant
density
(the case of noninteracting particles has been discussed in Sec.~3.4).  As the
perturbation series does not
yield self--energy contributions at all the parameter $r$ for ideal and
interacting Bose
systems is the same.  The value of the correlation length critical exponent
$\nu$, cited in Table 1, is greater than the exponent $1/\sigma$ of thermal
length $\lambda$. So in the zero-temperature limit
the ratio $(r/k_B T) \sim (\lambda/\xi)^{2}$ will tend to zero; c.f.  Eq.~(21).
Therefore, the low-temperature
critical behaviour described by GlFP will be classical which is true for
$T$--driven
transitions (Sec.~2.3).  The $\rho$--driven transitions will exhibit
quantum critical effects.  These results are consistent
with the general criterion (9).

In the same way one can show
that the critical effects near the phase transition points of interacting
bosons under the condition of a very low constant pressure
will be influenced by quantum effects (see also Sec.3.3).  The application
of the results from Sec.~5.3 to dilute Bose systems in the low density limit
$\rho \rightarrow 0$ will be briefly discussed in Sec.~5.7.

The NBG properties at finite low temperatures are very similar
to those of IBG at constant density
(Sec.~3.4).  It is then convenient to investigate
the corrections to the (spherical) Hartree limit by the
standard $1/n-$expansion~\cite{Ma:1976}.
The alternative approach~\cite{BU:1987} by a reduced Landau-Ginzburg free
energy function has been mentioned in Sec.~4.1.

{\em 5.6. Formulation by thermal wavelength }

An alternative RG treatment can be performed by the transformation
 \begin{equation}
\sqrt{\hbar^2/2m}\phi(q) \rightarrow \phi(q)
\end{equation}
in the Bose Hamiltonian (27)--(28).  In the new notations for the theory
parameters
the mass $m$ is absent and the Matsubara frequency $\omega_l$ is substituted by
$8\pi^2l/\lambda^2$, i.e., by $(1/\lambda^{2})$.  The parameters $r$ and $v$
are multiplied by
factors $(2m/\hbar^2)$ and $(2m/\hbar^2)^2$, respectively.  The important point
is that the recursion
relation (38) for the mass $m$ and the recursion relation (45) for
the temperature are now substituted by
 \begin{equation}
1 = b^{\eta}\;,
\end{equation}
and
\begin{equation}
 \lambda' = b^{-1}\lambda\;,
\end{equation}
respectively.  Eq.~(58) can be written in the form
\begin{equation} m'T' = b^{2}mT\;.
\end{equation}

The comparison of Eq.~(38) for $m$ and Eq.~(45) for $T$ with Eqs.~(57) and (59)
shows that the second formulation is not completely equivalent to the usual
one although the formal results for the critical behaviour are equivalent.
Within the present formulation
Eqs.~(57) and (59) do not imply  the mass invariance.  Rather the mass enters
as a factor in the parameters $\lambda$, $r$ and $v$ and, hence,
participates in the renormalization
scheme but in a way different from the suggested in
Ref.~\cite{Singh1:1975,Singh2:1975}.  From Eq.~(59) one cannot
conclude anything about the individual behaviour of $T$ and $m$ towards
the RG transformation. Rather the correct conclusion is that the thermal
wavelength $\lambda$ is renormalized, as given by Eq.~(58),
and that RG transformation will drive $\lambda$ to zero, unless it is at
the high-temperature FP value $\lambda^{\ast}_C = 0$;
another FP value of $\lambda$ is $\lambda^{\ast}_Q = \infty$.

Note that the renormalized thermal wavelength
$\lambda =\lambda(b)$ does not belong to the parametric space $(r,v)$ of
the Hamiltonian.  Hence, the temperature $T$ is also excluded from this space.
It is convenient to consider these parameters as describing HLTC.  The
renormalized wavelength $\lambda(b)$
drives the system from the high temperatures $(\lambda \ll a)$ to the zero
temperature $(\lambda \gg a)$ Gaussian--like
critical behaviour; see Eq.~(58).  The rescaling factor cannot be taken as a
crossover
parameter because of the RG restriction $\epsilon\;\mbox{ln}b < 1$.  So the RG
flows from the usual Heisenberg FP where $\lambda^{\ast}_C = 0$ to GlFP
where $\lambda^{\ast}_Q = \infty$ are produced by successive RG
transformations.  If the temperature
is equal to zero, the system will be exactly at GlFP and the FP Hamiltonian
is invariant
towards RG.  The system remains in the ground state of total BEC $(\rho_0 =
\rho)$.  For any
$T > 0$, the RG flow will drive the system to the usual classical behaviour.
Now it is not difficult to rederive
all RG results in the present formulation by avoiding the
unnecessary mass--invariance condition and the unnatural HT FP value
$T^{\ast}_C = \infty$.

{\em 5.7. Related topics and applications}

   As the systems undergoing zero-temperature phase transitions are numerous
 here we shall mainly discuss the applicability of the results from Sec.~5.3 to
 real systems and topics having some relationship with these results.

{\em 5.7.1. Magnetic and ferroelectric systems}

Experiments~\cite{Katsumata:1989,Date:1990,Lu:1991,Chiba:1991}
 on quasi-one-dimensional antiferromagnets with integer spin
 in an external magnetic field $\vec H$, mainly intended to
 a search of Haldane gap~\cite{Haldane:1983},
reveal a continuous phase transition to a nonzero ground state
 magnetization which occurs at the absolute zero $(T_c=0,H_c > 0)$.
Theoretical investigations~\cite{Schulz:1986,Tsvelik:1990,Mitra:1994}
 demonstrate that this phenomenon is a type of BEC of magnons with spin $S=1$.
The critical properties in the vicinity of the transition point $(0,H_c)$
 are described by  GlFP (Sec.~5.3), as shown in Ref.~\cite{Sach:1994},
 where low-temperature $(T\sim 0)$ features and the phase diagram of these
 spin chain antiferromagnets have been investigated.

The results for interacting real bosons are straightforwardly extended to
systems described by
the quantum XY model in a transverse field and, in particular, to XY magnets
 without a time reversal invariance.
RG calculations to the one--loop order in Refs.~\cite{GB:1977,Gold:1979}
have drawn the attention to the
possibility for a Gaussian
zero-temperature critical behaviour of this model.
In contrast to systems with different symmetry,
for example, uniaxial ferromagnets and ferroelectrics described the TIM
 (Sec.~6), in XY systems above the critical transverse field
the spins at $T = 0$ have maximal projections on the field axis and there are
no quantum spin fluctiations at all.
Below the critical field, in the ferromagnetic phase, such quantum
fluctuations exist;
for the ground state properties of spin systems see Ref.~\cite{KCh:1988}.

In Ref.~\cite{GB:1977} the real symmetry of XY systems has been neglected
and the obtained
RG equations depart from the correct Eqs.~(38)~-~(43) just as the RG equations
 in Ref.~\cite{BDe1:1980}(see Sec. 5.2).
 The ignored symmetry properties are related to the particular form of the bare
 correlation function $G_0(q)$ which
has been obtained in Refs.~\cite{Gold:1979,KCh:1988,Gold2:1979} exactly
of the form~(28).  The
main interaction term in the effective Hamiltonian of the XY model is of the
form given by Eq.~(27),
but there are also additional interaction terms of type
$|\phi|^{2m}$; $m > 2$.
It has been shown in Ref.~\cite{Chub1:1984},
that the
additional interaction terms do not change the critical behaviour
predicted with the help of the
$\phi^4$--Hamiltonian~(27)--(28) and, therefore, the RG results
from Secs.~5.2--5.6,
can be straightforwardly applied to XY systems; see, e.g.,
Ref.~\cite{KopK:1983}.

We should emphasize that the transverse field in the XY model plays the role
of the auxiliary parameter $X$, introduced in Sec.~2.
The $X-$driven transitions considered in Sec.~2.4
are easily performed in $XY$ systems by variations of the transverse field
around its critical value at a fixed low temperature.
Similar field variations in uniaxial ferroelectrics~\cite{Rech:1971,Bruce:1980}
 and ferromagnets~\cite{UZ:1993,Ma:1976} described by TIM
 are discussed in Sec.~6. The measurements in real substances
 should indicate the Gaussain--like critical behaviour in
XY systems and the standard zero-temperature critical behaviour
described by the universality Ising quantum FP of  TIM~\cite{Hertz:1976}.
The results for the quantum ctitical behaviour described by  GLFP
 have been applied~\cite{Sach:1994} also to low-dimensional quantum
 antiferromagnets.

Therefore, according to our considerations in Sec.~2 as well as
the concrete studies of nonuniversal characteristics of XY systems
~\cite{KCh:1987,KCh:1988,Chub2:1984}, quantum critical phenomena should be
 observed in
 low-temperature critical experiments on low dimensional XY ferromagnets.
The arguments presented so far demonstrate that the low-temperature
phase transitions in XY systems are a promising area
of experimental studies of the Gaussian--like critical behaviour. Besides,
 respective experiments on uniaxial ferromagnets and ferroelectrics
 described by TIM are expected to confirm the universal quantum critical
 behaviour.

{\em 5.7.2. Quantum Hall liquid and superconductivity}

 The results
 from Sec.~5.3 can be applied to the (fractional) quantum Hall liquids
(see. e.g.,~\cite{Nagaosa:1999, Zhang:1992}. This problem has been
 investigated by Schakel~\cite{Schak:1995, Schak2:1997}; see, also
Ref.~\cite{Sch:1999}. The quantum Hall liquid was
 considered~\cite{Schak:1995,Schak2:1997} in the framework of the effective
 (quasimacroscopic) Chern-Simons-Ginzburg-Landau (CSGL)
 theory which describes the fractional quantum Hall effect (FQHE)
~\cite{Nagaosa:1999,Zhang:1992}.
 This effective field theory is based on the $\phi^4-$theory given by
 Eqs.~(27) and (28) but the difference is that the Bose field $\phi(x)$
 interacts with the sum of the vector potential $\vec{A}(\vec{r})$
 of the external magnetic field $\vec{H}$ and the
 Chern-Simons gauge field which is related to internal degrees of
 freedom (composite particles featuring the FQHE). It has been shown that
 the basic features of the quantum phase transition in this system
 are described with the help of the exact solution presented in Sec.~5.3
 (see also Sec.~7.6.2 for a discussion of disorder effects).

The relationship between the superconductivity and the quantum
criticality seems to be a new important issue
 in strongly correlated materials,
 including heavy-fermion and high$-T_c$ cuprate superconductors,
 in particular, in cases of unconventional $d-$ and $p-$wave Cooper pairs
 (see, e.g., Refs.~\cite{Annett:1995, Uzunov:1990}).
 The problem has been mentioned
 for the first time within the framework of the
 resonating-valence-bond theory of high-temperature
 superconductors~\cite{Kivelson:1987,Anderson:1987};
 see also Ref.~\cite{Schn:2000} for a discussion of
 quantum critical points in superconductors. Here we shall draw the attention
to  several new results.

 Barzyukin and Gor'kov~\cite{Gorkov:2002} have recently shown
 that in two-dimensional superconductors,
 the phase transition line to the
 Larkin-Ovchinnikov-Fulde-Ferrell state may extend
 up to $T_c=0$ at a certain value of the external magnetic field. If the
theory is reliable, this phase transition line may give a new type of
 quantum (multi)critical point, or, a zero-temperature first-order transition.

Other magnetic field induced quantum phase transitions were
  established  experimentally in heavy-fermion superconductors such as
 Ce-based (CeCu$_{6-x}$Ag$_{x}$~\cite{Heuser1:1998,Heuser2:1998} and
 CeCoIn$_5$~\cite{Sidorov:2002}) and YbRh$_2$Si$_2$~\cite{Gegenwart:2002}
compounds (see also Ref.~\cite{Mathur:1998}). The bilayer ruthenate
Sr$_3$Ru$_2$O$_7$ exhibits a low temperature metamagnetism~\cite{Perry:2001}.
Assuming that for a  certain value of the external magnetic field
the phase transition line may reach the absolute zero,
theoretical arguments~\cite{Millis:2002} have been presented in favour
 of a weakly first  order quantum phase transiton in Sr$_3$Ru$_2$O$_7$.

A phenomenological model~\cite{Walker:2002}, that proposes a coexistence of
 the ferromagnetism and the unconventional superconducting
 state of spin-triplet Cooper pairs in heavy-fermion compounds, yields,
in accord with experiments
~\cite{Tateiwa:2001,Pfeiderer:2001,Aoki:2001}, a
two pressure-driven quantum phase transitions at the same zero-temperature
 multicritical point: a transition from ferromagnetic to paramagnetic phase
 and a transition from superconducting to normal metal.

The theory~\cite{Zaikin:1997} and the experiment~\cite{Giordano:1994} on
 ultrathin superconducting wires demonstrate the crossover from thermally
 activated phase slips (a product of thermal fluctuations) and quantum
 phase slips, activated by the quantum fluctuations. This crossover very much
resembles the general scaling picture of effective HLTC and, in particular,
  CQC at nonzero temperature,
  outside an unattainable by experiment, small asymptotic vicinity of the
critical point
(see the discussion in Sec.~2). Dissipation-driven quantum phase transitons
 in (quasi-)one-dimensional arrays of Josephson junctions~\cite{Bradley:1984}
 are a subject of intensive experimental~\cite{Miyazaki:2002} and
 theoretical~\cite{FisherMPA:1988, Scot:1996} research.

A Monte Carlo calculation~\cite{Schmid:2002} based on a two-dimensional
 boson Hubbard model, equivalent to an anisotropic spin-1/2 XXZ model
 with definite parameters, reveals a phase diagram with a quantum ($T=0$)
 superfluid-solid phase transition. As mentioned in Ref.~\cite{Schmid:2002}
 there is a striking qualitative similarity of this phase diagram to
those of fermionic $^3$He and bosonic$^4$He two-dimensional systems.

{\em 5.7.3. Dilute Bose fluids}

 At the end of this discussion let us come back
 to the systems of real boson particles. A potential application of the
 nonuniversal critical behaviour described in Sec.~5.3 is  to thin films
 of $^4$He~\cite{Bishop:1980,McQuenney:1984}. An interesting opportunity
 exists for an experimental investigation of the crossover of the critical
 behaviour of IBG to the critical behaviour of NBG by a careful
 controlled overall density $\rho$ of bosons up to the dilute limit
 $(\rho \ll 1)$ in $^4$He~\cite{Crooker:1983, Reppy:1984, Finotello:1987,
 Chan:1988, Finotello:1988} and spin-polarized hydrogen~\cite{Silvera:1982}.
The critical fluctuations of the superfluid order parameter $<\phi(x)>$ can
 be accounted by the experimental data for the behaviour of the superfluid
 density $\rho_s \sim <|\phi^2|>$.

 Experiments following this idea have been
 performed~\cite{Crooker:1983,Reppy:1984} for $^4$He condensed in Vicor
 which is a highly connected semiregular sponge-like glass. In other
 experiments~\cite{Finotello:1987,Chan:1988,Finotello:1988}
 another porous medium
 (Silica Gel) has been used. This is again a medium which consists of
 randomly distributed three-dimensional $(d = 3)$ networks of interconnected
 pores and channels of diameter approximately 0.005 $\mu$m (the pore size can
be varied from sample to sample). In these experiments the critical
 temperature $T_c$ of the superfluid phase transition is observed to
 decrease with
 the decrease of the density $\rho$ of $^4$He. The respective curves
can be reliably extrapolated to zero $T_c(\rho = \rho_c)$ for a nonzero
critical density ($\rho_c > 0$).

The experiments~\cite{Crooker:1983, Reppy:1984, Finotello:1987,
 Chan:1988, Finotello:1988} have been extensively analyzed in
 theoretical works~\cite{Wei:1986,FishW:1989, Weichman:1988, MHL:1986,
FF:1988, WeichmanFisher:1986, Rasolt:1984} based on scaling and RG methods.
All mentioned porous media are a source  of disorder, so that certain
properties of the
 Bose fluid in such media should be described with the methods of
 the theory of disordered systems (Sec.~7). However, some main
 features can be discussed by the means of pure systems where no disorder
 effects are present.

 As pointed out in  Ref.~\cite{FishW:1989} such systems undergo a
  superfluid onset transition as the density $\rho$ increases at $T=0$ and,
 moreover, this phase transition may be related with the theory
 presented in the preceding Secs.~5.2-5.6. A similar conclusion has been
 made in Refs.~\cite{MHL:1986, Rasolt:1984} where the empirical data is
  interpreted with the help of the crossover from the critical behaviour
 of NBG to that of IBG. These notes demonstrate that the experiments
are close to a confirmation of the Gaussian-like
 critical behaviour established in Sec.~5.3, provided the respective porous
medium can be considered as a factor
 leading to a decrease of the (effective) spatial dimensionality of the
 Bose fluid contained in it up to $ d \leq 2$. Similar experiments
 can be made on thin superfluid films.

   Following Ref.~\cite{FishW:1989} we
 shall briefly mention some properties of this experimentally observed phase
 transition in $^4$He placed in Vycor. While the experimental feasibility of
 the probing low-temperature behaviour stops short of $T=0$, the superfluid
 transition at sufficiently low temperatures will be dominated by
 GLFP until the very close vicinity of the critical density $\rho_c$ is
 approached. The crossover from the zero-temperature critical behaviour
 described by  GlFP and the finite-temperature critical behaviour
 corresponding to $ T_c > 0$, described by the standard XY FP$(n=2)$ has
 been treated in Ref.~\cite{FishW:1989} as a standard
 multicritical-to-critical crossover (within this interpretation the
 zero-temperature critical point is considered as a multicritical point).

 We should draw the attention to a
quantum $(T=0)$ phase transition~\cite{Matsumoto:1997}
 to superfluid state in $^3$He in aerogel
 at a nonzero pressure and a critical density of $^3$He $\rho_c > 0$. Unlike
 bulk $^3$He which is superfluid at all pressures (densities) between
 zero and the melting pressure, $^3$He in aerogel will be  not superfluid,
unless
 the density of $^3$He exceeds a critical value $\rho_c > 0$. Some of the
 problems of the quantum phase transition revealed
 in Ref.~\cite{Matsumoto:1997} are similar to that for
 the respective phase transition in $^4$He, discussed in a few lines above
 (for details, see Ref.~\cite{Matsumoto:1997}).

{\bf 6. Transverse Ising model}

{\em 6.1. Microscopic and field models}

Here we shall consider the quantum transverse Ising model with the
aim to reveal the properties of  HLTC.  We shall use MF, the  lowest-order
perturbation theory and Ginzburg criterion considerations and RG.

TIM describes essential features of phase transitions in ferromagnets
with a strong
uniaxial anisotropy~\cite{Genn:1963} and displacive phase transitions
in certain types of quantum ferroelectrics~\cite{Rech:1971, Beck:1975,
Schn:1976,Morf:1977,Bruce:1980}; for experiments, see, e.g.,
Refs.~\cite{Sam:1981, Rytz:1980} (for other
applications of TIM, see Refs.~\cite{Stinch:1973,Chak:1996}).
Here we shall mention the MF results~\cite{Genn:1963},
the cumulant expansion~\cite{Stinch:1973},
the RG investigations~\cite{Young:1975, Hertz:1976, Schn:1976,Morf:1977},
the description of CQC in Hartree limit~\cite{Law1:1978, Law2:1978}
and the formal interrelationship between the finite--size crossover in systems
of slab geometry and CQC in TIM~\cite{Law1:1978}.

TIM is given by the Hamiltonian~\cite{Genn:1963}
\begin{equation}
H = - \frac{1}{2} \sum_{ij}J_{ij} S^{z}_{i}S^{z}_{j} - \Gamma \sum_{i}\;
S^{x}_{i} \; ,
\end{equation}
where $S_{\gamma}$, $ \gamma = (x,y,z)$, are the components of spins with
a magnitude $S = 1$, $J_{ij}$ is the
exchange interaction and $\Gamma$ is
the transverse field magnitude.  We shall assume that TIM is defined on
a $d$--dimensional regular lattice with a lattice spacing $a = 1 $.  The
assumption for
nearest--neighbour $(nn)$ interactions which we shall use to define the
parameters
of the effective Hamiltonian, does not restrict the generality of results in
this Sec.

The fluctuation field Hamiltonian of TIM can be derived from the microscopic
 spin model~(60) by the Hubbard-Stratonovich transformation~\cite{Young:1975,
Law1:1978,Law2:1978}.  This Hamiltonian (${\cal{H}} = H/T$; $k_B = 1$) is
given in the form
\begin{equation}
{\cal{H}}[\phi] \; = \; \frac{1}{2}
\sum_{\alpha,q}G^{-1}_0(q)|\phi_{\alpha}(q)|^2 \; + \;
\frac{u_o}{V}\sum_{\alpha,\beta;q_1q_2q_3}\phi^*_{\alpha}(q_1)
\phi^*_{\beta}(q_2)\phi_{\alpha}(q_3)\phi_{\beta}(q_1+q_2+-q_3)\;,
\end{equation}
where $\phi(q)$ is a real scalar field with
a (bare) correlation function in the form
\begin{equation}
G_0^{-1}(q) \; = \; |\omega_l|^{2} + k^{2} + t_0\;.
\end{equation}
The parameters $t_0$ and $u_0$ are given by the expressions,
\begin{equation}
t_{0} = \left [1 - \frac{J}{\Gamma}\mbox{th} \left
(\frac{\Gamma}{T} \right ) \right ] \;, \end{equation}
and
\begin{equation}
u_{0} = \frac{J^{2}T}{8\Gamma^{3}} \left [\mbox{th} \left
(\frac{\Gamma}{T} \right ) - \frac{\Gamma}{T} + \frac{\Gamma}{T}\mbox{th}^{2}
\left (\frac{\Gamma}{T} \right ) \right ]\;.  \end{equation}
In Eq.~(64), $J \equiv J(0) = 2dJ_{0}$ is a product of the number $z = 2d$ of
nearest
neighbour {\it nn} spins and the constant $J_{0}$ of the single $(i-j)$ {\em
nn} interaction ($J_{0} = J_{ij}$ for {\it nn} sites {\it i} and {\it j}).

It is convenient to use units, in which all quantities in the effective
Hamiltonian ${\cal{H}}$ are dimensionless.  The dimensionless
wave components $k_{i}$ are given by $k_{i} = 2\pi\kappa l_{i}/L_{i}$, where
\begin{equation}
\kappa = \sqrt{\frac{J\mbox{th}(\Gamma/T)}{2d\Gamma}}\;\;.
\end{equation}
In the critical region $t_0 < 1$, $\kappa \sim 1$.
The upper cutoff for the wave
numbers $k$ is $\Lambda = \gamma\pi\kappa$, where $\gamma$ is a small number
($\gamma \ll 1$)~\cite{UZ:1996}.

The characteristic lengths $\xi$ and $\lambda$ are given by
 $\xi =1\sqrt{|t_0|}$
and $\lambda = (\Gamma/T)$.  We shall consider only the low temperature
 domain defined by $\lambda \gg a$.  In this case, $u_0 =
(J^2 T/8\Gamma^3)$.

{\em 6.2. Mean field approximation}

Within the standard MF
approximation we shall consider only the uniform mode $\phi(0)$ of the field
$\phi$, that is the $q-$dependent (classical and quantum) fluctuations will
 be ignored. It will be more convenient to use the MF free energy $\Omega =
T{\cal{H}}[\phi(0)]$
instead of the dimensionless MF free energy ${\cal{H}}[\phi(0)]$.  This choice,
after a change of the nonequilibrium order parameter from $\phi(0)$ to
$\phi_{0} = (T/V)^{1/2}\phi(0)$, will induce an extra factor $1/T$ in
front of $\phi_{0}^{4}$--term of free energy $\Omega$ which makes possible
to avoid difficulties in our further analysis connected with the definition of
the order parameter
$\phi_{0}$ at $T = 0$.  The problem is that the parameter
$u_{0}$ is proportional to $T$, see Eq.~(64), and the investigation with the
original parameter $\phi(0)$ will give for the equilibrium order
parameter $\phi(0) \sim (1/T)^{1/2}$ which is divergent for $T \to 0$.  This is
an example of order parameter HLTC.

The Gibbs thermodynamic potential in the form
\begin{equation}
\Omega = V\left [ \frac{t_{0}}{2}\phi^{2}_{0} + \frac{u_{0}}{T}\phi^{4}_{0}
\right ]\;
\end{equation}
allows a correct MF analysis at low temperatures.  The analysis of the free
energy (66) can be done straightforwardly  and
we shall not enter in details.  The critical line (Fig.~4a) is defined by
$t_{0}(T_{c}, \Gamma) = 0$, i.e.,
\begin{equation}
T_{c}(\Gamma) = \frac{\Gamma}{\mbox{Arth}(\Gamma/J)}
\end{equation}
or, equivalently, by $t_{0}(T,\Gamma_{c}) = 0$, which yields
\begin{equation}
 \frac{\Gamma_{c}}{J} = \mbox{th} \left (\frac{\Gamma_{c}}{T}
\right )\;.
\end{equation}

Despite of the simple form of Eq.~(66) the MF critical properties of TIM
cannot be investigated analytically for the whole curve $T_{c}(\Gamma)$
because of the quite complex dependence of parameters $t_{0}$
 and $u_{0}$ on $T$ and $\Gamma$.

\begin{figure}
\begin{center}
\epsfig{file=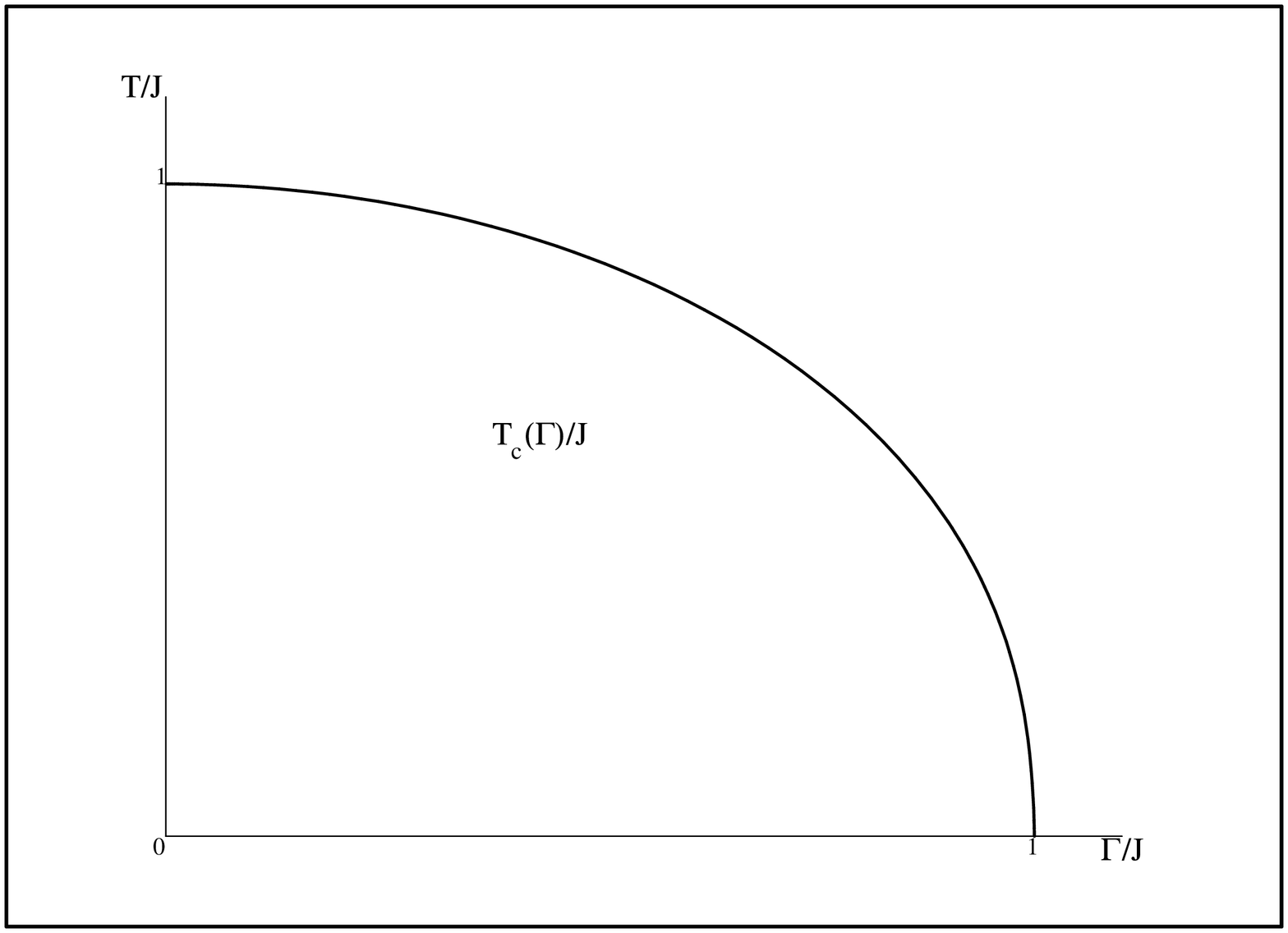,angle=-90, width=7.5cm}
\epsfig{file=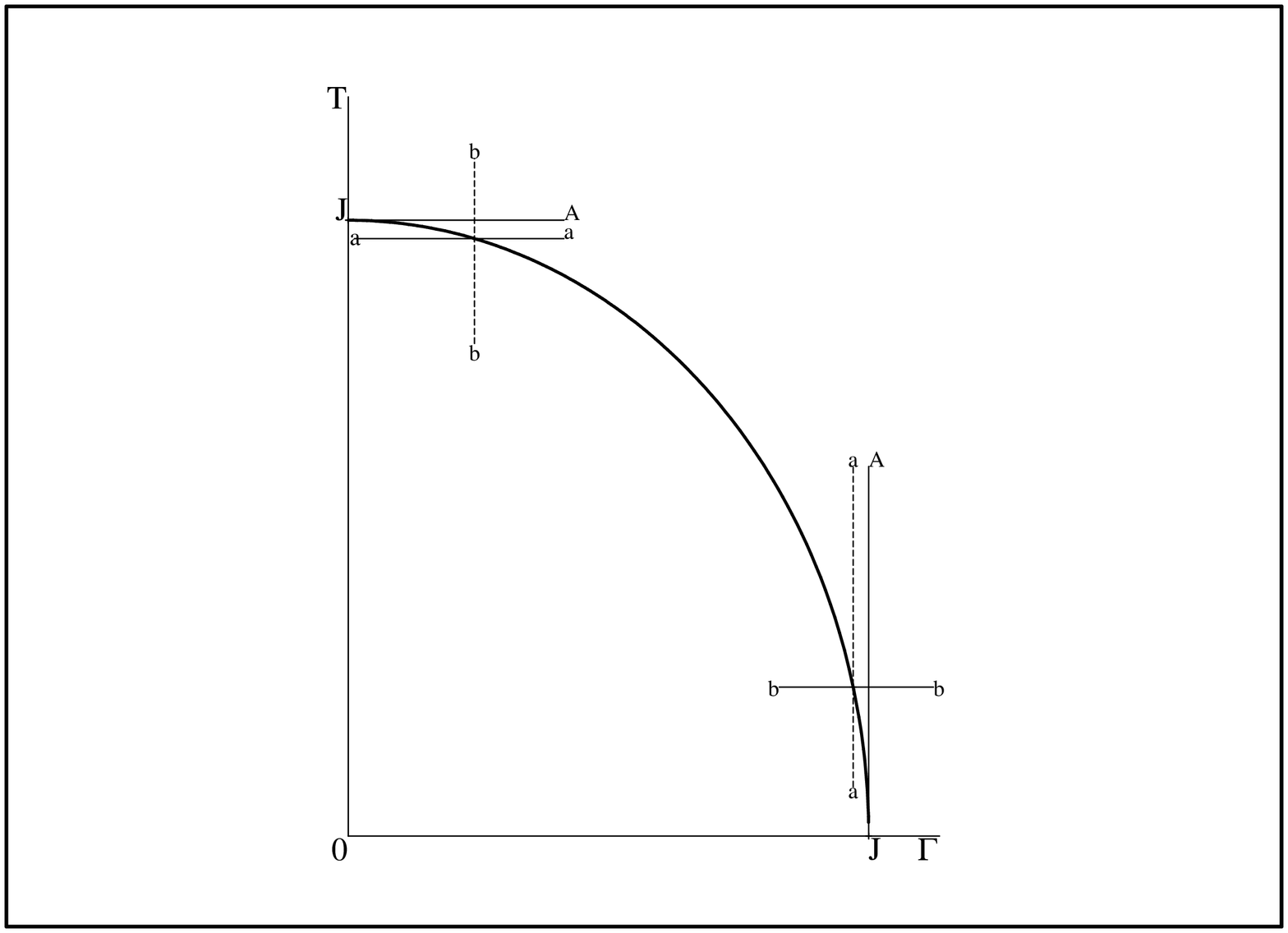, angle=-90, width=7.5cm}\\
(a)\hspace*{7cm}                   (b)
\end{center}
\caption{(a) The graphical representation of Eq. (67) for $J=1$. (b)
The high- and low-temperature parts of the curve in Fig.4a with an
indication of
 $T-$ and $\Gamma-$ transitions (see the text).}
\label{QPTf4.fig}
\end{figure}

We shall consider the low temperature critical behaviour in the vicinity
($|\Gamma - J| < J$) of zero temperature critical point
 [$ T_{c}(J) = 0, \Gamma_{c}(0) = J $]
and the neighbour critical points with coordinates
 $T_{c} \ll J$ and $\Gamma_{c} \sim J$. In this low temperature region we can
 distinguish four types of phase transitions along the lines:  {\it 0J},
{\it AJ}, {\it aa}, and {\it bb}; see Fig.~4b.  We shall suppose that the
couples of parallel lines are very near to each other.
 Following the terms introduced in Sec.~2,
the transitions along the lines {\it AJ} and {\it aa} are $T$--transitions
whereas the transitions along the lines
{\it 0J} and {\it bb} can be thought of as $\Gamma$--transitions.
The results can be compared with those for the high temperature
$T-$ and $X-$transitions in the vicinity of the critical point $T_c(0)$;
see the respective lines $0J$, $bb$, $AJ$, and $aa$ in Fig.4b.

In MF we shall consider the critical exponents $\beta$,
$\gamma$, and $\nu$, of the order parameter $|\phi_0| = (T|t_0|/4u_0)^{1/2}$,
the susceptibility
$\chi = 1/|\bar{t}_0|$, and the correlation length $\xi = 1/|\bar{t}_0|^{1/2}$,
respectively,
where $\bar{t}_0 = t_0$ for $t_0 > 0$, and $\bar{t}_0 = 2t_0$ for $t_0 < 0$.
Obviously, the behaviour of these
quantities with respect to variations of $T$ and $\Gamma$ depends on the
behaviour
of $t_0$.  Note, that parameter $t_0$ as given by Eq.~(63) does not change in
the Gaussian approximation for the
quantum and classical fluctuations.  This makes possible to extend the
consideration of the
parameter $t_0$ in the paraphase $(\phi_0 = 0)$, where the results should be
interpreted as pure fluctuation effects.

The summary of results reads:

{\it Line 0J}:  We obtain $t_{0} \approx (\Gamma - J)/\Gamma$, $\phi_{0}
\sim |t_{0}|^{1/2}$, $\chi \sim \Gamma/|\Gamma - J|$ and, therefore,
$\gamma = 1$ and $\beta = 1/2$.
There exists a full
coincidence between the standard MF behaviour with respect to
variations of $T$ around $J$ at $\Gamma = 0$ and the present critical
behaviour due to variations of $\Gamma$ around $J$ at $T = 0$.  This
correspondence can be written as $T \leftrightarrow \Gamma$.

{\it Line bb:} The parameter $t_{0}$ is
 \begin{equation}
t_{0} \approx \frac{J(\Gamma - \Gamma_{c})}{\Gamma_{c}^{2}}(1 -
\frac{4\Gamma_{c}}{T}e^{-2\Gamma_{c}/T} + ...)\;.
\end{equation}
Here $t_{0}$ tends to the value $(\Gamma - J)/\Gamma$ when
$\Gamma_{c}(T) \to J$ for $T \to 0$.  The critical exponents are the same as
those along the line {\it 0J}.
 The exponential correction in Eq.~(69) can be neglected.

{\it Line AJ:} There is no equilibrium ordering along this line but there is a
criticality in the paraphase.  The susceptibility above the zero-temperature
critical $(\Gamma_c = J)$ is
\begin{equation}
t_0 = 2e^{-2J/T}\;.  \end{equation}
This exponential behaviour corresponds to the critical
 exponents $\gamma = \nu = \infty$.

{\it Line aa:} Along this line
 \begin{equation}
t_0 \approx \frac{4J|T - T_{c}|}{T_{c}^{2}}e^{-2\Gamma/T_{c}}
\end{equation}
and, therefore, for both $T < T_c$ and $T > T_c$, the susceptibility $\chi \sim
 1/|t_0|$ obeys the scaling law
$\chi = \chi_0/|T - T_c|$ with an exponentially increasing scaling amplitude
$\chi_0$.
There is a smooth crossover between the pure exponential behaviour (70) and the
power law
(71).  As $ (u_0/T) \approx J^2/8\Gamma^3$ for $ \Gamma \sim J \gg T$, we have
$|\phi_0| \sim |t_0|^{1/2}$, where $t_0 $ is given by Eq.~(71).  The order
parameter exponent $\beta$ has the classical value $\beta = 1/2$ but the
scaling
amplitude $[\sim T^{-1}_c\mbox{exp}(-\Gamma/T_c)]$ of the order
 parameter $\phi_0$
exponentially decreases for $T_c \rightarrow 0$; for a comparison of these
low-temperature MF properties with high-temperature ones; see
Ref.~\cite{DeC:1998}.

The Eqs. (7),~(70)~and~(71) yield a crossover exponent $\nu_0 = \infty$.
Perhaps, this value of $\nu_0$ will be different provided the
 fluctuation phenomena are taken into account.

{\em 6.3. Lowest order perturbation theory}

The $(T,\Gamma)$ domains of validity of the MF results in Sec.~6.2 can be
investigated by the Ginzburg criterion; see, e.g., Ref.~\cite{UZ:1993}.
Here we shall use the
standard derivation of this criterion from the first--order perturbation
contribution to the ``self--energy'' $t_{0}$:
\begin{equation}
\tilde{t}_{0} = t_{0} + 12u_{0}A_{1}(t_{0}) \end{equation}
with
\begin{equation} A_{1}(t_{0}) = \int \frac{d^{d}k}{(2\pi)^{d}}S_{1}(t_{0},k)\;,
\end{equation}
where
\begin{equation} S_{1}(t_{0},k) =
\frac{\Gamma\;\mbox{cth}[(\Gamma/T)\sqrt{k^{2} + t_{0}}]}{T\;\sqrt{k^{2} +
t_{0}}}\;.  \end{equation}
Neglecting a term of order $u^{2}_{0}$ we can substitute $t_{0}$ in the
intergal $A_1(t_0)$ by
$\tilde{t_0}$.  The Ginzburg criterion for the validity of the MF results will
be given in the general form
\begin{equation}
|\tilde{t}_{0}| > 12u_{0}[A_{1}(0) - A_{1}(\tilde{t}_{0})]\;.
\end{equation}
The same criterion defines the validity of the Gaussian approximation of
noninteracting fluctuations above and below the critical point.
The general criterion (75) is valid for both the paramagnetic $(\tilde{t}_{0} >
0)$ and the ferromagnetic
$(\tilde{t}_{0}< 0)$ phases; in the ferromagnetic phase a factor $1/2$ should
be added to the r.h.s.  of (75).

Obviously the results from this first--order perturbation approximation depend
on
the properties of the integral (73).  We shall write this integral in the form
\begin{equation}
A_{1}(\lambda, \xi) = K_d\lambda^{2 - d}\int\limits_{0}^{\lambda\Lambda}dy
y^{d-1} \frac{\mbox{cth}\sqrt{y^{2} + \varrho^2})}{\sqrt{y^{2} + \varrho^2}}\;,
\end{equation}
where $ y = \lambda k$, and $\varrho$ is given by Eq.~(3).
For high temperatures ($\lambda < 1$, $\varrho \ll 1)$
the main contributions in the integral are from the small wave numbers $y =
\lambda k \ll 1$, and
the approximation $\mbox{cth} y \sim 1/y$ yields the standard perturbation
integral $A_1$ known from the classical theory (the Curie--Weiss limit).
One can rederive the classical theory ($\omega_l = 0$) by using
the high-temperature value of the interaction parameter:  $u_0 = (J^2/12T^2)$;
see Eq.~(64) for $\Gamma \ll T$.

In the low-temperature
range of temperatures ($\lambda \gg 1, \lambda\Lambda \sim 1$) we can
distinguish between
the quantum limit $\varrho \gg 1$ and the classical limit $\varrho \ll 1$.  In
the quantum limit
$(\varrho \gg 1)$ the $coth$ can be approximated with unity by neglecting
exponentially small correction terms.  They are very
similar to those given by Eq.~(69)--(71).  These corrections enter in the
renormalized parameter
$\tilde{t}_0$ and give small ($\sim u_0$) corrections to the coefficients of
the corresponding exponential terms in $t_0$; see Eqs.~(70) and (71).

Let us denote the integral which is obtained from
$A_1$ for $\mbox{cth}y \sim 1$ by $A_{01}$.
The difference $u_0[A_1 - A_{01}]$ has been
estimated~\cite{Rech:1971} to be of order $\lambda^{-2} \sim T^2$ for
$3d$--systems; note that in the low-temperature limit, $u_{0} \sim T$.
This type of temperature corrections to the pure quantum
limit has been widely used in interpretations of
experimental results for quantum ferroelectrics~\cite{Sam:1981,Rytz:1980}.
The same
corrections were also derived in Refs.~\cite{Schn:1976,Morf:1977}; for a
calculation in the Hartree limit, see Ref.~\cite{Opp:1975}.

At extremely low but finite temperature $(T \sim 0)$ one may try to take
 into account the temperature corrections to the critical behaviour
 at the absolute zero $(T=0)$ as an alternative to the present consideration,
where we take the low temperature limit in the general self-energy
 integral~(76). Taking into account the temperature corrections
 to the zero-temperature limiting case is equivalent to an
estimation of the difference arising from the substitution
of summation over the Matsubara frequencies with an integration according to
the rule~(35). It is  clear that such corrections will come from the
 counterterm series in powers
of $\varrho^{-2}$ in the Euler--Maclaurin summation formula.  Therefore, these
corrections
could not be given by $d-$dependent powers in $T$ as is in the Hartree limit
considered in Ref.~\cite{Opp:1975}.

The temperature
corrections are important for the phase transition properties along lines like
$AJ$ and $aa$, where $\Gamma \sim J$.  Within the present lowest order
perturbation theory the corrections in
powers of $T$ should be considered small.  However, it has been proven by the
``parquet" summation~\cite{Rech:1971}
that they are quite big and essentially influence the critical behaviour in
the classical low-temperature region ($\varrho < 1, \lambda \gg 1 $).

Now we shall consider the Ginzburg criterion (75).  In the quantum limit we
must substitute
the integral $A_1$ with $A_{01}$.  As a result of CQC, the integral $A_{01}$
yields
the upper and lower borderline dimensionalities:  $d_U = 3$ and $d_L = 1$.  In
order to simplify the
calculations we shall consider the case $d = 2$ which shows the main features
of the quantum critical behaviour for all dimensionalities $1 < d < 3$.

The straightforward calculation gives the criterion (75) in a simple form:
\begin{equation}
|t_{0}| > \left (\frac{3}{4\pi} \right )^{2} \left (\frac{J}{\Gamma} \right
)^{4}\;.  \end{equation}
This criterion cannot be applied to the $T$--transitions along the lines $AJ$
and $aa$ where
the behaviour is classical ($\lambda < \xi$).  The reason is that the strong
quantum condition
($\varrho \gg 1$) has been used in the derivation of inequality~(77).

Along the line {\it bb} we obtain the criterion
\begin{equation} |\Gamma - \Gamma_{c}| > 0.06\Gamma_{c} \left
(\frac{J}{\Gamma_{c}} \right )^{3} \sim 10^{-2}\Gamma_{c}\;\;.  \end{equation}
If we set in (78) $\Gamma_{c} = J$, we shall find a criterion along the line
{\it 0J}.  Because of the very sharp slope of the transition curve
$T_{c}(\Gamma)$
near $\Gamma = J$, the quantum criterions along the lines {\it bb} and {\it 0J}
are practically the same. The Ginzburg critical region $(\sim
10^{-2}\Gamma_{c})$ given by (78) is well established.  It enlarges at $T > 0$
by $T$--correction terms.

The quantum condition (3) along the line {\it bb} can
be written in the simple form $J(|\Gamma - \Gamma_{c}|) \gg T^{2}$ provided
$(\Gamma/\Gamma_{c})^{2} \approx 1$.  As $J \gg T$, this condition
becomes
\begin{equation}
|\Gamma - \Gamma_{c}| > T\;, \end{equation}
which is obviously consistent with (78).
The condition (79) shows that the
quantum region approaches the critical point $\Gamma_{c}$ when $T$
decreases to zero.  The whole surrounding of the zero-temperature
critical point $(\Gamma_{c} = J)$ is influenced by
quantum effects which produce the quantum fluctuation region.

A more detailed analysis
including the onset of thermal fluctuations can be obtained by the exact
calculation of the integral $A_{1}$ at $d = 2$:
\begin{equation}
A_{1} =
 \frac{1}{2\pi}\mbox{ln}\frac{\mbox{sh}\sqrt{\lambda\Lambda + \varrho^2}}
{\mbox{sh} \varrho}\;.
\end{equation}
The approximation $\mbox{sh}y \sim \mbox{exp}(y)/2$ corresponds to $A_1
\approx A_{01}$.  The thermal
corrections are obtained in powers of $(\varrho^2/\lambda\Lambda)$ which,
together with the factor
$u_{0} \sim T$, yields the lowest order correction term to $u_0A_{01}$ of the
type $T^2|t_0|$.

The asymptotic critical behaviour corresponding to the
$T-$transitions will not exhibit quantum critical phenomena
but rather a low-temperature classical behaviour which is different
from the high-temperature one.  The only quantum effect is related
with CQC at zero-temperature $\Gamma-$transitions
(Sec.~6.4).  A study of the Ginzburg critical region for zero-temperature
structural phase transitions has been presented in Ref.~\cite{UZ:PSS83}.
The results are consistent with the present analysis.

{\em 6.4. Renormalization group arguments}

The RG recursion relation for the interaction parameter $u_{0}$ in the
one--loop approximation will be
\begin{equation}
u'_{0} = b^{4-d}u_{0}[1 - 36u_{0}A_{2}(0,b)]\;,
\end{equation}
where
\begin{equation}
A_{2}(t_{0},b) = - \frac{\partial A_{1}(t_{0},b)}{\partial
t_{0}}\;.
\end{equation}
In Eq.~(82) $A_{1}(t_{0},b)$ is the integral (73) with lower ($0 < b^{-1} < 1$)
and upper ($\Lambda = 1$) cutoffs of the wave number $k$.
The other recursion relations are
~\cite{Bruce:1980}
\begin{equation}
t'_{0} = b^{4-d}[t_{0} + 12u_{0}A_{1}(t_{0},b)]\;, \end{equation}
and
\begin{equation}
\lambda' = b^{-z} \lambda\;, \end{equation}
where the dynamical critical exponent $z$ is
equal to unity in this order of the theory.

In the high-temperature
region ($\lambda < a$) the integrals $A_1$ and $A_2$ can be
substituted with the classical integrals by setting $\mbox{cth}(y) \sim 1/y$.
In this case the integral
$A_{2}(0,b)$ has a logarithmic infrared divergence at the upper critical
dimensionality $d_{U} = 4$.  For $d = 4$, $A_{2}(0,b) = K_{d}\mbox{ln}b$.

Further, using the standard RG analysis~\cite{Hertz:1976} one reveals the usual
universality class of the critical behaviour of the classical Ising model
($\Gamma = 0$). Besides, there is a possibility to perform a calculation of
dynamical critical exponent $z$ and, hence, to reveal the quantum dynamics of
TIM~\cite{MU:1983}.
The $\epsilon$ corrections to $z$ are calculated from the
$q$--dependent self--energy diagrams in two-- and higher--loop approximations.
These corrections are small compared with the $\epsilon$--corrections to the
static exponents, for example, $\nu$ and $\gamma$.

The low temperature limit is treated as shown in Sec.~6.2.  In this case we
should make the approximation
$\mbox{cth}y \sim 1$ in Eqs.~(81)--(83).  Thus we recover the dimensional
CQC:  $d \rightarrow (d + 1) = D$.
The analysis in $\epsilon = (3 - d)$ yields results
for the quantum critical behaviour corresponding to the nontrivial Ising
universality  class $(D, 1)$~\cite{Hertz:1976}.
For all dimensionalities $d \geq 3$ the QC behaviour will be
described by the classical MF universality class,
while for  dimensionalities $1 < d < 3$ the QC behaviour
will be nontrivial.  This result is straightforwardly generalized
for a $n$--component real field with a bare correlation function of the
form (62).

Thus we obtain that in TIM CQC satisfies the universality property
discussed in Sec.~4.2.
An important point in the mechanism of this crossover is that the interaction
parameter $u_0$ in the recursion
relation (81) is changed to the parameter $v_0 = (u_{0}/\lambda)$:
\begin{equation}
v'_{0} = b^{3-d}v_{0}[1 - 36v_{0}K_{3}\mbox{ln}b]\;.  \end{equation}

The quantum effects play the crucial role for CQC in TIM.  If the RG equations
are treated
in the classical scheme, as shown in Refs.~\cite{Beck:1975, Morf:1977},
the transformation of Eq.~(81) for $u_{0}$ to that for $v_0$ cannot be
performed
in the way shown by Eq.~(85).  An extra--factor $\lambda^{-1}$ that remains in
the second term of
Eq.~(85) leads to the prediction of a Gaussian critical behaviour at low
temperatures for dimensionalities $d > 2.$

The RG investigation~\cite{Schn:1976,Morf:1977}
 of the critical behaviour along the
 onset $T-$transition at $T_c=0$ yields critical exponents
$\gamma = 2\nu = 2$ instead
of classical exponents $\gamma = 2\nu = 1$
 (here we ignore $\epsilon-$corrections).  These exponents
do not depend on the dimensionality $d$ as is in the Hartree
limit~\cite{Opp:1975}.
For $t_0 = 0$, i.e., on the line $AJ$, the integral $A_1$
has a logarithmic divergence which
is an evidence of strong thermal fluctuations. Perhaps this nonuniversal
  critical exponetns of the zero temperature $T-$transition are the outcome of
 the simultaneous effect of classical and quantum fluctuations.

The HLTC of the $T$-transitions from classical critical exponents at high
temperatures to low-temperature critical exponents at the zero temperature
phase transition $(T_c=0)$,
$\gamma = 2\nu = 2$, shows that $z \nu_0 = 1$. Therefore, within this accuracy
 of the theory one cannot conclude whether the cirterion~(9) is satisfied or
 not.  These bare values of $z$ and $\nu_0$ have
$\epsilon$ corrections from RG and we should have in mind that the correction
to $z$ is of order $O(\epsilon^2)$ whereas that to $\nu$ is of first order in
$\epsilon$ with a positive sign. Therefore the criterion (9) is not satisfied
 and, hence, the asymptotic critical behaviour along the onset $T-$transition
to the absolute zero $(T_c=0)$ is classical.

Another important feature of the quantum critical
 behaviour is that it is unstable with respect to any
perturbation of the temperature from zero.  This feature is common to
the quantum critical phenomena
in all systems.  Of course, the quantum critical phenomena
will belong to the nontrivial ($d$--dependent) class of
critical behaviour in the quantum critical region shown by Eq.~(78).
The zero-temperature transitions outside
this narrow region will exhibit the usual MF behaviour.
It becomes clear from this picture that like
in the XY model discussed in Sec.~5.6, the uniaxial ferroelectics and
ferromagnets described by TIM,  are convenient
for the  experimental observation of quantum critical phenomena
near $\Gamma$--transitions at low and extremely low temperatures.

{\bf 7. Disorder effects}

Here we shall review investigations on the effects of quenched disorder
of type ``random critical temperature" which is caused by randomly
distributed (quenched) impurities and inhomogeneities
~\cite{UZ:1993, Ma:1976, Hertz:1985, Grinstein:1985, UZ:1987}.
The main results~\cite{UKM1:1985,UKM2:1985} for the quantum critical behaviour
in systems with disorder of type ``random field" will also be mentioned
 (for the random field disorder see, e.g., Refs.~\cite{UZ:1993, Hertz:1985,
Grinstein:1985}).

The critical behaviour in disordered systems obeys
the Harris criterion derived by rather general
arguments~\cite{Harris:1975}.  This criterion states that the disorder will be
irrelevant to the (classical) criticality, if the following inequality between
the specific heat exponent $\alpha$ and the correlation length exponent $\nu$
takes place:  $(\alpha \nu - 2) > 0$.  One can check that
this criterion is satisfied in all examples considered below.

The first RG investigation~\cite{Kor:1983,Kor:1984} of the quantum critical
behaviour in disordered quantum
systems reveals an instability of the usual quantum critical phenomena at
$T = 0$ towards the disorder of type random impurities.
The instability
has been proven~\cite{Kor:1984} for a quite general quantum field model
 which describes the quantum phase transitions in
 almost all known quantum systems. This problem is closely related with
the description of the localization in real superfluids.
 Here we shall consider the instability phenomenon,
 RG methods of stabilization of the quantum criticality and the related
 problems of localization and superfluid-insulator transition as well as the
 formation of ``glass-like'' phases ~\cite{FishW:1989,MHL:1986} in
 disordered Bose fluids.

{\em 7.1. Random impurities}

The disorder of type random impurities changes the local interaction
 responsible
for the phase transition and, hence,
the critical temperature which depends on the sites of the crystal
lattice.  In the continuum limit the local (nonequilibrium~\cite{Hertz:1976})
critical temperature
depends on the spatial vector $\vec{r}$.  This fact is taken into
account in the effective Hamiltonian
by an additional $\phi^2$--term containing a random
function $\varphi(\vec{r})$ that obeys the Gaussian distribution
\begin{equation}
g(\vec r , \vec r\:^\prime) \equiv [\varphi_\alpha(\vec r)
 \varphi_{\alpha^\prime}(\vec r\:^\prime)]_R =
 \bar{\Delta}\delta_{\alpha \alpha^\prime} \delta(\vec r -
\vec r\:^\prime)\;,
\end{equation}
or, in the $\vec{k}$--space,
\begin{equation}
 g(\vec k, \vec k^\prime) \equiv [\varphi_\alpha(\vec k)
\varphi_{\alpha^\prime}(\vec k^\prime)]_R = \Delta\delta_{\alpha \alpha^\prime}
\delta (\vec{k} +\vec{k}^\prime, 0)\;.
 \end{equation}
Here $[\;]_R$ denotes the operation of averaging (hereafter,
 the suffix ``R'' will stand for quantities related to disordered systems).
We shall often use the notations $g_{\alpha \alpha^\prime} =
 \delta_{\alpha \alpha^\prime}g$ and $\tilde{g}_{\alpha \alpha^\prime} =
 \delta_{\alpha \alpha^\prime}\tilde{g}$.

The distribution function (86)   or respectively,
 $\tilde{g}(\vec r,\vec r\:^\prime)$, which is
 often called ``random correlation function,''
describes quenched impurities with the
so--called short--range random correlations. Sometimes, this type of random
impurities  is referred to as ``$\delta-$correlations'' or, even,
 ``$\delta-$impurities''  because of their mathematical expression
 by the $\delta-$function valid in the continuum limit.

For the long--range random correlations,
the $\delta$--function  in Eq.~(86) should be substituted by a function of type
$f(R) = 1/R^{\Theta}$, where $R = |\vec{R}| = |\vec{r} - \vec{r^\prime}|$,
 $ 0 < \Theta < d$.  In this case, the random distribution (correlation)
function is given by
$\tilde{g}(\vec r, \vec r\:^\prime) \equiv g(R)$ and~\cite{WH:1983}
\begin{equation}
g(R) = \frac{\bar{\Delta}}{R^{\Theta}}\:.
\end{equation}

  A justification of this definition of long-range correlations is
presented in an investigation~\cite{UKM1:1985} of the random-field
 problem~\cite{UZ:1993,Hertz:1985,Grinstein:1985}. The Fourier
 transformations of various shapes of random correlation functions $g(R)$
 with powerwise or more complex dependence on the distance $R$
 are considered in details, and the Fourier transforms
 $g(k, \vec k^\prime)= \delta (\vec k + \vec k^\prime) g( k)$ are
 given for various shapes of $g(R)$~\cite{UKM1:1985,UKM2:1985}.
Note, that in the general case of random correlations,  different from
the $\delta-$correlations given by Eq.~(86), the parameters $\bar{\Delta}$ and
$\Delta$ in Eqs.~(86) and (87) can be different from each other but
both of them should be nonnegative
on account of the requirement for the stability of the  Gaussian distribution.

An alternative way of representation~\cite{UKM1:1985,UKM2:1985}
 of the correlation function $g(R)$ from Eq.~(88)
 seems to be more convenient for
 several considerations (see Sec.~7.4.1). This representation is given by
 the exponent $\Theta^\prime = (d - \Theta)$, $d \geq \Theta^\prime > 0$ and
  is suitable for the scaling and RG analysis in the $\vec{k}-$space.

In order to describe the disorder of type random impurities a new term should
 be added to the quantum Hamiltonian (27) in the form
\begin{equation}
{\cal{H}}_R[\phi] \; = \; \frac{1}{\sqrt{V}}
\sum_{\alpha,\omega_l;\vec{k}_1,\vec{k}_2 } \varphi(\vec{k}_1 -
\vec{k}_2)\phi^{\ast}_{\alpha}(\omega_l,\vec{k}_1)\phi_{\alpha}
(\omega_l,\vec{k}_2)
\;.  \end{equation}
This term can be derived by the Hubbard--Stratonovich transformation
from the microscopic model of the concrete system or, alternatively,
postulated and used in further investigations of disorder effects.

The sum of the Hamiltonian parts (27) and (89), i.e., the total Hamiltonian
\begin{equation}
\cal{H}_{\mbox{\footnotesize tot}}(\phi) = {\cal{H}}(\phi) +
{\cal{H}}_R(\phi)\:,
\end{equation}
describes a great amount of quantum systems with quenched
 (randomly distributed) impurities,  often called ``impure systems.''
As usual, we shall discuss also the critical behaviour
 of the corresponding classical systems ($\omega_l \equiv 0$). The
 respective pure systems are those, for which the Hamiltonian part
 ${\cal{H}}_R(\phi)$ is equal to zero [$\varphi(\vec{k}) \equiv 0$, or,
 $g(R) = 0$, which gives the same results for the thermodynamic and
 correlation properties].

The thermodynamic behaviour strongly depends on the properties of the
random potential represented by the random function $\varphi(\vec{r})$.
The random potential $\varphi(\vec{r})$ is not a thermodynamic variable,
that is why, the theoretical treatment will include two steps. At first one
should try to calculate
the thermal averages as functionals of $\varphi(\vec{r})$. As a second step
the averaging over the random function is done with the help of
Eqs.~(86)--(88).  As a result one obtains
the thermodynamic quantities as functions of the disorder parameter $\Delta$;
see.  e.g., \cite{UZ:1993, Ma:1976,Hertz:1985,Grinstein:1985}.

Alternatively, an ansatz called the  ``replica-trick''
can be used (see, e.g.,~\cite{Hertz:1985, Grinstein:1985, UZ:1987}).
In perturbative
and perturbative-like investigations as, for example RG, this trick should
always give the same results as the method of direct perturbation treatment
of disorder.  This fact has a simple explanation.  The replica index
($m$) appears in the perturbation
terms through polynomials and, hence, there is no problem in the  analytical
continuation of the results in the limit $m \rightarrow 0$. The limit
must be taken at the end of the replica-trick
calculation in order to retrieve the original disordered
system~\cite{Hertz:1985, Grinstein:1985, UZ:1987}. In some non-perturbative
calculations, however, the same analytical continuation and
the justification of the limit $m \rightarrow 0$ are not very easy;
see, e.g., Ref.~\cite{Hertz:1985,Grinstein:1985, UZ:1987}. A recent theory of
 of the thermodynamic behaviour of disordered superfluids
below the finite-temperature critical point $(T_c > 0)$ has been developed by
Lopatin and Vinokur~\cite{Lopatin:2002} on the basis of the Beliaev
 theory~\cite{Beliaev1:1958,Beliaev2:1958} and the standard replica trick
treatment of $\delta-$correlated random impurities.

The results reviewed in the remainder of this paper are obtained by RG
and provide an information about the asymptotic critical behaviour, namely, the
critical behaviour in an infinitesimally close vicinity of the critical
point.  Note, that less information is available about the MF behaviour
of systems with quenched disorder and the pre-asymptotic critical behaviour.

{\em 7.2. Results for classical and quantum models at finite temperatures}

The RG investigation of quantum effective Hamiltonians (27)--(29) with the
additional term (88)
for all possible values of $m, m'$ and $\sigma$ was published for the first
time in Refs.~\cite{Kor:1983,Kor:1984}.  Here we shall review the results from
Refs.~\cite{Kor:1983,Kor:1984} together with
the necessary information from other papers; see,
e.g., Refs.~\cite{Ma:1976,Hertz:1985,Grinstein:1985, UZ:1987}.

{\em 7.2.1. Some results for classical systems}

In systems with symmetry index $n > 4$ the disorder effects
are irrelevant to the critical behaviour at finite temperature
critical points ($T_c > 0$).  The finite temperature critical behaviour
of systems
with symmetry indices $n > 4$ is described by the usual Heisenberg FP
of the corresponding pure systems
$(\Delta = 0)$ (for $n =1$ this FP is usually called Ising FP)
~\cite{UZ:1993,Ma:1976,Bin:1993}.  Here we shall often refer to this FP as
``pure" FP or, shortly, PFP and even P.  Note, that PFP exists also for
some values $n < 4$ but for such systems this FP is unstable towards disorder
($\Delta > 0$).

For systems with $n < 4$ the disorder effects essentially influence
the critical behaviour at finite critical points $(T_c > 0)$;
see.  e.g., Ref.~\cite{Ma:1976}.  In this case the critical behaviour is
described by a``random'' FP established for the first time by
Lubensky~\cite{Lub:1975};
hereafter referred to as RFP, or Lubensky RFP, or shortly denoted by ``R''.

The effect of the random impurities on the critical behaviour in
Ising-like systems ($n = 1$) is quite peculiar (see,  e.g., Refs.
~\cite{Ma:1976, Grinstein:1985}).  The RG equations exhibit a special form of
degeneration and, as a consequence, the usual perturbation expansion
breaks down and should be conveniently modified.  For example, the usual
$\epsilon-$expansion should be substituted with an expansion in non-integer
powers $(\sim \epsilon^{p/2}\:,\; p=1,2,...)$ of $\epsilon = (d_U - d)$.

The modified $\epsilon-$expansion yields  special FP which attracts the RG
flows through spiral paths.  Such FPs are often called ``focuses"" or focal
FPs.  These FPs give complex values for the correction-to-scaling
exponents~\cite{Ma:1976} which result in an oscillatory behaviour of the
corrections to the scaling laws.  Perhaps, this
extraordinary aspect of the impure critical behaviour
is a feature of systems where
an anisotropic ordering appears below the critical temperature or other
anisotropy effects are present, as shown
in Ref.~\cite{LMU:1987} for a wide class of anisotropic impure systems.  We
shall not dwell on the case $n = 1$
(for this case, see, e.g., Refs.~\cite{UZ:1993,Ma:1976})
rather we shall discuss systems with
continuous symmetry $(n > 1)$.  Let us, however, mention that several
FPs of focal type are discussed in Secs.~7.4 and 7.5 for other reasons
 which are associated with the effective increase of the upper critical
 dimensionality $d^R_U$ of the disordered system $(d^R_U > d_U)$.

{\em 7.2.2. Quantum systems at finite temperature}

The picture outlined in
Sec.~7.2.1 for the asymptotic critical behaviour
of classical systems does not change for quantum systems at finite critical
points $(T_c > 0)$.  The additional information which
can be obtained for the finite temperature critical behaviour
of disordered quantum systems is about the quantum critical dynamics,
in particular, about the value of dynamical critical exponent $z$.
The results~\cite{Kor:1983,Kor:1984} for
the dynamical critical exponent $z$ of finite temperature
critical points can be written in the form:
\begin{equation}
z = \sigma + \frac{(4-n)}{8(n-1)}\epsilon \;, \;\;\;\;\;\; 1 < n < 4 \;,
\end{equation}
where $\epsilon = (4 - d)$.  Eq.~(91) has been obtained for the $XY$ model and
Bose systems
which are described by the bare correlation function (28).  In the case of
models that have
Hamiltonians (27) with the correlation function (29),
the dynamical exponent has been obtained~\cite{Kor:1983,Kor:1984} in the form
\begin{equation}
z = \frac{\sigma}{m} + \frac{(4-n)}{8(n-1)}\epsilon \;,\;\;\;\; m^\prime = 0,
\;\;\;\;\;\; 1 < n < 4 \;.  \end{equation}
For $m^\prime > 0$ the exponent $z$ is given by Eq.~(36).
In the latter case the critical exponent $z$ has no correction to the first
order in $\epsilon$~\cite{Kor:1984}.  These results correspond to  RFP, namely,
when the disorder is relevant.  The same results~\cite{Kor:1984}
have been confirmed for disordered itinerant antiferromagnets
~\cite{Kirkpatrick1:1996,Kirkpatrick2:1997}

Eq.~(91) coincides with that obtained in Ref.~\cite{Gri:1977} for
disordered classical systems
within an approach based on a time dependent Landau--Ginzburg
equation~\cite{Ma:1976,HH:1977}.  The results (91) and (92) demonstrate that
the dynamical
critical exponent $z$ in systems with random impurities has an
 $\epsilon-$correction in the one--loop
(first order in $\epsilon = 4 - d$) approximation whereas the dynamical
exponent of
the corresponding pure system ($\Delta = 0$) has an $\epsilon$--correction
of order $O(\epsilon^2)$.
This is a direct consequence of the fact that the random function
$\varphi(\vec{k})$ does not depend on the Matsubara frequency $\omega_l$.

{\em 7.3. Instability of the quantum critical behaviour in disordered systems}

The zero-temperature
critical behaviour exhibits an
instability with respect to the quenched
disorder~\cite{Kor:1983,Kor:1984}.  The mechanism of the
instability is related to CQC.  Let us consider the RG equations
corresponding
to this problem~\cite{Kor:1984}.  In the limit
$T_c \rightarrow 0$ the parameters of the Hamiltonian obey the
lowest order RG transformations of the type~\cite{Kor:1983,Kor:1984}:
\begin{equation}
v^\prime \; = \; b^{2\sigma - d - z_0}v \;, \end{equation}
and
\begin{equation}
 \Delta^\prime \; = \; b^{2\sigma - d}\Delta \;,
 \end{equation}
where $z_0$ denotes the bare value $z(0)$ of the dynamical exponent
$z(\epsilon)$.  The "tree" approximation used in Eqs.~(93) and (94) is
sufficient
for our present discussion (the one-loop RG equations have been derived
in Ref.~\cite{Kor:1984}).

Because of
CQC the interaction parameter $v$ is relevant to the zero-temperature
critical behaviour in spatial dimensionalities
$d < (2\sigma - z_0)$, whereas the disorder parameter $\Delta$ is relevant to
$d < 2\sigma$.  This is readily seen from Eqs.~(93) and (94).
The
difference in the upper critical dimensionalities of the parametes $v$ and
$\Delta$ is produced by CQC and the lack of $\omega_l$--dependence of the
random function $\varphi(\vec{k})$.  On one side, at the classical borderline
dimensionality $d_U = 2\sigma$, the parameter $v$ is irrelevant and,
hence, the RG equations in $d = (4 - \epsilon)$ dimensions
describe a simple Gaussian instability towards the disorder parameter
$\Delta$.  On the other side, the zero-temperature
$\epsilon$--expansion in terms of $\epsilon_0 = (2\sigma - d - z_0)$ yields
pure FPs of the
type $\Delta^\ast = 0$ that are unstable with respect to the disorder
parameter $\Delta$~\cite{Kor:1984}.
In these two variants of the theory  FPs of corresponding pure system are
unstable with respect to disorder effects for dimensionalities
less than the upper borderline dimensionality $d_U$~\cite{Kor:1984}.

Moreover, the disorder itself does not generate new stable FPs in the quantum
limit ($T_c \rightarrow 0 $).  This is a
clear indication for the lack of a standard (pure or random) zero-temperature
critical behaviour and, therefore, one should expect either some unconventional
(multi)critical behaviour in these systems at zero temperature
or a first--order phase transition.

\begin{figure}
\begin{center}
\epsfig{file=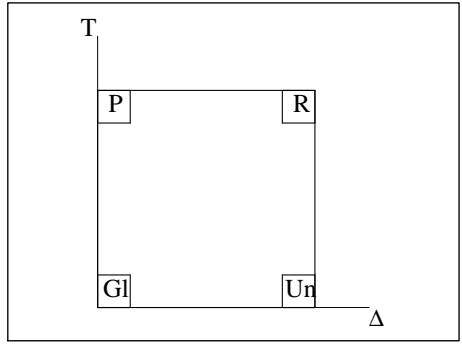, width=6cm}
\end{center}
\caption{A scheme representing four relevant FPs (P, R, Gl, Un)
in disordered systems.}
\label{QPTf5.fig}
\end{figure}

The zero-temperature and, perhaps, low-temperature instability of
 the standard critical behaviour in pure systems against disorder effects of
 type quenched impurities is probably related with the instability of the
 usual ground state of the respective system and, hence, with localization
 effects in real disordered superfluids (see Sec.~7.7.3). Therefore,
 it seems important to have a good notion about the relevant FPs which
 govern the impure quantum systems.
Using the results discussed up to now we can distinguish between four types
of critical behaviour of pure and disordered
Bose fluids and XY systems described by the
Eqs.~(27),~(28), and (90).  They are given by four FPs:  (i) Pure FP (P)
 corresponding to $T_c >
0$ and $\Delta = 0$ (lack of disorder), (ii) Gaussian-like FP (Gl),
corresponding to $T_c = 0$ and $\Delta = 0$, Random FP (R) - for $T_c >0$ and
$\Delta > 0$, and Unstable FP (Un) corresponding to the case of
zero-temperature critical point $(T_c = 0)$ of disordered system $\Delta > 0$.
These main four FPs are shown in Fig.~5 at their places in the parameter
space $(\Delta,T)$.

For other quantum systems, for which  Eq.~(29) should
be used instead of Eq.~(28), the diagram in Fig.~5 remains valid, provided the
Gl FP is substituted with the usual zero-temperature
FP corresponding to
CQC in pure systems~\cite{Hertz:1976}.  Thus we have two types of instability
of the quantum critical behaviour towards random impurities:
(a) instability of the quantum critical non-universal behaviour represented
by GlFP, and (b) instability of the quantum critical universal behaviour
described by the universality PFP (see Sec.~5.4).

The effect of quenched (random) inpurities on the quantum tricritical
behaviour has been also investigated~\cite{BCU:1984}.  The RG results
demonstrate an instability of the pure quantum critical behaviour towards
the disorder and lack of any new stable FP of the RG
equations~\cite{BCU:1984}.  Therefore, the usual critical and
tricritical points exhibit an instability due to  CQC which takes place in
the pure subsystem (the subsystem described by $\Delta = 0$) but does not
affect the ``disorder subsystem," described by the Hamiltonian part~(90).

The random field disorder is itself a source of a dimensional
crossover~\cite{UZ:1993, Ma:1976, Hertz:1985,Grinstein:1985}.  This is another
type of quenched disorder that may occur in real experiments.
It has been
shown~\cite{UKM1:1985,UKM2:1985} that the outcome of the simultaneous action
of quantum and random field effects at zero temperature is again an
instability of the quantum critical behaviour below the upper
borderline dimensionality (the latter dimensionality is higher than $2\sigma$
for the presence of short- or long-range correlations of the random
fields~\cite{UKM1:1985,UKM2:1985}).

{\em 7.4. Long-range random correlations and extended impurities}

{\em 7.4.1. Long-range random correlations}.

 The quantum critical behaviour is
unstable also towards disorder effects produced by random impurities with
long-range random correlations considered for the first time by
Weinrib and Halperin~\cite{WH:1983} (for the calculation
of the critical exponent $\eta$ in this case, see Ref.~\cite{KorU:1984}).  The
quenched disorder is described again by the Hamiltonian part (90) but the
correlation function $g(R)$ is given by Eq.~(88).

The Fourier transform of the
random correlation function (88), to the leading order in the wave vector
$\vec{k}$, has the form $g(k) = \Delta_1 > 0$ for $\Theta \geq d$, and
\begin{equation}
g(k) = \Delta_2k^{\Theta -d} \end{equation}
for $\Theta < d$.  These two cases have been combined in Ref.~\cite{WH:1983},
where the RG analysis is carried out by the correlation function
\begin{equation} g(k) = \Delta_1 + \Delta_2k^{\Theta -d}\:, \end{equation}
which contains both short-range ($\Delta_2 = 0$) and long-range ($\Delta_1 =
0,\Theta < d$) random correlations between the quenched impurities.

For classical models
$(\omega_l = 0)$~\cite{WH:1983} the RG analysis gives  new stable FP of a
focal type with complex eigenvalues, leading to oscillating corrections to
scaling laws and consistent with the Harris criterion~\cite{Harris:1975}.
Obviously, the same features will remain valid for quantum
systems at finite critical temperatures $T_c > 0$.
As there is no spatial anisotropy in this model and, moreover,
 no anisotropy is present at all in the $n-$ dimensional space of the
 order parameter vector, we are faced with an example when  focal FP,
 describing oscillating corrections to the scaling laws can appear also in
 cases of a total lack of anisotropy. The only reason, for which such a focal
 FP appears in this class of systems is the disorder-driven
 dimensional crossover associated
with the upper borderline dimensionality $(d_U^R > d_U)$ as explained
 in more details at a next stage of our consideration.

Let us consider why the classical impure behaviour is proven to be  stable by
RG investigations, at least for a rather large class of
systems~\cite{WH:1983}.  This is a result
of an ansatz, by which a double $\epsilon-$expansion is introduced in
the RG  studies~\cite{WH:1983}. However, without this ansatz, namely,
by following the
 standard RG scheme, one should obtain an instability of the critical
 behaviour of the impure system as a result of a lack of any stable FPs of the
 RG equations in the nontrivial dimensionality range $ d_L^R < d < d_U^R $.

 The reason
 for the instability is in the dominant role of the disorder parameter
 $\Delta_2$ associated with the long-range random correlations. The presence
 of long-range random correlations simply quells the the fluctuation
 interaction effects represented by the parameter $v$ as well as the
 short-range random correlations, represented by the parameter $\Delta_1$.
 The behaviour of the system very much resembles the behaviour of
 noninteracting fluctuations (free field; $v=0$) in a random potential.

In order to clarify this problem one may perform a simple dimensional
analysis~\cite{UZ:1993, Bin:1993} of the Hamiltonian,
or, equivalently, to perform the RG
rescaling within the simple ``tree" approximation.  The result of the RG
rescaling procedure is
\begin{equation}
v' = b^{2\sigma - d}v, \;\;\;\;\;\;\; \Delta_1 = b^{2\sigma - d}\Delta_1 \:,
\end{equation}
and
\begin{equation} \Delta_2 = b^{2\sigma - \Theta}\Delta_2\:.  \end{equation}
If the difference $(d - \Theta) > 0$ is of order unity, as  is for real
systems, an instability of the classical critical behaviour corresponding to
the pure system ($\Delta_1 = \Delta_2 = 0$) will occur and,
moreover, a new stable critical behaviour will not appear at all.
 The reason is that the parameter $\Delta_2$ will always be relevant and
 the perturbation contributions which are supposed to be of order
 $\epsilon$ cannot compensate the growth of this parameter to infinity.

 The ansatz introduced for the first time in Ref.~\cite{WH:1983}
for this type of systems consists in the assumption that,  at
 least during the formal derivation and analysis of the RG equations,
$(2\sigma - \Theta) = \delta$ will be a small quantity of order
 $\epsilon = (2\sigma - d)$; remember, that the upper critical dimensionality
 of the pure system is $d_U=2\sigma$. Then one performs a double
 $(\epsilon,\delta)$ which yields a new stable FP of focal
 type~\cite{WH:1983} and complex exponents for the correction-to-scaling
laws which describe
 the critical behaviour of systems with long-range random correlations.

The formal assumption that
 $ \delta \equiv (2\sigma - \Theta) \ll 1$ makes possible to develop
  a double $\epsilon-$expansion but the form of Eq.~(98) does not allow
 an easy way to obtain the upper borderline dimensionality $d_U^R$ of
 the disordered system, which seems to be higher than that ($d_U = 2\sigma$)
 of the pure system on account of the particular effect of the parameter
$\Delta_2$.
 This is a peculiar feature of the above described approach because usually
the tree  approximation immediately yields the upper critical dimensionality.
 The reason is in the lack of a $d$-dependence in the
 scaling transformation~(98). Thus one is left with the opportunity to
determine the dimensionality $d_U^R$ from the singularities of the
 one-loop integrals rather than from the lowest order tree approximation.
 Therefore, one begins with the derivation of the
 RG equations for a general dimensionality $d$
 and only after that one has to determine the critical
 dimensionality $d_U^R$, at which RG equations are to be expanded.

The formal comparison of the scaling ($b-$) factors in Eqs.~(97) and (98)
 shows that $(2\sigma - \Theta) = [\epsilon - (d-\Theta)]$ and, hence, one may
 speculate that the upper dimensionality $d_U^R$ of the disordered system
 is raised with respect to $d_U$ by $(d-\Theta)$ but this argument,
 although correct (see below), cannot be used to determine $d_U^R$ as a
 number independent of $d$. The reason for this difficulty is in the
 choice of the exponent $\Theta$ that decribes the $R-$dependence in Eq.~(88).
 If we use the exponent $\Theta^\prime = (d - \Theta)$, which has been
 already introduced in our analysis, the Eqs.~(88), (95), (96), and (98) will
 change in an obvious way. For example, the $k-$dependence in the second term
 in the r.h.s. of Eq.~(96) will be written  as $(1/k^{\Theta^\prime})$,
 and the $b-$ factor in Eq.~(98) will take the form
 $b^{(2\sigma +\Theta^\prime - d)}$. Now we can be certain that $d_U^R$ is
exactly equal to ($2\sigma + \Theta^\prime$). The difference $(d_U^L - d_U)$ is
 equal to $\Theta^\prime = (d - \Theta)= (\delta - \epsilon) > 0$. Note,
 that the focal FP revealed in Ref.~\cite{WH:1983} has a singularity just at
$\delta = \epsilon$.

The main result of the RG analysis within the double $\epsilon-$expansion --
the stabilization of the quantum critical behaviour in disordered systems of
special type, has been achieved by an ansatz which does not affect the
$(1/T)-$axis of the total $(\tau,\vec{x})$ space of the critical events.  It is
easy to see that CQC in the limit $(T \rightarrow 0)$ will cause
an instability of all available finite temperature FPs in these systems.

{\em 7.4.2. Extended impurities}

Up to now we have
discussed random ``point" impurities, i.e.  impurities localized at
 single points
$\vec{r}$.  The disorder of type extended (non-point) random impurities is
described by infinitely-ranged random correlations along one or more spatial
dimensions, say, $\bar{d} < d$, and short--range random correlations along
the other $(d - \bar{d})$
dimensions~\cite{Dor1:1980,Dor2:1980, BK:1982,Stolen:1984}.

For example, in systems with a slab geometry the appropriate
choice of extended impurities is the one--dimensional (linear) impurities
oriented along the direction of the small size $L_{0}$ (the slab thickness)
and randomly distributed along the other spatial directions~\cite{Craco:1999}.
The extended quenched impurities are described by a modification of the
model for short- and long-range correlated point impurities,
in which these finite-range correlations along the ``small size''
$L_0$, for instance, along the $z-$axis, are substituted with infinite--range
correlations.  The random correlations along the long dimensions of the slab,
i.e.,  along the $x-$ and $y-$axes, are kept of the short-range type.
The length scale of ``the infinite-range" correlations is much larger than the
correlation length $\xi$ and the thickness $L_{0}$ of the slab.  So,
the strongly correlated along
the small size point impurities behave like continuous uniform strings.  In
regard to the critical behaviour this disorder acts like point impurities with
a short--range random distribution along the large (infinite) dimensions
$L_{i}$ and an uniform distribution along the small size
$L_{0}$~\cite{Craco:1999}.

In infinite systems one may consider:  (i) straight lines of impurities or
straight dislocation lines of random orientation, (ii) two-dimensional
(plane, $\bar{d}= 2$) defects, and finally, (iii) randomly distributed
impurities or defects of dimensionality $\bar{d} < d$.

Now the important question is about the way of description of infinite-range
correlations.  This is simple because one  takes the exponent $\Theta$
of the correlation function (88) to tend to
zero $(\Theta \rightarrow 0)$ along the directions
in the $\bar{d}-$dimensional subspace of extended impurities.  Then
the correlations along this directions will be of infinite range but uniform.
The body will be homogeneous (aligned) along this directions and the disorder
will be present only along the rest part of spatial directions
$(d - \bar{d})$.  Now we can suppose either short or long range random
correlations along these directions and let us choose, as in many papers, the
former opportunity.  Then we shall have no defects (ideal order) along chosen
axes of number $\bar{d}$ and a short-range disorder of type random impurities
along the rest $(d-\bar{d})$ axes of the $d-$dimensional coordinate
space.  The respective correlation function of type (86) is given in the
form~\cite{Dor1:1980,Dor2:1980,BK:1982,Stolen:1984}
\begin{equation}
g(\vec r, \vec r\:^\prime) = \bar{\Delta}\delta^{d - \bar{d}}(\vec r - \vec
r\:^{\prime})\:.  \end{equation}
The case when $d = 2$ and $\bar{d}=1$ corresponds to the exactly solvable
McCoy-Wu lattice model~\cite{McCoy1:1968, McCoy2:1968, McCoy3:1973} of
disorder along one of the spatial directions for the two-dimensional
Ising model. We must emphasize that the mentioned  correspondence is only in
some general features.

In all cases the
extended impurities will generate a critical behaviour instability for
dimensionalities less than $d < d^R_U = (2\sigma + \bar{d})$.  The mechanism
of this instability lies in the appearance of the finite
 shift $\bar{d} = (d_U^R - d_U)$ of the upper borderline dimensionality,
 i.e., the reason is similar to that producing instability by long-range
 random correlations (Sec.~7.4.2). In fact, the raised dimensionality
 $d_U^R$ corresponds to the disorder parameter $\bar{\Delta}$ and the
 critical dimensionality associated with the fluctuation interaction $v$
 remains the same as in the pure system ($d_U = 2\sigma$). This means that
 the disorder effect
represented by the parameter $\bar{\Delta}$ suppresses the fluctuation
 interaction $\sim v\phi^4$ in Eq.~(27) just like in the case of long-range
 random correlations described in Sec.~7.4.2 and, as a result,
 stable FP is lacking and the critical behaviour is unstable.

The instability is
overcomed again by an ansatz of
a suitably chosen double
$\epsilon-$expansion~\cite{Dor1:1980, Dor2:1980, BK:1982, Stolen:1984},
namely
under the
supposition that $\bar{d}$ is a small quantity $\bar{d} \sim \epsilon$ and
can be treated on the same footing as the expansion parameter $\epsilon =
(2\sigma - d)$.  Thus using $\bar{d} = \epsilon_d$ as
a second small parameter in the double $(\epsilon, \epsilon_d)-$expansion,
the RG
theory~\cite{Dor1:1980,Dor2:1980,BK:1982} gives a stable finite temperature
critical behaviour by stable FP of a focal type. This (Dorogovtsev)  FP is
different  from  focal FP discussed in Sec.~7.4.1 and describes the
critical  behaviour
with complex corrections corresponding to the extended impurities
 with a random distribution~(99). Despite of the number of similarities
 between the critical phenomena, from one hand, in systems with point
 impurities with long-range random correlations and, from the other hand,
 in systems with extended inpurities with short-range random correlations
 there are also essential differences. These differences are mainly in
  theoretical aspects, namely, in particular features which create
 methodical problems. From an experimental point of view, however,
 the effects of these types of disorder on the critical behaviour
 can hardly be distinguished.

 The stability of the critical behaviour in
 disordered systems with extended impurities vanishes when
the quantum limit $(T \rightarrow 0)$ is taken and CQC occurs.
 The reason is again in the appearance of CQC which produces
 an effective suppression of the fluctuation interaction represented by
 the parameter $v$.

{\em 7.4.3. Summary}

We can conclude that the quantum critical behaviour
is unstable against all relevant forms of disorder used in the theoretical
physics
for the description of real systems. The instability of the quantum critical
behaviour in pure systems with
respect to disorder effects of several types considered
in Secs.~7.3 and 7.4
is a result of dimensional CQC that suppresses the thermal fluctuation
interactions but does not affect the disorder.  The reason is simply in the
fact that the quenched disorder is considered as time $(\tau-)$
independent.

The similarity between dimensional
CQC and the dimensional shift $(d_U^R-d_U)$ of the upper borderline
 dimensionality established in certain types of disordered classical
systems can be used in the development of a unified approach to the
 description of dimensional crossover  phenomena in quantum systems.
It has been pointed out~\cite{BK:1982}
that the related with the time $(\tau-)$ extra dimensionality
of the quantum Ising model can be supposed throughout
a double $(\epsilon, \epsilon_d)-$expansion
equal to $\epsilon_d$ in a complete analogy with the case of
linear random impurities in classical models.

{\em 7.5. The stability ansatz about the quantum criticality in impure systems}

The idea of Boyanovsky and Cardy~\cite{BK:1982} mentioned at the end of
Sec.~7.5, has been applied by Weichman and Kim~\cite{WeiKim:1989} for solving
the problem for the instability of the
critical behaviour at $T = 0$.  In Ref.~\cite{WeiKim:1989}
disordered superfluids with random impurities have been investigated.
These systems are described by the sum the Hamiltonian parts
(27) and (88), and the bare Green function, given by Eq.~(28).  But
the results in Ref.~\cite{WeiKim:1989} can be applied
beyond the model of disordered Bose fluids and include other
quantum systems.  Note, that an artificial upper cutoff $\Lambda_\omega$
for the Matsubata frequencies $\omega_l$ has been used in this
investigation~\cite{WeiKim:1989} but this cutoff is
avoided in a later paper~\cite{Schak2:1997}.

Weichman and Kim~\cite{WeiKim:1989} considered linear impurities along the
time $(\tau-)$ axis instead of the spatial ones.  Thus the impure
strings along the time direction
are of size approximately given by $\beta = 1/T$.
For $\lambda \sim 1/\sqrt{T} \rightarrow 0 $ (for simplicity, here we set
$\theta = 1/2$), when the
temperature time ($\tau-$) dimension reduces to zero and one can take the
continuum limit along the ``$\beta-$axis", these lines shrink to randomly
distributed points in the $d$--dimensional volume of the system.
For $T \rightarrow 0 $, however, the ``length'' $\beta $ of these lines tends
to infinity.
Note, that the consideration based on the exponent
 $\theta = 1/2$ corresponds to a dynamical exponent $z = 2$ and, hence,
 to all systems described by the Hamiltonian (27) with
 exponents $\sigma $, $m^\prime $, and $ m $, that satisfy  Eq.~(36)
for $z = 2$. But the genaral features of this considerations are valid for all
 systems described by Eqs.~(27), (28), and (29) and the respective disordered
systems given by the Hamiltonian (90).

In fact, within the framework of the problem,
considered in Ref.~\cite{WeiKim:1989} and later in Ref.~\cite{Schak2:1997},
the dynamical critical exponent $z$ is equal to two and, hence,
the term ``lines'' should be used, if we consider a (d + 1)--dimensional
space--time($\tau $).  However, if we consider this space--time
representation
equivalent to $(d + 2)$ dimensional space, as is in CQC dimensional crossover
for Bose fluids, we should stretch the
``lines'' (or strings) of ``length'' $\beta \sim \lambda^2$ in planes
of area $\lambda \times \lambda$.  In general, we shall have hyperplanes of
dimensionality
$z$ and ``area'' $ \sim\lambda^z$.  For $\lambda \rightarrow 0$ the randomly
distributed objects will
shrink to randomly distributed point impurities.  These are two different
geometrical representations of the same phenomenon.
The formal analysis can be performed
with the additional small parameter $\epsilon^{\prime}_{d}$ which runs from
zero to unity and
thus describes the classical ($\epsilon^\prime_{d} = 0$) and the quantum
($\epsilon^\prime_{d} = z$) limits of
Bose system with point impurities.  Alternatively one may consider the small
parameter
$\epsilon_d = \epsilon^\prime_{d}/z$~\cite{WeiKim:1989,Schak2:1997}. The double
$\epsilon-$expansion of the RG equations again yields  stable focal FP.

In result, the following interesting picture
of the Bose systems critical behaviour is obtained~\cite{WeiKim:1989}.
In the high-temperature range, where
$\lambda < a$, the critical behaviour is described
by the high-temperature RFP~\cite{Lub:1975} corresponding to
short--range (point--like) impurities,
i.e, Lubensky RFP (see also our notations defined in Sec.~7.2.1).
In the zero-temperature limit ($T \rightarrow 0$), the impure critical
behaviour described by  Lubensky RFP undergoes
a crossover to a critical behaviour
governed by extended (plane) impurities.  This zero-temperature
critical behaviour is  described by the corresponding zero-temperature
variant of Dorogovtsev RFP with complex stability
exponents~\cite{Dor1:1980,Dor2:1980}. In the Hamiltonian parameter space
$(\beta, v, \Delta)$, Lubensky RFP is
conjugate to Heisenberg PFP $(\Delta^{\ast} = 0)$ and both of them
are far from the
zero-temperature plane $\beta = \infty$. Dorogovtsev RFP is conjugate
to GlFP $(\Delta_{Gl} = T_{Gl} = 0)$ considered
in details in Section~5.3.  Like Heisenberg PFP, GlFP is stable only for
pure Bose fluids (for real bosons, $n = 2$).
For any $\Delta > 0$ and $T > 0$, the RG
flows tend to the finite-temperature Lubensky RFP,
whereas the zero-temperature Dorogovtsev RFP will be attainable, if only
 the system is at $T = 0$.

These four conjugate FPs describe the critical phenomena
in disordered superfluids and all disordered quantum systems presented by
the Hamiltonian given by Eqs.~(27), (28), and (90).
 One may visualize this scheme by a picture similar to that given in Fig.~5.

For the rest part of
disordered quantum systems, namely, those described by the bare Green function
 from Eq.~(29), the same scheme works again provided one substitutes the
 nonniversality GlFP with the respective universality PFP
 (see also Sec.~5.4). Then the zero-temperature pure
 critical behaviour obeys the exact
universality rule (Sec.~4), that is, the $d$--dimensional
zero-temperature critical behaviour is described
exactly by the finite-temperature universality class $(d + z, n)$,
to which the system belongs (see also Sec.~5.4). The same is valid for
the respective disordered systems. In this case, the disordered quantum
 system enters in the high-temperature universality class $(d + z, n)$ of the
 respective classical system $(\omega_l = 0)$.

{\em 7.6. Related problems}

{\em 7.6.1. Disordered ferro- and antiferromagnets}

An approach identical
 to the Weichman-Kim approach~\cite{WeiKim:1989}
 (Sec.~7.5) has been used in an investigation of  the quantum phase transition
in disordered itinerant quantum antiferromagnets~\cite{Kirkpatrick1:1996,
Kirkpatrick2:1997}. In the notations given by Eqs.~(27)--(28)
these systems correspond to Eq.~(29) with
$m^\prime = 0, m = 1$. Note, that in Ref.~\cite{Kirkpatrick1:1996}
 the result (92) for the dynamical critical exponent $z$ has been generalized
with the help of $(\epsilon, \epsilon_d)$--expansion.

The picture of the quantum phase transition properties
 of pure and disordered itinerant ferromagnets given in
the theoretical treatment~\cite{BK:1996, KB2:1996,VBNK:1997} is worth
mentioning.  The energy spectrum
$\varepsilon (k)$ of the fluctuation mode (the magnetization) in the effective
 field Hamiltonian depends on the spatial dimensionality $d$:
$\varepsilon (k) \sim k^{(d-2)}$.
 The quenched disorder of type random impurities
 produces a diffusive electron dynamics which induces an effective long-range
 electron spin interaction of the form $1/R^{2(1-d)}$ and this gives
 a $d-$dependent spectrum of the type $1/k^{(2-d)}$. The bare Green function
 $G(q)$ of the same magnetization fluctuating modes is described with the
 help of the terms given by Eq.~(29) for $m^\prime = 2$, $m = 1$, and
 $\sigma = \mbox{min}(2,d-2)$ and an additional term of type $k^2$ which
 is often redundant. The $d-$dependence of the energy spectrum of
 the fluctuating order parameter leads to a completely new, and in some
 aspect, quite non-universal picture of the critical behaviour
 in the most interesting domain of spatial dimensionalities $(d \leq 4)$.

In order to evaluate this result one should be acquainted with the
 behaviour of the respective pure system, where the disorder is not present.
 By the same method of treatment - a derivation of an effective field
Hamiltonian  from the microscopic electronic Hamiltonian with the help of
the Hubbard-Stratonovich transformation and by a
 generalization of the Hertz~\cite{Hertz:1976} approach the respective
 energy spectrum of the critical magnetic modes
 in pure itinerant ferromagnets was found again to be
 $d-$dependent~\cite{VBNK:1997}.
Now the bare Green function $G_0(q)$ will be given by Eq.~(29), provided
 the exponents have the values $m^\prime =1$, $m=1$, $\sigma = 2$ and,
 moreover, an additional term of $c^\prime k^{(d-1)}$ with $c^\prime > 0$
 is added to the r.h.s. of the same expression for $G_0(q)$. As a result, a new
 universality class of critical behaviour arises at dimensionalities
 $ d \leq 3$. This problem has a further development in
 Ref.~\cite{Belitz:2002} with the conclusion for a fluctuation-induced first
 order phase transition at dimensionalities $ d \leq 3$.

 For details about the motivation and
 the results of this challenging direction of research,
 see the original papers~\cite{BK:1996, KB2:1996,VBNK:1997}. One should
 have in mind that the Hubbard-Stratonovich transformations as a general
 method of establishing a correspondence between a microscopically
 formulated Hamiltonian and its effective field (quasi-macroscopic)
 counterpart have shortcomings unless one is interested only on
 the asymptotical long-wavelength limit (see, e.g.,
 Refs.~\cite{UZ:1996, Brout:1974}). On the other hand let us mention that
the integration out of critical or auxiliary-to-critical modes in complex
 models almost always generates $d-$dependent terms in the spectrum of the
principal fluctuating mode describing the phase transition
 of interest to the particular study. In usual cases of
 effective field Hamiltonians with more than one fluctuating
 fields this procedure of a further reduction of the description is
 rarely correct.

A density matrix RG theory~\cite{Peschel:1999} was applied to
 the one-dimensional (chain)
 random-exchange spin-1/2 XXZ  model~\cite{Hamacher:2002}. The interplay
 of quantum fluctuations and disorder results in a disorder-induced
 quantum phase transition that exhibits a nonuniversal behaviour
 of the spin correlations.

{\em 7.6.2. Disorder in quantum Hall liquids}

 A RG investigation of the
 effect of point random impurities with short-range correlations of
 FQHE systems was done by Schakel~\cite{Schak2:1997} (see, also,
 Ref.~\cite{Sch:1999}). The effective field CSGL theory, discussed in
 Sec.~5.7.2, has been considered in Ref.~\cite{Schak2:1997} in a reduced
 variant, in which the gauge fields are omitted and the Bose field $\phi(q)$,
 which interacts only with the random mode $\varphi(\vec k)$
 subjected to $\delta-$correlations
 is described by Eqs.~(27), (28), (86), (89) and (90). In this case one
 retrieves the instability~\cite{Kor:1984} (Sec.~7.3) and in order to
 ensure a delocalization transition, one is forced to look for
a stabilization of the ground state. Note, that the gauge fields
(the Chern-Simons field and the vector potential of the magnetic field) cause
 other fluctuation effects (see, e.g.,
 Ref.~\cite{UZ:1993, Halperin:1974, FSU:2001}) that additionally lead
to an instability of  RG FPs. Thus one cannot easily distinguish
 between the effects producing the instability of the quantum critical
 behaviour within the framework of the complete CSGL theory.
 Note, that the reduction to a $\phi^4$ theory makes possible
 the investigation of the net effect of the disorder on  FQHE.

In order to restore the stability of the quantum critical behaviour in this
 disordered system, Schakel~\cite{Schak2:1997} has proposed a method
 which departs from the Weichman-Kim
 approach~\cite{WeiKim:1989} (see also Sec.~7.5). The main disadvantage of the
Weichman-Kim treatment, as mentioned in Ref.~\cite{Schak2:1997}, is the
 introduced by these authors an upper cutoff for the Matsubara frequencies.
 Note, that such a cutoff, after the paper of
 Hertz~\cite{Hertz:1976}, is used by a number of authors. On the one hand,
 this cutoff is considered as irrelevant for the asymptotic critical behaviour
which is governed by low-frequency fluctuation modes, but on the other hand,
 as mentioned in Ref.~\cite{Schak2:1997}, the same cutoff is difficult to
 justify as it would imply a notion for a ``discrete'' rather than continuous
imaginary time $(\tau-)$ variable.

In the Schakel approach the frequency
 cutoff is avoided and the convergence of the frequency sums (35) in
 the perturbation terms is saved by introducing a noninteger imaginary
 time $(\tau-)$ dimensionality $\epsilon_d$: $0 \leq \epsilon_d \leq 1$. The
 value $\epsilon_d = 1$ describes the quantum limit ($T \rightarrow 0$),
 when the imaginary time $\tau \in [0,1/T]$ runs over the values
 in the whole interval
 $[0, \infty]$. In this limit one can apply the standard rule (35).
 When $\epsilon_d = 0$  we deal with the
respective classical system where the temperature is sufficiently high
 and the $\tau-$dependence of the fluctuation field $\phi(x)$ can be
 safely ignored. The intermediate values $0 < \epsilon_d <1$ simulate
 a form of an approximate interpolation between the two limiting cases.

The respective double $(\epsilon, \epsilon_d)-$expansion reveals  stable FP
 corresponding to the quantum regime $(T =0)$ of the
 disordered system (for details, see Ref.~\cite{Schak2:1997}). In some aspects
this FP is similar to that found by Weichman and Kim~\cite{WeiKim:1989}.

 The Schakel approach works also for pure systems. In this case, the
 variations of the dimensionality $\epsilon_d$ can be used to describe the
 classical-to-quantum crossover of the critical behaviour. Note, that
 this interpolation procedure is rather approximate, and the
 description is exact only for the asymptotic cases ($\epsilon_d = 0, 1$).
 The approximation is a result of the substitution of the frequency sums in
the perturbative terms by $\epsilon_d-$dimensional frequency integrations.
This problem is important for all studies based on double
 $\epsilon-$ expansions and has been investigated in details in
 Ref.~\cite{Craco:1999} (see also Sec.~7.6.3).
 Perhaps, this approximation is one of the reasons
 for the singularity separating the quantum and classical regimes within
 the Schakel approach~\cite{Schak2:1997}.

 {\em 7.6.3. Thin films with quenched impurities}

 The formal conformity
 between  CQC in quantum systems and the finite size crossover in
 systems of slab geometry~\cite{Fisher:1972,Barber:1983,USUZ:1994,
USUZ:1995} mentioned
 in Sec.~4.3 as well as ideas discussed in Secs.~7.4-7.5 have
 been used in Ref.~\cite{Craco:1999} in the study of disordered thin
 films with quenched point and extended (line) impurities.
 The method of work is similar to  the Weichman-Kim~\cite{WeiKim:1989} and
 Schakel~\cite{Schak2:1997} methods. The investigation has been carried out
 by a double $\epsilon-$expansion and reveals stable FPs which describe
the critical behaviour of the disordered films.

 The second small parameter in the double
 $\epsilon-$expansion is the quantity $\delta \in [0,1]$ which comes from
 an ansatz, discussed also in Sec.~7.6.2, and is used to define a substitution
 of lattice sums over the wave vector components with a $\delta-$fold
 integrals. The ansatz is not
 new and defines a $\delta-$integration that has two slighty different
 variants as shown in Ref.~\cite{Craco:1999}.

The method of integration over wave vectors with noninteger dimensionality
 lies in the basis of all double $\epsilon-$expansions
 discussed in this review. In Ref.~\cite{Craco:1999} the two small
 parameters $\tilde{\epsilon}$ and $\delta$ have been chosen in the form:
$\tilde{\epsilon} = (4 - D_{\mbox{\scriptsize eff}})$ and
 $\delta = (D_{\mbox{\scriptsize eff}} - d)$, where $d_U = 4$,
 $D_{\mbox{\scriptsize eff}} = (d +\delta)$ is the effective
 (not the physical) spatial dimensionality of the slab (film). Obviously,
 the quantuty $D_{\mbox{\scriptsize eff}}$ should depend on
 the ratio $y = (L_0/\xi)$, where $L_0$ is the thickness of the film. This
 dependence  can be ascribed to $\delta$: $\delta(y)$. Thus the RG
 treatment in Ref.~\cite{Craco:1999} has been performed by a double
 $(\epsilon,\delta)-$expansion, where the small parameters are defined
 as deviations from the effective dimensionality
$D_{\mbox{\scriptsize eff}}$ of the system.

 As shown in Ref.~\cite{Craco:1999} the
 $\delta-$integration as a method of calculation of perturbation
 terms introduces a systematic error of a
 finite magnitude, except for the limiting cases $\delta = 0, 1$.
While the RG studies are related with infrared divergences this finite error
 produces only incorrect values of the FP coordinates but does not affect the
physically measurable quantities as the critical exponents. Since the
 precise location of the FP coordinates is important for specific
 intermediate RG
 calculations but is not important for the physical predictions,
 the RG results obtained so far by double $\epsilon-$expansions are reliable.

Furthermore, one may try to
find a functional relation between the noninteger dimensionality $\delta-$ and
the ratio $y = (L_0/\xi)$. Obviously,
 the dependence $\delta(y)$ should have the properties:
 $\delta(y) \rightarrow 0$ for $y \rightarrow 0$,
 and $\delta(y) \rightarrow 1$
for $y \rightarrow \infty $. The method used in Ref.~\cite{Craco:1999}
 confirms this properties but the attempt to obtain an analytical
 dependence $\delta(y)$ have not been successful. Moreover, the method
 does not provide a sufficient  accuracy of the results for the
 intermediate values $y \sim 1$  correspondong to $\delta \sim 0.5$.

The comparison of the RG results for disordered thin films
~\cite{Craco:1999} with the respective results for models of
 disordered quantum systems
 shows that there should be a formal equivalence between
 the universality features of the critical behaviour in thin films and that
 in quantum systems such as ferroelectrics and magnets described by TIM;
see Eq.~(29) for $\sigma = 2$, $m = 2$, and $m^\prime = 0$.
This correspondence is valid for both pure~\cite{Law1:1978} and
impure films~\cite{Craco:1999}, and is given by the formal replacement
 of the film thickness $L_0$ with $1/T$.

{\em 7.6.4. Superfluid-insulator and metal-insulator phase transitions}

The instability of the quantum critical behaviour with respect to  the
disorder  can be interpreted as a signal for phenomena of localization
~\cite{Huang:1992,Giamarchi:1988,Scalettar:1991}. Huang and
 Meng~\cite{Huang:1992} have shown by the methods of the pseudopotential
 and the Bogoliubov transformation that at zero temperature the random
 impurities can deplete the Bose condensate, though not completely.
 On the other hand, the random impurities generate an amount of normal
 fluid equal to a part of the condensate depletion.

A paper by Giamarchi and Schulz~\cite{Giamarchi:1988} also provides
 reliable theoretical
 predictions about localization-delocalization phase transitions in
 disordered superfluids. This paper is mainly devoted to the localization
 problem in one-dimensional electronic systems (metal-insulator
 transition~\cite{BK:1994, Abrahams:2001})
 but the analogy with the localization-delocalization problem
 (superfluid-insulator transition~\cite{FishW:1989,MHL:1986})
 in superfluids is also considered.
 A variant of RG is developed~\cite{Giamarchi:1988} that is
 convenient for the treatment of
 the simultaneous effect of  interelectron interactions and disorder. The
phase diagram of the one-dimensional metal is deduced from the stability
 properties of the RG equations.

 Remember that the interelectron interaction alone can produce
 a localization of electrons and this is the Mott
 localization. On the other hand, the disorder
 alone can induce the Anderson localization
 of the eigenstates. The interplay between these two localization mechanisms
 is a matter of an intensive research. The analogy between the
 superconducting-insulator transition in metals and the superfluid-insulator
 transition in Bose superfluids was discussed in Refs.~\cite{FishW:1989,
MHL:1986, Giamarchi:1988}. The superconducting-insulator phase
 transition in disordered thin films and wires was investigated in
 Refs.~\cite{Haviland:1989, Giordano:1989}.
 As noted in Ref.~\cite{FishW:1989} the outcome
 of the competition between the interparticle interactions and the disorder
 may result, under certain circumstances, in a gapless insulating
 ``Bose-glass'' phase. It is supposed that this may happen
 when the interparticle interaction is  sufficiently strong~\cite{FishW:1989}.

A generalization of the Halperin-Lubensky-Ma
 paper~\cite{Halperin:1974} for magnetic fluctuation effects on the order of
 the phase transition in pure superconductors was made for the
zero-temperature superconductor-insulator phase transition in
 one- and two-dimensional pure~\cite{FisherGrinstein:1988} and
 disordered~\cite{Girvin:1990} superconductors (a $\delta-$correlated disorder
 of type random impurities). It has been  found that
the zero-temperature phase transition is continuous and the system behaves
 like a normal metal up to the phase transition point (the resistence has
 a finite nonzero value at $T=0$). A recent theoretical
 study~\cite{Dalidovich:2002} of the superconductor-insulator transition
 in a two-dimensional array of Josephson junctions with random
 couplings demonstrates that the previously predicted
 Bose-glass phase~\cite{FishW:1989, Fisher:1990} is a metal state that has
a well defined zero-temperature limit for the conductivity.

In Ref.~\cite{Scalettar:1991} one-dimensional Bose system described by
a Hamiltonian of type (27) has been studied by Monte Carlo techniques with
the aim to clarify the competition between the strong interaction and the
random impurities. The model considered in
 Ref.~\cite{Scalettar:1991} is a lattice version of the field theory
 given by Eqs.~(27), (28), (89), and (90), only the
 random distribution function is somewhat different. The main result from
 this Monte Carlo investigation is the numerical justification of the
  Bose-glass phase proposed in Ref.~\cite{FishW:1989} and the prediction of
a new Anderson-type insulating phase (``Anderson glass'') in the weak
 coupling regime. The existence of two insulating phases
 of glassy type was proposed for the first time
 in Ref.~\cite{Giamarchi:1988}.

{\bf 8. Classification of quantum phase transitions}

Here we shall enumerate the main types of quantum phase transitions.
The general classification in two types:  first order (discontinuous) and
continuous phase transitions known from the classical theory remains valid
also for the zero temperature phase transitons.  Both basic types of phase
transitions are possible at zero temperature in accord with
the Nernst theorem.

We should have in mind that
the quantum effects affect more essentially the close vicinity of low- and
zero-temperature equilibrium points of first order phase transitions rather
than low- and zero-temperature critical points of continuous phase
transitions.  While the correlation length at critical points is infinite and
the classical fluctuations dominate at any finite $(T_c > 0)$ critical point,
at the equilibrium points of the first order transitions this length is finite
and
the quantum effects may compete the thermal fluctuations up to an infinitesimal
vicinity of the transition point.  The outcome of this competition between the
classical and quantum fluctuations depends on nonuniversal properties of
the system and should be checked for each particular investigation.

The low- and zero-temperature phase transitions may be either quantum or
classical ones,
depending on whether the quantum correlations have an essential effect
on the phase transition properties or the classical effects prevail.
General thermodynamic and scaling arguments
(Sec.~2) as well as statistical considerations (Secs.~3 and 6) show that
the zero temperature limit itself does not
guarantee a quantum effect on the zero-temperature phase transition.
This is valid both for discontinuous and continuous phase transitions.

Moreover, the zero
temperature phase transitions can be classified as $T-$driven and $X-$driven
transitions and the properties of these two types of transitions are
quite different.  Note, that there is a wide class of systems as, for example,
Bose fluids, which do not exhibit $X-$driven transitions (Secs.~3 and 5).

The continuous quantum phase transitions, i.e.  the quantum critical phenomena
can be classified in two large groups:

(1) universal, and

(2) nonuniversal.

If all universal characteristic features (critical exponents,
including the correction to scaling exponents) of a quantum system (model)
at spatial dimensionality $d$
are identical to the respective quantities characterizing the correspondent
classical system (model) at spatial dimensionalities $(d + z)$, where $z$ is
the dynamical critical exponent, then the quantum critical phenomena in this
quantum system are from the group (1).  The ``correspondent classical
model" of
a quantum field model of the type (27) is that given by setting all Matsubara
frequencies in Eq.~(27) equal to zero.  Therefore, the correspondent model is
the same (the usual $\phi^4$-theory) for
the whole set of quantum models defined by Eqs.~(27), (28), (29), and the
possible values of the exponents $m$, and $m^\prime$.
For quantum lattice models, one should
take the classical limit in the proper way in order to obtain the
correspondent classical model.  Within RG, the phenomena of the
type (1) are described by zero-temperature
FPs which are identical in properties to some
respective (conjugate) finite temperature FP and CQC is the only quantum
effect on the zero-temperature critical behaviour.

If the quantum critical phenomena are not in group (1) they belong to the
group (2).  The examples discussed in details in this review are the quantum
critical phenomena in Bose fluids and XY magnets,
described by the Gaussian-like FP (Secs.~5.3-5.4).

To the best our knowledge this classification scheme is published for the
first time in this review.  The nonuniversality of quantum critical phenomena
in Bose fluids and XY systems has been pointed out for the first time in
Ref.~\cite{Uzun:1982} and then applied in Ref.~\cite{MU:1983} for the
 interpretation of results for a particular problem of  structural phase
 transitions.  With the help of various convenient
 terms of the authors choice the phenomenon of nonuniversality is
 confirmed by researchers in this field as already mentioned in Sec.~5.1.

{\bf 9. Concluding remarks}

The following results and notes can be enumerated at the end of our review:

1.  The complete investigation of quantum phase transitions
naturally includes the phase
transitions at low- and extremely low temperatures and cannot be restricted
to zero-temperature phase transitions only.  The quantum phase transtions
are a selected part of
the low-temperature (including zero-temperature) phase transitions.

2.  The properties of $T-$driven and $X-$driven phase transitions at low and
zero temperature differ substantially from each other.
The quantum effects on the
phase transition properties are mainly exhibited at $X-$driven phase
transitions at extremely low or zero temperature.  There is, however, a
conformity between the low-temperature $T-$driven continuous  phase transitions
and the  high-temperature $X-$driven continuous phase transitions, as noticed
in  Ref.~\cite{DeCU:1999} and Sec.~2.4.

3. The first-order quantum phase transitions with sufficiently low  equilibrium
transition temperatures $T_{\mbox{\scriptsize eq}}$ have not been
 studied  enough although they occur more frequently in crystal bodies and
should be  drastically affected by the quantum fluctuations.

4. The phenomenological scaling theory of continuous quantum
 phase transitions enters in the class of theories of dimensional crossover
 and takes advantage from similar theories as, for example,
 the scaling theory of finite size systems. Note, that the
 crossover exponent $\nu_0$, introduced in Sec.~2.3 can be compared with
 similar exponents in alternative approaches to finite-size
 scaling and quantum criticality.

5.  Two types of crossover phenomena can be defined in the low-temperature
limit:  HLTC and CQC.  The former does
not necessarily include quantum effects on the critical behaviour.  CQC
is the product of quantum effects.

6. The performance of zero-temperature limit in the scaling equations
does not itself guarantee quantum effects on the phase  transition properties.
Rather a phase transition can be considered  quantum  provided
the general criterion (3), or, equivalently, the criterion (9) is satisfied
 in the transition region. Asymptotic quantum critical behaviour occurs
 unless the respective criterion, (3) or (9), is obeyed in the asymptotic
 vicinity of the critical point. This is quite strong requirement and may be
 fulfilled in certain systems with suitable nonuniversal parameters
only at their zero temperature (multi)critical points $(T_c=0)$. However, for
the finite experimental accuracy, apparent quantum effects may be observed
 also at sufficiently low-temperature but finite transition points $(T_c>0)$.

7.  The scaling behaviour of basic models of statistical physics does not
exhibit quantum effects in the zero temperature limit
but rather follows the criteria for classical critical behaviour.  This is
valid for certain regimes of performance of the phase transition (Secs.~3
and 6).

8.  The interacting Bose fluids and XY magnets exhibit a
nonuniversal quantum critical behaviour.  The RG
investigations indicate a break down
of a well definite form of universality in these
systems, i.e., the phenomenon of quantum universality breaking  takes place.

9.  A quantum non-universality occurs also in disordered quantum systems,
where it appears as an instability of the
zero-temperature critical behaviour towards all known
mechanisms of disorder. All RG predictions for the critical
properties of disordered systems at
finite temperatures break down at zero temperature where the dimensional
crossover lowers the critical dimensionality of the fluctuation
interactions.

10.  There is a formal analogy between three physically different types of
crossover phenomena:   CQC, the (finite-size) crossover from two-to-three
dimensional systems, and the crossover from point to extended random
impurutes in disordered systems.

11.  The available results about quantum phase transitions
are not enough to build up a satisfactory picture of these
 intriguing phenomena.

12.  In this review we have tried to show some of the outstanding problems and
present a unified and consistent framework of the available results.  It
has been shown that the understanding of the quantum phase transitions as
``zero-temperature phase transitions" is a restricted and incomplete point
of view.  Besides, the concept of CQC as a main result of the
quantum effects on the critical behaviour is also unsatisfactory
and should be complemented by the concept of the quantum universality breaking.

13.  A substantional interest from
both theoretical and practical points of view are the so-called $T-$driven
phase transitions in the ultra-low temperature scale ($T\sim 0$).  Their
description is an outstanding problem.  Among the unresolved problems
are the equation of state for quantum phase transitions, in particular, in
Bose fluids, where one should represent the behaviour of the system in terms
of the density of particles and the temperature in the quantum limit
$(T \rightarrow 0)$.

{\bf Acknowledgements}

One of us (D.I.U.) thanks the hospitality of the MPI-PKS-Dresden
 and ICTP-Triest during the work on this article.

\newpage

\vspace*{1cm}

{\bf Figure captures:}

Fig. 1. (a) Low temperature part of a critical line with zero temperature
critical point $T_c(X_0) = 0$.  The shaded part ($a-X_0-b$) of the transition
region, marked by the lines 1 and 2, corresponds to a low temperature classical
behaviour.  (b) Low temperature critical line with $X_0 = 0$;
domains ($a-X_0-b$) and $1-0-2$ coincide.

Fig.2. (a) The function $f(\varphi)$ for a standard second order phase
transition for: $r_0 = 1.0$ (curve 1), $r_0 = 0.5$ (2), $r_0 = -0.5$ (3),
 $r_0 = -1.0$ (4).
 (b) The function $f(\varphi)$ for the IBG at constant density
(the curves 1-4 correspond to the same values of $r_0$ as given for Fig.2a).

Fig.3. (a) A tree diagram denoting the interaction part of the Bose
Hamiltonian. (b) The compact self-energy diagram which is equal to zero in the
limit $T \rightarrow 0$ [the thick loop denotes the full (renormalized) Green
function $G(q)$]. (c) An example of a diagram from the perturbation series for
the interaction vertex which gives a zero contribution in the
zero temperature limit. (d) The infinite ladder series of diagrams which
yields the geometric progression (51).

Fig.4. (a) The graphical representation of Eq. (67) for $J=1$. (b) The
high- and low-temperature parts of the curve in Fig.4a with an indication of
 $T-$ and $\Gamma-$ transitions (see the text).

Fig.5.  A scheme representing four relevant FPs (P, R, Gl, Un)
in disordered systems.
\end{document}